\documentclass{jfm}

\usepackage[usenames,dvipsnames]{color}
\usepackage{graphicx}
\usepackage{natbib,bm}
\usepackage{epstopdf,epsfig}
\usepackage{adjustbox}
\usepackage{appendix,amsmath}

\usepackage{lscape}
\usepackage{rotfloat}
\usepackage{ulem}

\ifCUPmtlplainloaded \else
  \checkfont{eurm10}
  \iffontfound
    \IfFileExists{upmath.sty}
      {\typeout{^^JFound AMS Euler Roman fonts on the system,
                   using the 'upmath' package.^^J}%
       \usepackage{upmath}}
      {\typeout{^^JFound AMS Euler Roman fonts on the system, but you
                   dont seem to have the}%
       \typeout{'upmath' package installed. JFM.cls can take advantage
                 of these fonts,^^Jif you use 'upmath' package.^^J}%
      }
  \else
  \fi
\fi

\ifCUPmtlplainloaded \else
  \checkfont{msam10}
  \iffontfound
    \IfFileExists{amssymb.sty}
      {\typeout{^^JFound AMS Symbol fonts on the system, using the
                'amssymb' package.^^J}%
       \usepackage{amssymb}%
         \let\leq=\leqslant
         
      }{}
  \fi
\fi

\ifCUPmtlplainloaded \else
  \IfFileExists{amsbsy.sty}
    {\typeout{^^JFound the 'amsbsy' package on the system, using it.^^J}%
     \usepackage{amsbsy}}
    {}
\fi

\newcommand\pt{\partial}
\newcommand\la{\langle}
\newcommand\f{\frac}
\newcommand\ra{\rangle}

\newcommand{\ben}{\begin{enumerate}[1)]}
\newcommand{\een}{\end{enumerate}}
\newcommand{\be}{\begin{equation}}
\newcommand{\ee}{\end{equation}}
\newcommand{\bc}{\begin{center}}
\newcommand{\ec}{\end{center}}
\newcommand{\bq}{\begin{quote}}
\newcommand{\eq}{\end{quote}}
\newcommand{\bmat}{\begin{pmatrix}}
\newcommand{\emat}{\end{pmatrix}}

\renewcommand{\d}{\hbox{\rm d}}

\title{Laboratory Investigation of Entrainment and Mixing in Oceanic Overflows}

\author[P. Odier, J. Chen, and R. E. Ecke]
{P\ls H\ls I\ls L\ls I\ls P\ls P\ls E\ns O\ls D\ls I\ls E\ls R\ls $^{1,3}$
J\ls U\ls N\ns C\ls H\ls E\ls N\ls$^{1,2}$,\ns
\and R\ls O\ls B\ls E\ls R\ls T\ns E.\ns E\ls C\ls K\ls E\ls $^1$}

\affiliation{$^1$Condensed Matter and Thermal Physics Group and
    Center for Nonlinear Studies\\ Los Alamos National Laboratory,
    Los Alamos, New Mexico, 87545, U.S.A.\\[\affilskip]
$^2$School of Mechanical Engineering, Purdue University, West Lafayette, Indiana, 47907, U.S.A.\\[\affilskip]
$^3$Laboratoire de Physique, ENS Lyon, 46 all\'ee d'Italie, 69364 Lyon cedex 07, France}

\begin{document}


\maketitle

\begin{abstract}
We present experimental measurements of a wall-bounded gravity current, motivated by characterizing natural gravity currents such as oceanic overflows.  We use particle image velocimetry and planar laser-induced fluorescence to simultaneously measure  the velocity and density fields as they evolve downstream of the initial injection from a turbulent channel flow onto a plane inclined at 10$^\circ$ with respect to horizontal.  The turbulence level of the input flow is controlled by injecting velocity fluctuations upstream of the output nozzle. The initial Reynolds number based on Taylor microscale of the flow, R$_\lambda$, is varied between 40 and 120, and the effects of the initial turbulence level are assessed. The bulk Richardson number $Ri$ for the flow is about 0.3 whereas the gradient Richardson number $Ri_g$ varies between 0.04 and 0.25, indicating that shear dominates the stabilizing effect of stratification.  Kelvin-Helmholtz instability results in vigorous vertical transport of mass and momentum.  We present baseline characterization of standard turbulence quantities and calculate, in several different ways, the fluid entrainment coefficient $E$, a quantity of considerable interest in mixing parameterization for ocean circulation models. We also determine properties of mixing as represented by the flux Richardson number $Ri_f$ as a function of $Ri_g$ and diapycnal mixing parameter $K_\rho$ versus buoyancy Reynolds number $Re_b$. We find reasonable agreement with results from natural flows.
\end{abstract}


\section{Introduction}\label{sec:introduction}

Gravity currents are observed in many geophysical flows \citep{Simpson:ARFM:82,Huppert:JFM:06} including the important category of oceanic overflows such as the Denmark-Strait/ Faroe Bank-Channel overflows in the North Atlantic \citep{Hansen00,Girton01,Girton03} or the Mediterranean Outflow \citep{Price93}.  The North Atlantic overflows from the Norwegian-Iceland-Greenland Sea are important components of the global thermohaline circulation which plays a key role in overall climate evolution \citep{Broecker:Science:97,Wunsch:ARFM:04}.  In such currents, the flow is stably stratified but the shear between the gravity current and the quiescent fluid lying above the current can produce instability (e.g., Kelvin Helmholtz) and vertical mixing \citep{Morton56, Istweire:JPO:93}.  Modeling of these overflows \cite[see, e.g., ][]{Bacon98, Willebrand01}  relies on understanding and characterizing the mixing and entrainment of the ambient fluid into the denser gravity current which occurs at smaller spatial and temporal scales than can be captured in large scale numerical simulations.  Capturing the essential physics in sub-grid parameterizations of entrainment \citep{Legg06,Jackson08} is critical in determining the uncertainty of net transport of the thermohaline circulation with its associated impact on overall climate evolution \citep{Willebrand01}.  

There are a number of important non-dimensional measures that characterize gravity currents and the more general problem of stratified shear flow turbulence.  We introduce these parameters and the equations they enter before discussing results about such systems.  A gravity current, in the incompressible Boussinesq approximation and in non-dimensional form,  is described by equations for velocity and dynamic density $\theta(\mathbf{x},t) \equiv \left(\rho_1-\rho(\mathbf{x},t)\right)/\left(\rho_1-\rho_0\right)$ 

\begin{eqnarray}
\frac{\partial \mathbf{u}}{\partial t} + \mathbf{u}\cdot\nabla\mathbf{u} & =& -\nabla P_d + \frac{1}{Re_0}\nabla^{2}\mathbf{u}-\left(1-\theta\right) Ri_0\hat{\mathbf{z}}_g\\ \label{equ:NS}
\frac{\partial \theta}{\partial t} + \mathbf{u}\cdot\nabla\theta & = & \frac{1}{Re_0 Sc}\nabla^{2}\theta \label{equ:scal}
\end{eqnarray}

\noindent where $\hat{\mathbf{z}}_g$ is the direction of gravity (magnitude $g$), $P_d$ is the normalized pressure, and $\rho_0$ ($\rho_1$) is the initial density of the light (heavy) fluid such that $\theta$ has values between 0 (ambient fluid, $\rho=\rho_1$) and 1 (non-diluted gravity current, $\rho=\rho_0$). The non-dimensional parameters in these equations are the Reynolds number $Re_0=U_0H/\nu$, Schmidt number  $Sc=\nu/D$ and Richardson number $Ri_0= gH \left(\rho_{d_0}/\rho_0\right)/U_0^2$ resulting from choosing characteristic initial values of velocity $U_0$,  current thickness $H$,  density difference $\rho_{d_0} = \rho_1-\rho_0$, and  reference pressure $P_0 = \rho_0 U_0^2$ combined with fluid parameters $\nu$ (kinematic viscosity) and $D$ (mass diffusivity).  The Reynolds and Richardson numbers provide non-dimensional measures of the relative importance of inertia to dissipation and buoyancy to shear (potential to kinetic energy), respectively.

The fundamental problem of mixing and entrainment in gravity currents has roots in early theory and experiments \citep{Morton56,Ellison59,Turner86} which motivated detailed experimental characterization of stably-stratified shear flows \cite[][etc.]{Stillinger83, Strang01} and to increasingly detailed numerical simulations~\citep{Ozgokmen:JPO:06}.  The main global quantity is the global entrainment rate $E$ defined by the ratio of the velocity of the flow perpendicular to the current direction, $w_e$, to the downstream gravity current velocity $U$, i.e., $E = w_e/U$. The Morton-Taylor-Turner entrainment 
assumption empirically relates $E$ as a function of $Ri$ where smaller values correspond to more unstable flows: $E=\left(0.08-0.1Ri\right)/\left(1+5Ri\right)$ for $Ri<0.8$ and 0, otherwise~\citep{Turner86}. Extensions of this early work include research on eddy formation and 
mixing structures \cite[e.g.,][]{Hallworth93,Kneller99,Baines01,Sutherland02,Baines02,Baines05}, the effects of rotation \cite[][etc.]{LaneSerff98, Shapiro97,Cenedese04,Cenedese08}, and the development of the gravity current front and its speed, \cite[e.g.,][]{Lowe02,Thomas07}.  
Other laboratory experiments are described in \cite{Ivey08}, \cite{Ilicak08}, \cite{Simpson87}, and references therein.  In several of these experiments, flow visualization techniques were used to obtain global measurement of parameters such as the current thickness, whereas others have used point-wise measurement techniques (e.g., hot-wire anemometer or Laser Doppler Velocimetry) for more local quantitative studies. Our work applies techniques of particle image velocimetry (PIV) and planar laser-induced fluorescence (PLIF) to simultaneously determine velocity and density fields.

Some considerations are important in connecting laboratory experiments and in situ field measurements.  In particular, many laboratory experiments on the stability of stratified flows  \cite[e.g., ][and others]{Koop79, Strang01}  start with laminar gravity currents, very different from the turbulent conditions of ocean currents where the bulk Reynolds number is of order $10^8$.  In overflows, a relatively stable gravity current accelerates over a ridge or sill and the stability conditions change but the internal turbulence properties of the gravity current persist.  Achieving similar conditions in the laboratory is challenging because establishing a steady-state turbulent boundary layer takes time, i.e., distance downstream (using water with $U_0 \sim 10 $ cm/s, an isolated boundary layer would need about 5~m to undergo turbulent boundary layer transition \citep{Pope00}).  We create a turbulent channel flow over a short distance using a turbulent grid which acts to trip the boundary layer so that the output flow is characteristic of a developed channel flow with $Re \approx 3500$.  Without the injectors, the bulk $Re$ is the same (same $U_0$ and $H$) but the turbulent fluctuations are reduced because the output flow has not reached a turbulent steady state.  Over the relatively short extent of our experiments of about 1/2 m, these considerations are important.  With downstream distance, fluctuations generated by Kelvin-Helmholtz instability will begin to populate the gravity current, reducing the difference between the two initial conditions, but this process will depend sensitively on $Ri$.

Although existing laboratory experiments  provide significant information about gravity currents, our simultaneous velocity and density field measurements on an initially turbulent flow allows access to new properties of gravity currents with better correspondence to high $Re$ environments such as occur in the ocean.  Elsewhere, we described a Prandtl mixing length representation of the momentum and buoyancy fluxes~\citep{Odier09}. We also discussed the dependence of various length scales on $Ri$, included a detailed analysis of fluxes including their probability distributions, and presented a local interpretation of entrainment and detrainment \citep{Odier:PhysD:12}.

In the present paper, we report experimental measurements of mixing and entrainment of a less dense gravity current flowing into a dense non-rotating environment with a controlled initial density difference of 0.26\%. Our objective is to quantitatively test to what degree common assumptions such as the isotropy of the flow, the steady-state nature of the turbulence, or using bulk parameters as opposed to parameters determined from the turbulent dynamics (e.g., Re versus $R_\lambda$, etc.) affect measures of mixing and entrainment.  In particular, the structures and energetics of the turbulent gravity current are examined to quantify the interaction of stratification and turbulence and comparison is made with other laboratory experiments on stratified turbulence and with results from geophysical systems.  Details about the experimental apparatus are presented in section \ref{sec:facility}. Instrumentation techniques are introduced in section \ref{sec:instrumentation}. The experimental conditions are characterized in section \ref{sec:characterization}. In section \ref{sec:results}, we detail the overall flow properties of the current to provide proper context for our measurements and analysis of entrainment, presented in section~\ref{sec:entr} and mixing properties, presented in section~\ref{sec:mixing}. Section \ref{sec:conclu} provides summary remarks and a discussion of the connection of our work to geophysical situations.

\section{Experimental Facility}\label{sec:facility}

\begin{figure*}                                                     
\centering                                                              
\includegraphics[width=1.0\textwidth]{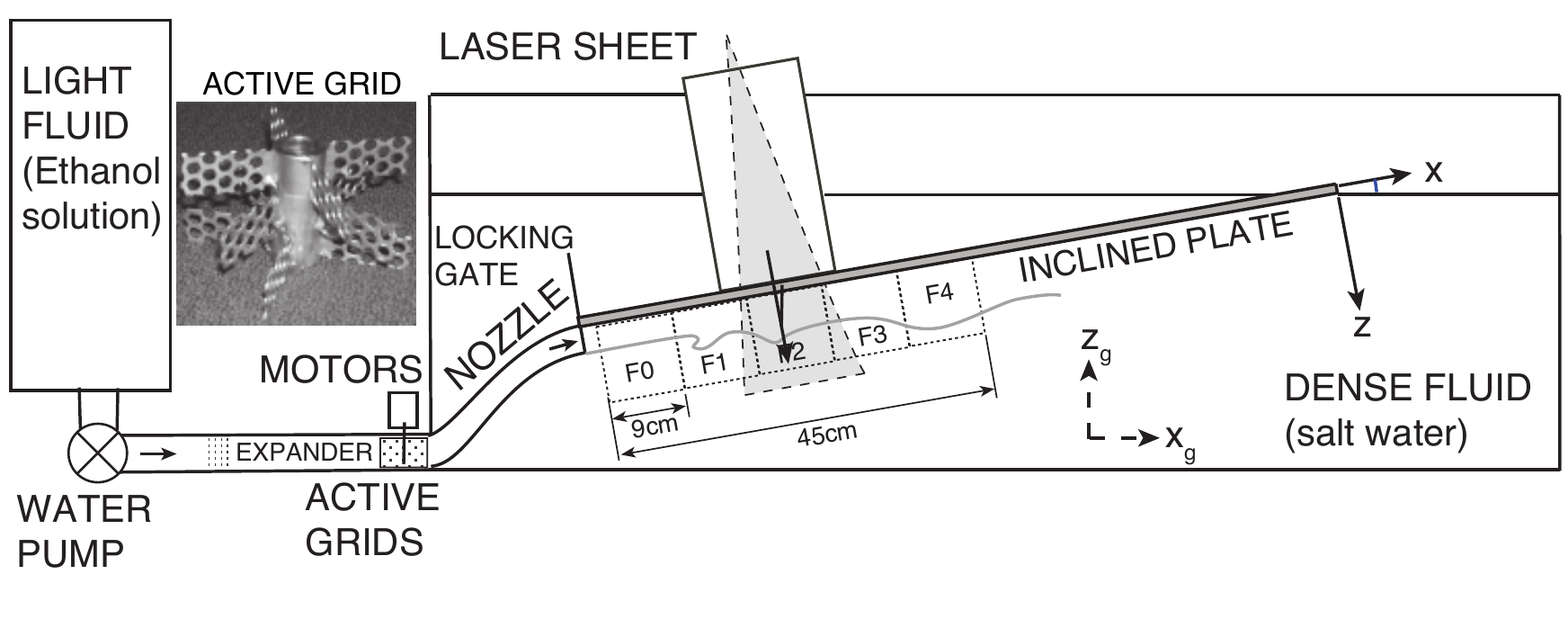} 
\caption{\label{fig:facility}  Schematic illustration of the experimental apparatus.}
\end{figure*}

A schematic of the experimental facility is shown in figure \ref{fig:facility}. The central part of the facility is a water tank, with  dimensions of 180 cm (length) $\times$ 48 cm (height) $\times$ 52 cm (width), made of acrylic side walls and an aluminum bottom plate. Prior to each experiment, this main tank is filled with dense fluid (salt water, 308 liters). A secondary tank is filled with light fluid (ethanol solution, 101 liters). The density difference between the fluids in the main and secondary tanks is precisely controlled as discussed below.
An acrylic plate, 1.2 cm thick and 150 cm long, spans the entire width of the tank and is inclined at an angle $\alpha=10^{\circ}$ with respect to horizontal. A glass plate is nested seamlessly in the acrylic plate to provide better optical access to the test section. The origin of the coordinate system is set on the bottom face of the inclined plate at the nozzle exit. The $x$ direction points downstream along the inclined plate, and the $z$ direction is down, perpendicular to the plate. 
The coordinate system, \{$x_g$, $y_g$, $z_g$\}, where $z_g$ is vertical upwards, is also shown in figure \ref{fig:facility}. 
This system, adopted for analyzing buoyancy-related terms as discussed in section \ref{sec:mixing}, has $\hat{\mathbf{x}}_g = \cos\alpha\;\hat{\mathbf{x}} + \sin\alpha\;\hat{\mathbf{z}}$ and $\hat{\mathbf{z}}_g = \sin\alpha\;\hat{\mathbf{x}} - \cos\alpha\;\hat{\mathbf{z}}$. Parameters in this system are denoted by the subscript $g$.

A pump injects the light fluid with adjustable initial speed from the secondary tank into the main tank through an expander-nozzle combination (width in $y$ direction: 48 cm and height $H=5.0$ cm).  Thus, a gravity current with less density is introduced into a denser environment. Twelve layers of 0.5 mm thick plates perforated with 0.6 cm holes, placed vertically inside the expander, generate a fairly uniform flow pattern along the $y$ direction.  One end of the inclined plate is attached to the nozzle. The curvature of the nozzle is designed to provide a smooth transition and match with the inclined plate. A locking gate, installed near the tip of the nozzle, is closed during the experiment preparation process to prevent fluid exchange between the main tank and the secondary tank. It is opened just before the experiment starts. The temperature in both tanks is the same (20 to 22$^{\circ}$C) to within 0.1$^{\circ}$C. 
 
The present setup is  reversed compared to most geophysical gravity currents: light fluid moves underneath an upwardly inclined plate into a dense environment whereas in geophysical contexts the dense fluid flows down inclined topography into a less dense environment. The physics for these two scenarios is the same within the Boussinesq approximation, which is valid for our conditions and for oceanic applications. Each experimental run lasts for 40-80 seconds, depending on the injection speed. After each data set is acquired, both tanks are emptied, thoroughly rinsed and refilled with dense and light fluids for the next data set. Independent runs (between 5 and 10) were taken for each set of given experimental conditions.

To increase turbulent fluctuations in the gravity current, four active grids (see figure \ref{fig:facility}), driven at 200~rpm, are located along the $y$ direction between the expander and the nozzle. The respective direction of rotation for each of the four grids is
optimized to minimize the creation of lateral (in the $y$ direction) large scale velocity gradients. Similar active grid techniques were used to obtain nearly homogenous and isotropic turbulent velocity fields with relatively high Reynolds number based on the Taylor microscale of the flow, $R_{\lambda}$ (see definition in Table~\ref{tab:parameters}, section~\ref{sec:characterization}) \cite[e.g., ][]{Makita91, Mydlarski96, Kang03, Chen06}.

In the present paper, we consider three sets of data. The first, ``turbulent-stratified'' (TS) conditions,  is acquired when the active grids are deployed and a density difference is present. In the second,  ``laminar-stratified" (LS) conditions, the active grids are absent (thus, the turbulence level is lower) but the density difference is the same. The third is referred to as ``not stratified'' (NS), where the active grids are used but with no density difference between the fluids. Comparing TS  and LS helps understand the role of Reynolds number, whereas comparing TS and NS helps elucidate the influence of stratification.



\section{Instrumentation and Methods}\label{sec:instrumentation}

\begin{figure} 
\centering
\small{(a)}
\includegraphics[width=.4\textwidth]{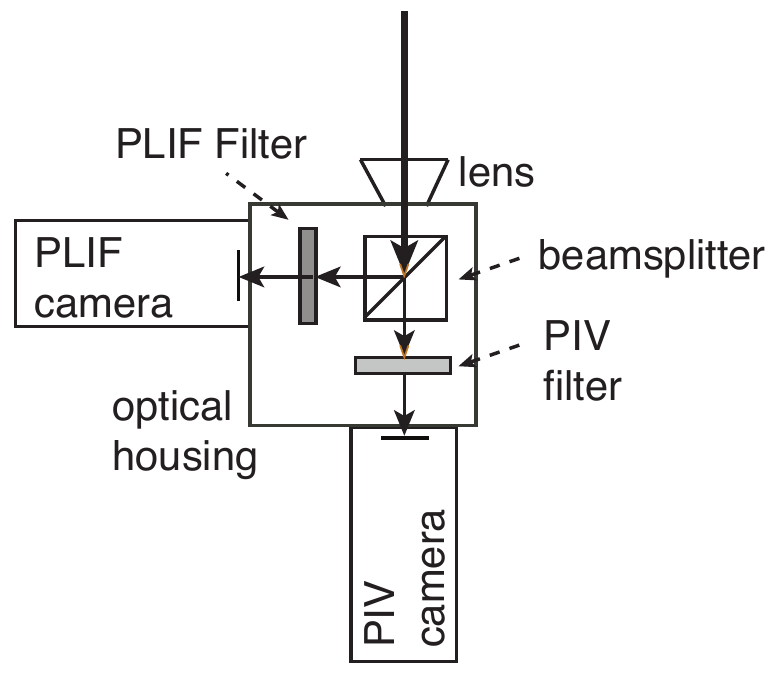}
\small{(b)}
\includegraphics[width=.5\textwidth]{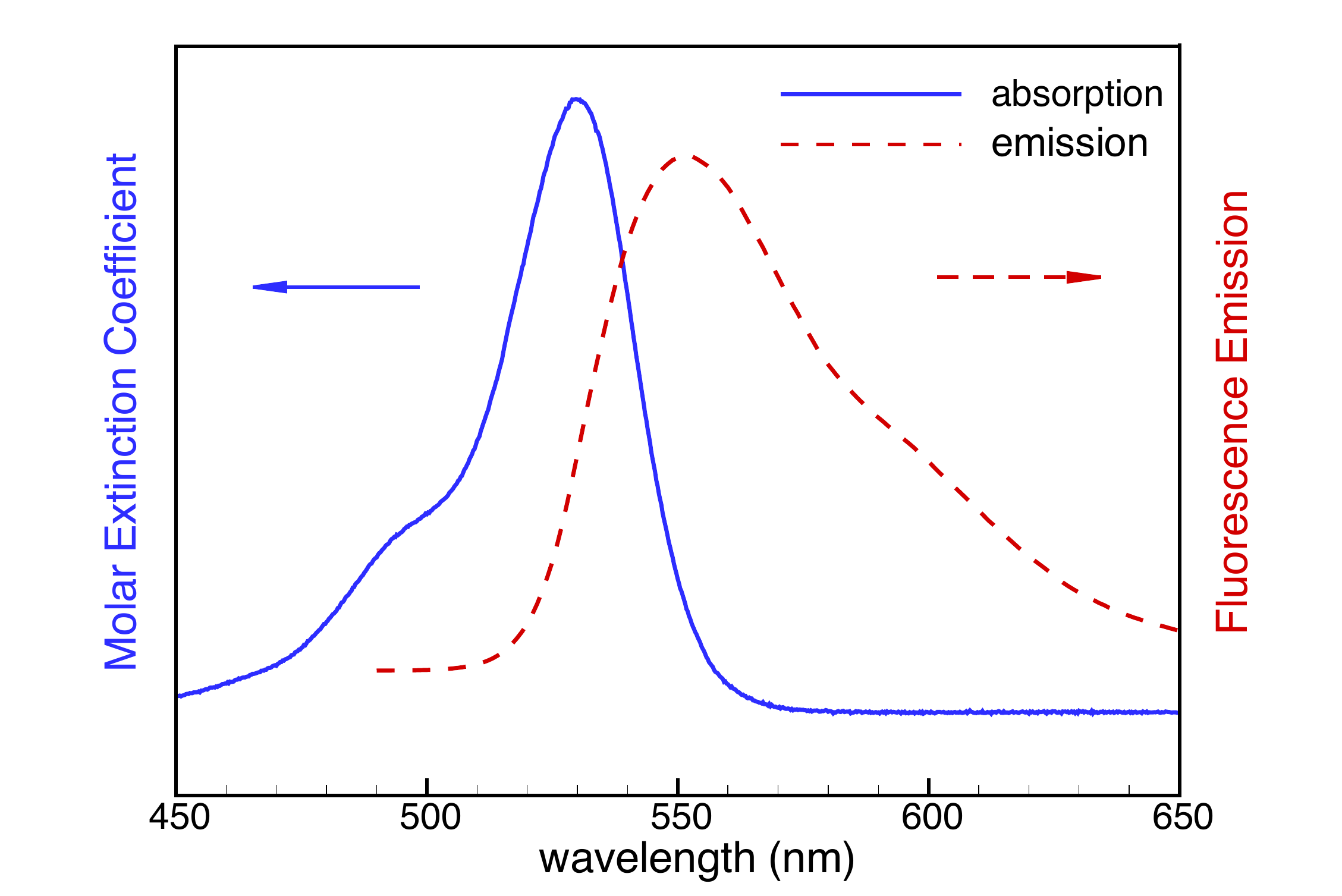}
\caption{\label{fig:PIVPLIF} (a) Schematics of combined PIV-PLIF system. (b) Absorption and emission spectra of Rhodamine 6G using data in \cite{Du98}.}
\end{figure}

A combination of Particle Image Velocimetry (PIV) and Planar Laser Induced Fluorescence (PLIF) is used for simultaneous velocity and density measurements \citep{Hu00, Borg01, Hjertager03, Feng07}, as illustrated in figure \ref{fig:PIVPLIF}(a). A dual-head Nd:YAG pulse laser (532~nm, maximum intensity 90~mJ/pulse) is used for both PIV illumination and PLIF excitation.  Through PIV optics, the laser beam is expanded into a 1~mm thick laser sheet illuminating the sample area in the $x-z$ plane along the center line of the tank. To avoid interaction with the free surface, the laser enters the water through a partially immersed acrylic box (see figure~\ref{fig:facility}).
To implement the simultaneous PIV/PLIF measurement, the camera lens, beamsplitter, PIV filter, PLIF filter, and two cameras, are mounted in  an optical housing shown in figure \ref{fig:PIVPLIF}(a). The PIV filter (bandpass, 525$\pm$20~nm) blocks most of the fluorescence and passes  scattered light from PIV seeding particles.  The PLIF filter (high pass in wavelength with cut-off 550~nm) blocks the scattered light and only passes the fluorescence signal. The housing allows the two cameras to be aligned to record the same sample area ($9.0\times 9.0~{\rm cm}^2$) to within $\pm$5 pixels (45~$\mu$m). A Schneider lens combination (Componon-S 150/5.6 50 mm mount and Unifoc 58 focus mount) is used to image the test section. We measure at five consecutive downstream frames from 3 cm to 48 cm in the $x$ direction (labeled as $F_0$ to $F_4$ in figure \ref{fig:facility}) by translating the cameras and optical housing in front of the tank at an inclination angle $\alpha$ matching the plate inclination angle. The laser sheet optics is similarly adjusted. An image acquisition and laser control system synchronizes the measurements with a sampling rate of 4~Hz.

\subsection{Particle Image Velocimetry}

For velocity measurements, both the dense fluid and the light fluid are uniformly seeded with hollow glass beads (median diameter $d_p$=10 ${\mu}$m, specific gravity $\rho_p$=1.1).  The Stokes number is $St = {\tau_p}/{{\tau_\eta}_K}\sim0.0001$,
where the particle response time \cite[e.g.,][]{Raffel07}  is $\tau_p = d_p^2{\rho_p}/{18\mu}$ and ${\tau_\eta}_K$ is the Kolmogorov time scale (see table \ref{tab:parameters}). Since $St\ll1$, the seeding particles follow turbulent fluid motion with good fidelity.

A RedLake ES~4020 digital camera, with 
$2048\times2048$ pixel$^2$ resolution and operated in triggered double-exposure mode, is used to record two consecutive particle images (12-bit grayscale) separated by 4.000  ms. A corresponding velocity field is obtained using standard PIV analysis \citep{Roth01} with a PIV interrogation window of $48\times48$~pixels with $67\%$ overlap. The resulting vector spacing is 0.53~mm (0.1 pixels), and $163\times163$ vectors are obtained from each image pair. 

An estimate of uncertainty of an instantaneous PIV velocity measurement is $0.1$ cm/s in the present experiments; $2\%$ using a characteristic velocity $U=5.0\ \rm{cm/s}$. The uncertainty in variables involving the mean downstream velocity computed from an ensemble set of 500 instantaneous measurements is about 0.09\%, whereas rms values of downstream velocity have an uncertainty of $0.6\%$. 
More on uncertainties in PIV analysis can be found in \citet{Keane90, Huang97, Raffel98, Roth01, Chen06}. 


\subsection{Planar Laser Induced Fluorescence}
When PLIF is used to measure the concentration field, the fluorescent dye is a surrogate for the stratifying scalar of interest (e.g., concentration of  ethanol solution). The dye must have approximately the same diffusivity ($\kappa$, usually measured by Schmidt number $Sc = \nu/\kappa$) as the ethanol solution or the mismatched diffusivities cause the dye concentration to differ from the concentration of the ethanol solution (the so-called ``double-diffusivity'' problem, see, e.g., \cite{Troy05} and references therein). The Schmidt numbers for our water solutions are 600 for Rhodamine 6G, 540 for ethanol, and 770 for salt water. For a typical fluid element staying at the most during a time $t$ of a few seconds in the field of view, the corresponding molecular diffusion length is $l \sim \sqrt{\kappa t}\sim 0.1~\rm{mm}$ and the differential diffusion is at most 20\% of that value, much smaller than the interpolated velocity and density grid spacing of 0.5 mm.  Thus, the effects of differential molecular diffusion can be ignored in this measurement technique. 

The fluorescent dye is uniformly mixed into the light fluid in the secondary container prior to each experiment. The dye absorbs excitation light around 530 nm and emits broadband fluorescence with a peak at 555 nm (see figure \ref{fig:PIVPLIF}(b)). The procedure used to convert the recorded grayscale image to a density field is described below. Photobleaching is negligible because of the 3 ns duration of the laser pulses~\citep{Crimaldi:EF:97}. A photodiode placed in the laser path showed no systematic shot-to-shot power variations.

\subsection{Refraction Index Matching}
Optical flow diagnostics techniques in a stratified environment require closely matching the refraction indices of the dense and light fluids~\cite[see][]{Daviero01, Alahyari94, Hannoun88, McDougall79}. To accomplish this matching, we use salt water for the heavy fluid and ethanol water mixture for the light fluid. The density and index of refraction of the fluids at different concentrations are measured using a densitometer (Anton Paar DMA5000) and a  refractometer (Milton Roy Company). The densities of the dense fluid $\rho_1$ and the light fluid $\rho_0$ are selected to obtain a density difference of $\rho_{d_0}=\rho_1-\rho_0=2.6~ \rm{g/L}$. The matching of refractive indices is accurate to about 0.01\%: circulation pumps facilitate the mixing of salt and ethanol in tanks and, because density is also dependent on temperature, temperature differences between the fluids are maintained at less than 0.1$^{\circ}$C.


\subsection{Density Field from Grayscale PLIF Image}\label{subsec:PLIF2Den}
For an excitation illumination $I_o(x,z)$ and with no absorption along the laser path, there is a linear relationship between the intensity of fluorescence emission and the dye concentration~\citep{Borg01}, consistent with our measurements at dye concentrations ranging from 0 to 100 $\mu$g/L. For low enough laser intensities (the so-called ``weak excitation assumption''~\citep{Crimaldi:EF:08}, which we have checked, see later), the pixel grayscale values at each location in the fluorescence image, $g_o(x,z,t)$, are linearly related to the dye concentration at that location, $c(x,z,t)$, and to $I_o(x,z)$ via
$g_o(x,z,t)=\Gamma\cdot I_o(x,z)\cdot~c(x,z,t)+g_b(x,z)$, where $g_b(x,z)$ is the background noise or camera dark-response. The constant $\Gamma$ accounts for the system-specific optical collection efficiency (conversion of photons into digital signal recorded by the camera) and the effective quantum yield of the fluorescent dye. Here the subscript ``o'' denotes the non-absorption case. The equation above accounts for inhomogeneous laser illumination. The values of $\Gamma\cdot I_o(x,z)$ and $g_b(x,z)$ are determined from a calibration process discussed below. 

\begin{figure}                                                     
\centering   
\includegraphics[width=.8\textwidth]{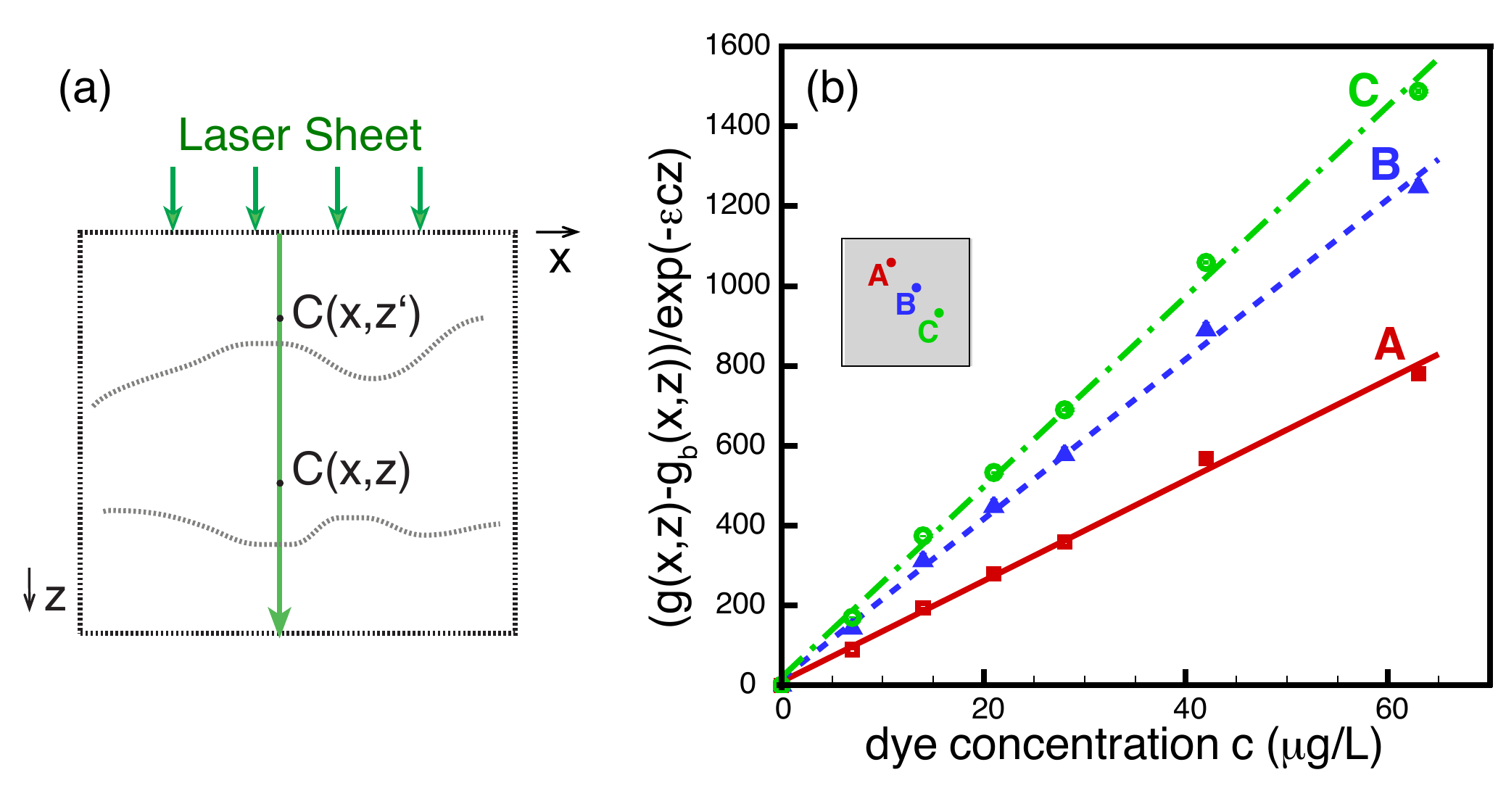}
\caption{\label{fig:attenuation} (a) Laser intensity attenuation correction inside the field of view. (b) PLIF calibration curves at three representative pixel locations (A, B, and C) in the field of view. The inserts shows the location of the three points. Symbols show the mean value  of the grayscale distribution in a 50 snapshots ensemble set with statistical uncertainty less than the symbol size. The lines are linear fits to the data.}
\end{figure}

When laser light passes through the test section the dyed solution attenuates the laser intensity and this attenuation must be corrected to obtain accurate dye concentrations from PLIF images \cite[see, e.g.,][]{Ferrier93, Atsavapranee97, Karasso97}. In the present setup, the propagation of the laser sheet in the field of view is nearly parallel to the $z$ axis (maximum angular deviation less than $1.5^{\circ}$), as illustrated in figure \ref{fig:attenuation}(a).

According to the Bouguer-Lambert-Beer law, the laser intensity in such an absorbing medium is

\begin{equation}
I(x,z,t)=I_o(x,z)\exp\left(-\varepsilon\int_0^z c(x,z',t)dz'\right) ,
\label{eq:attenuation}
\end{equation}

\noindent where $\varepsilon$ is the extinction coefficient of the dyed solution. With this attenuation, the recorded grayscale value is $g(x,z,t)=\Gamma\cdot I(x,z,t)\cdot~c(x,z,t)+g_b(x,z)$. Substituting~\eqref{eq:attenuation} into this last equation leads to

\begin{equation}
g(x,z,t)=\Gamma\cdot I_o(x,z)\exp\left(-\varepsilon\int_0^z c(x,z',t)dz'\right)\cdot~c(x,z,t)+g_b(x,z) .
\label{eq:PLIF3}
\end{equation}

To do PLIF calibration, the main tank is filled with Rhodamine 6G solution with uniform concentrations ($c$ = 0.0, 7.0, 14.0, 21.0, 28.0, 42.0, and 63.0 $\mu$g/L). The zero concentration measurement gives the camera dark-response $g_b(x,z)$. The laser optics and recording cameras are not changed between the calibration and the experimental runs. For each concentration, 50 calibration images are recorded, and the ensemble averaged grayscale information is analyzed. For a uniform concentration $c$, from~\eqref{eq:PLIF3}, we get $(g(x,z)-g_b(x,z))/{\exp\left(-\varepsilon c z\right)} =\Gamma\cdot I_o(x,z)\cdot c$, where $g_b(x,z)=g(x,z)|_{c=0}$. The extinction coefficient, $\varepsilon=1.6\pm0.3\times10^{-4}\ \rm{(cm\cdot\mu g/L)^{-1}}$, is also determined from the calibration data. Figure \ref{fig:attenuation}(b) shows three typical PLIF calibration curves at different locations inside the field of view. The slope of each curve represents the corresponding local value of $\Gamma\cdot I_o(x,z)$. The linearity of the curves is consistent with our PLIF system operating in the weak excitation limit~\citep{Crimaldi:EF:08}.

In a real measurement, from~\eqref{eq:PLIF3} and the values of  $\Gamma\cdot I_o(x,z)$, $g_b(x,z)$ and $\varepsilon$ measured through calibration analysis, one obtains the concentration field:

\begin{equation}
c(x,z,t)=\frac{g(x,z,t)-g_b(x,z)}{\Gamma\cdot I_o(x,z)\exp\left(-\int_0^z\varepsilon c(x,z',t)dz'\right)} .
\label{eq:PLIF2Conc}
\end{equation}

\noindent The determination of $c(x,z,t)$ implies knowing $c(x,z',t)$ for $z'<z$.
This is done iteratively for each line of the grid along $x$,
assuming that there is no absorption for the first line ($z\approx$ 0).
The validity of this assumption is ensured by a transparent box located above the
inclined plate, on top of the field view (see figure \ref{fig:facility}), preventing any
dyed fluid from intercepting the laser sheet above the inclined plate. One
then has: $c(x,0,t)=(g(x,0,t)-g_b(x,0))/{\Gamma I_o(x,0)}$. The non-dimensional dynamic density is related to the measured dye concentration by $\theta(x,z,t) ={c(x,z,t)}/{c_o}$, where $c_o$ is the dye concentration in the injected light fluid.

The aforementioned static calibration procedure is applied independently to each of the 5 measurement frames $F_0$ to $F_4$. In an additional dynamic calibration step, we extract dyed samples next to the bottom surface of the inclined plate ($z\simeq 0$) using syringe needles. Direct density measurements of these samples show that the fluid close to the plate (within a few mm in the wall boundary layer) is not diluted by the ambient fluid in the main tank, i.e., $\theta(x,z=0)=1$ within the downstream locations of the present experiments. If the measured $\theta(x, z=0)$ displays variations from unity, a global correction factor is applied along the path of each light ray (nearly parallel to $z$ direction) to ensure that $\theta(x,z=0)\simeq1$ (within 5\%) for each frame separately.  This procedure results in a consistent global density field from frames $F_0$ to $F_4$.

The uncertainty of the PLIF measurement includes the jitter of laser intensity, background noise drift of the CCD sensors, nonlinear effects of the fluorescence emission, etc. We estimate the statistical uncertainty by analyzing the calibration data. At a characteristic dye concentration of $c\sim 50 \rm{\mu g/L}$, the standard deviation of the calibration data is about 1\%. Based on averaging 50 images, the relative error of an instantaneous measurement under the same experimental conditions is about $1\%\times\sqrt{50}\simeq7\%$. 


\section{Characterization of Experimental Conditions} \label{sec:characterization}

\begin{figure}
\centering
\includegraphics[width=.99\textwidth]{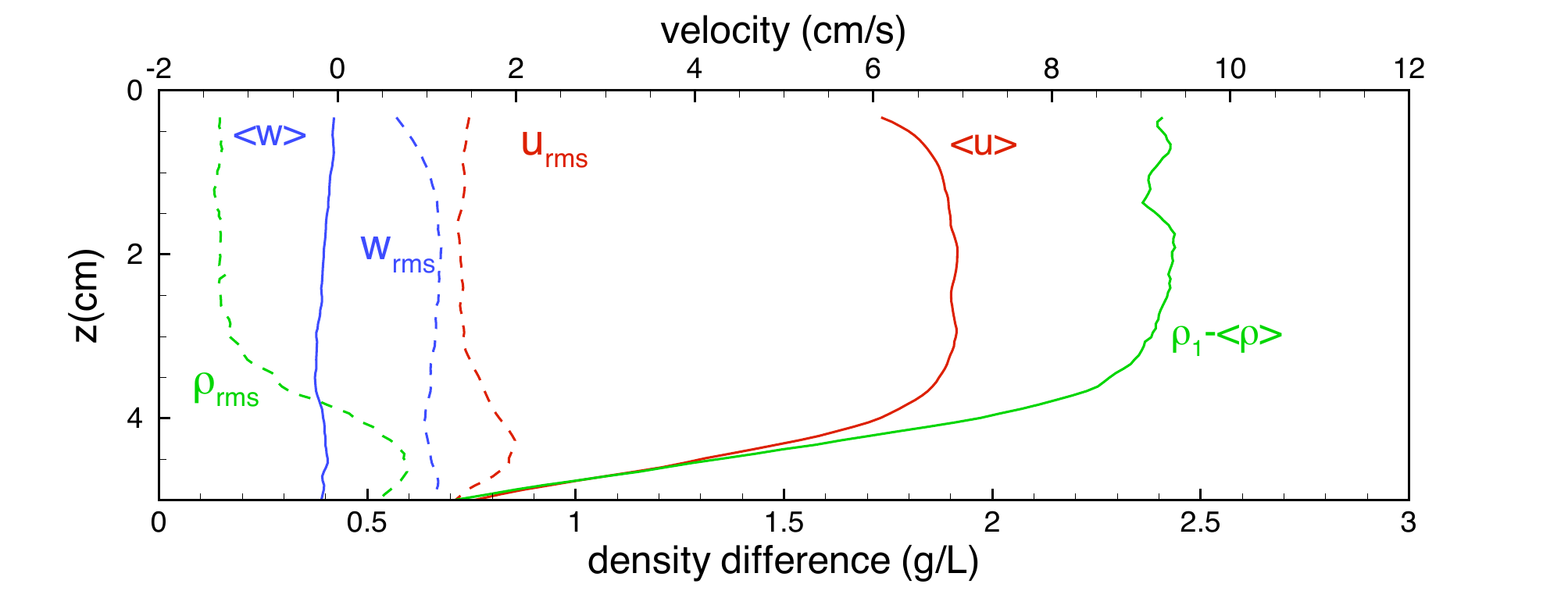}\\
\vspace{3mm}
\caption{\label{fig:injmeanrms} Mean velocity components and mean density anomaly of the injected current and their rms counterparts measured for TS evaluated at $x=3.0$ cm. }
\end{figure}

The $x-z$ planar components of velocity and density fields are obtained from PIV and PLIF measurements, respectively. We focus on the quasi-steady regime: data acquisition is triggered after the front of the gravity current moves out of the sample area.
The velocity and non-dimensional dynamic density are then ensemble averaged to get the mean and fluctuating parts:
$u_i(\mathbf{x},t) = \la u_i\ra(\mathbf{x}) + u'_i(\mathbf{x},t)$ and 
$\theta(\mathbf{x},t) = \la \theta\ra(\mathbf{x}) + \theta'(\mathbf{x},t)$
where  $\langle \cdot \rangle$ denotes ensemble averaging and superscript $'$ denotes the corresponding fluctuating part\footnote{In our previous publication, \cite{Odier09},  $\overline{~\cdot~}$ represents ensemble averaging and $\langle~\cdot~\rangle$ denotes spatial averaging. This adjustment in nomenclature is made to be consistent with popularly used notation, e.g., \cite{Pope00}.}. The ensemble averaging procedure consists of averaging, for each spatial grid-point, all the images of a given experimental run, then averaging over all runs with the same experimental conditions. This procedure yields statistically-steady quantities. Table \ref{tab:parameters} lists initial averaged parameters of the current (next to the nozzle outlet) of TS , LS and NS, respectively, including the mean velocity and root-mean-square fluctuations. These parameters are calculated from measurement data at the downstream location $x=3.0$ cm ($x/H=0.6$); although the actual injection starts at $x=0.0$ cm, the laser sheet could not be placed closer than 3 cm from the nozzle outlet owing to other design considerations. The mixing zone {(region of strong gradients, we will define it more precisely later)} has increased, typically by about 1.0 cm, by this location in TS, but the majority of the injection gravity current has not been disturbed, as shown in figure \ref{fig:injmeanrms}, which represents vertical profiles of various quantities in TS , at $x=3.0$ cm. One can indeed observe that the injected gravity current for TS is nearly homogenous in the $z$ direction with the density at its maximum value. This is true in the range of $0.1 < z/H < 0.8$, i.e., over most of the current height. The profiles are similar for the non-stratified case (not shown) but for LS (not shown), the $z$-profile of $\langle U\rangle$ displays more variation near the nozzle outlet (about $20\%$ across the entire nozzle height), but its rms fluctuations do not.
As can be seen in table~\ref{tab:parameters}, the injected current is fairly isotropic, i.e., the degree of anisotropy is $\overline{u}_{rms}/\overline{w}_{rms} \sim1.3$. 
The Taylor microscale $\lambda$ given in table~\ref{tab:parameters} is estimated by fitting the autocorrelation function of the downstream velocity $u$, calculated near the nozzle outlet, by a parabola around zero.

\begin{table}
\begin{center}
\begin{tabular}{lcccc}
Parameter/Case && TS & LS & NS\\
\hline
Mean Velocity & $U_0\equiv\overline{\langle u\rangle}$ & $7.0\ \rm{cm/s}$ & $6.5\ \rm{cm/s}$ & $7.5\ \rm{cm/s}$\\
&  $\overline{\langle w\rangle}$ &  $-0.1\ \rm{cm/s}$ &  $-0.1\ \rm{cm/s}$ &  $-0.2\ \rm{cm/s}$ \\
Velocity Fluctuation (RMS) & $\overline{u}_{\rm{rms}}$ & $1.6\ \rm{cm/s}$ & $0.7\ \rm{cm/s}$ & $1.8\ \rm{cm/s}$\\
 & $\overline{w}_{\rm{rms}}$ & $1.2~\rm{cm/s}$ & $0.55\ \rm{cm/s}$ & $1.5\ \rm{cm/s}$ \\
Initial Density Difference & $\rho_{d_0}=\rho_{1}-\rho_{0}$ & $2.6\ \rm{g/L}$ & $2.6\ \rm{g/L}$ & $0.0\ \rm{g/L}$\\
Density Fluctuation (RMS) & $\overline{\rho}_{\rm{rms}}$ & $0.15~\rm{g/L}$ & $0.11\ 
 \rm{g/L}$ & \\
Characteristic Velocity Fluctuation & $u_c=\frac{1}{2}(\overline{ u}_{\rm{rms}}+\overline{ w}_{\rm{rms}})$ & $1.4\ \rm{cm/s}$ & $0.6~\rm{cm/s}$ & $1.7\ \rm{cm/s}$\\
Degree of Anisotropy & $\overline{ u}_{\rm{rms}}/\overline{ w}_{\rm{rms}}$ & 1.3 & 1.2 & 1.2\\
Reduced Gravity & $g'=g\rho_{d_0}/\rho_0$ & $2.6\ \rm{cm/s^2}$ & $2.6\ \rm{cm/s^2}$ &  $0.0\ \rm{cm/s^2}$  \\ 
Richardson Number (bulk) & $Ri_0=g'H/ U_0^{2}$ & $0.27$ & $0.31$ & 0\\
Turbulence Kinetic Energy & $K_0=\frac{1}{2}\left(\overline{u}_{\rm{rms}}^2+2 \overline{w}_{\rm{rms}}^2\right)$ & $2.3\ \rm{cm^2/s^2}$ & $0.7\ \rm{cm^2/s^2}$ & $3.6\ \rm{cm^2/s^2}$\\
Bulk Reynolds Number & $Re_0=U_0 H/\nu$ & $3,500$  & $3,300$  & $3,800$ \\
Turbulence Dissipation Rate & $\epsilon_0$ & $1.1\ \rm{cm^2/s^3}$  & $0.6\ \rm{cm^2/s^3}$  & $1.8\ \rm{cm^2/s^3}$ \\
Integral Scale & $l_0\sim  K_0^{3/2}/\epsilon_0$ & $3.3\ \rm{cm}$ & $1.0\ \rm{cm}$ & $3.8\ \rm{cm}$ \\
Taylor Microscale & $\lambda_{0}$ & $0.70\ \rm{cm}$ & $0.65\ \rm{cm}$ & $0.70\ \rm{cm}$ \\
Kolmogorov Length Scale & $\eta_K\sim(\nu^3/\epsilon_0)^{1/4}$ & $0.3\ \rm{mm}$ & $0.4\ \rm{mm}$ & $0.3\ \rm{mm}$ \\
Kolmogorov Time Scale & $\tau_{\eta_K}\sim \sqrt \nu/\epsilon_0$ & $0.1\ \rm{s}$ & $0.2\ \rm{s}$ & $0.1\ \rm{s}$\\
Integral Scale Reynolds Number & ${Re_l}_0=\sqrt{\frac{2}{3}}\frac{K_0^{2}}{\nu\epsilon_0}$ & 400 & 67 & 590\\
Microscale Reynolds Number & ${R_\lambda}_0=u_c \lambda_{0}/\nu$ & $100$ & $42$ & $120$\\
\\
\end{tabular}
\caption{Parameters of the injected gravity current. Parameters are defined in terms of measured quantities. Subscript ``$0$'' denotes the parameter near injection. $\overline{~\cdot~}$ represents a spatial averaging operation over the height of initial injection in the $z$ direction. $H=5$ cm is the nozzle height.}
\label{tab:parameters}\end{center}
\end{table}

Some values given in table~\ref{tab:parameters} rely on the estimation of turbulence dissipation rate 
$\epsilon$.  There are two major difficulties in obtaining $\epsilon$ directly from PIV data \cite[see, e.g., ][and references therein]{Doron01, Tanaka07} :  (i) the spatial resolution of PIV measurement must be close to the smallest turbulence scale or $\epsilon$ is significantly underestimated, and (ii) the unresolved third component ($y$-component in the present study) terms must be approximated using resolved terms. Our PIV resolution ($\Delta = 0.53\ \rm{mm}$) is close to the Kolmogorov length scale ($\eta_K\sim0.3\ \rm{mm}$, estimated to first approximation using $\epsilon \sim u'^3/H$ with $\eta_K \sim \epsilon^{1/4}$ being very insensitive to this rough approximation). Thus, a direct calculation of $\epsilon$ is feasible.  Assuming local isotropy \cite[e.g., ][]{Monin71,Tennekes72} yields $ \epsilon=15\nu\left\langle\left(\partial u'/\partial x\right)^2\right\rangle$, where only one gradient of velocity fluctuation is needed.  Because we have velocity gradients in both $x$ and $z$, we use all the available velocity gradients in calculating $\epsilon$ \citep{Doron01}:

\begin{equation}
\small {\epsilon = \nu\Bigg\la4\left(\frac{\pt u'}{\pt x}\right)^2 + 4\left(\frac{\pt w'}{\pt z}\right)^2 + 3\left(\frac{\pt u'}{\pt z}\right)^2 + 3\left(\frac{\pt w'}{\pt x}\right)^2
+4\frac{\pt u'}{\pt x}\frac{\pt w'}{\pt z} + 6\frac{\pt u'}{\pt z}\frac{\pt w'}{\pt x} \Bigg\ra}
\label{eq:TurbDissipationApx3}
\end{equation}

{The quantitative difference between assuming total isotropy and using the less constrained expression in~\eqref{eq:TurbDissipationApx3} can be appreciable. For example, at the inlet of the injected gravity current (from $z/H$= 0 to 1), the isotropic approximation gives an averaged turbulent dissipation rate $\epsilon_0=2.2~\rm{cm^2/s^3}$ whereas~\eqref{eq:TurbDissipationApx3} (anisotropic) yields $\epsilon_0 = 1.1~\rm{cm^2/s^3}$.  Thus, at least for the cases considered here, the  isotropy assumption significantly overestimates $\epsilon$.}
We use~\eqref{eq:TurbDissipationApx3} to estimate $\epsilon$ in the present study and discuss in Appendix~\ref{app:consist} how this anisotropic approximation remains roughly consistent with other estimates in the text that are based on the isotropic, homogenous turbulence assumption.

The initial turbulent dissipation rates $\epsilon_0$ for the three flow conditions reported here, calculated using~\eqref{eq:TurbDissipationApx3} at $x/H=0.6$ and averaged over the height of the nozzle ($0\leq z/H\leq 1$), are listed in table  \ref{tab:parameters}. The values obtained are consistent with other experimental data of a similar nature~\citep{Rohr:JFM:88,Hult:JGR:11b}. The performance of the active grids is reflected in several parameters listed in table \ref{tab:parameters}. Comparing the two turbulent cases and the laminar case, the characteristic velocity fluctuations, the turbulent dissipation and the Taylor-microscale Reynolds number increase by about $200\%$, whereas the bulk Reynolds numbers are nearly the same.

\section{Kinematic description of the flow}\label{sec:results}

\begin{figure}
\centering
\includegraphics[width=.95\textwidth]{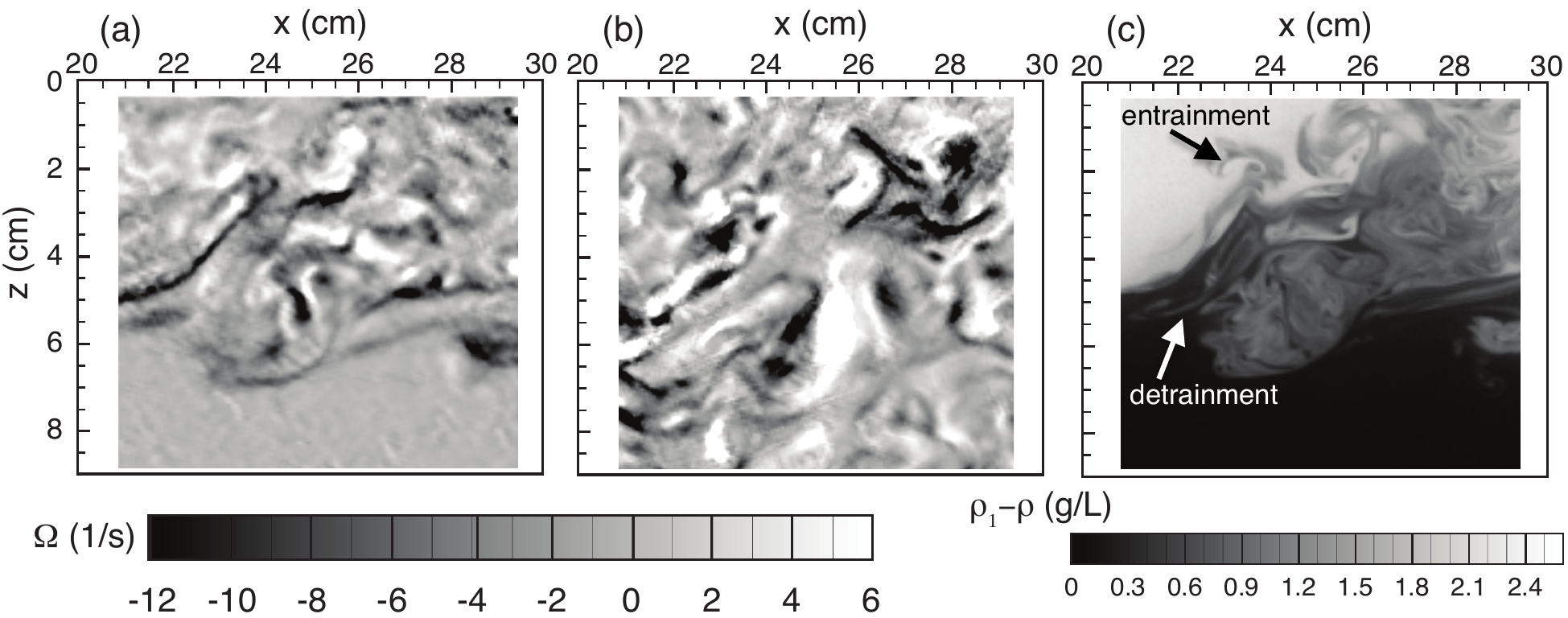}
\caption{\label{fig:snapshotComp} Instantaneous vorticity $y$-component for :(a)  TS, (b) NS. (c) Dynamic density for TS . Arrows highlight examples of entrainment and detrainment events.}
\end{figure}

We first present our results using mean quantities measured in our flow to give a kinematic description of our gravity current. This picture serves as background for the rest of our analysis. Individual snapshots of vorticity, shown in figure~\ref{fig:snapshotComp}(a) and \ref{fig:snapshotComp}(b), reveal the evolution of the gravity current in the quasi-steady range. A typical vorticity field of TS  (figure~\ref{fig:snapshotComp}(a)) is less active than that of NS (figure~\ref{fig:snapshotComp}(b)) because of the stabilizing effect of stratification (light fluid moving on top of dense fluid) which reduces the mixing activity. The mixing process is observed in the density field (figure~\ref{fig:snapshotComp}(c)) with Kelvin-Helmholtz instability in which ambient dense fluid is entrained into the main current while light fluid filaments detach (detrain) from the main current. A local measure of entrainment and detrainment is described elsewhere~\citep{Odier:PhysD:12}.

\begin{figure}
\centering
\includegraphics[height=.20\textheight]{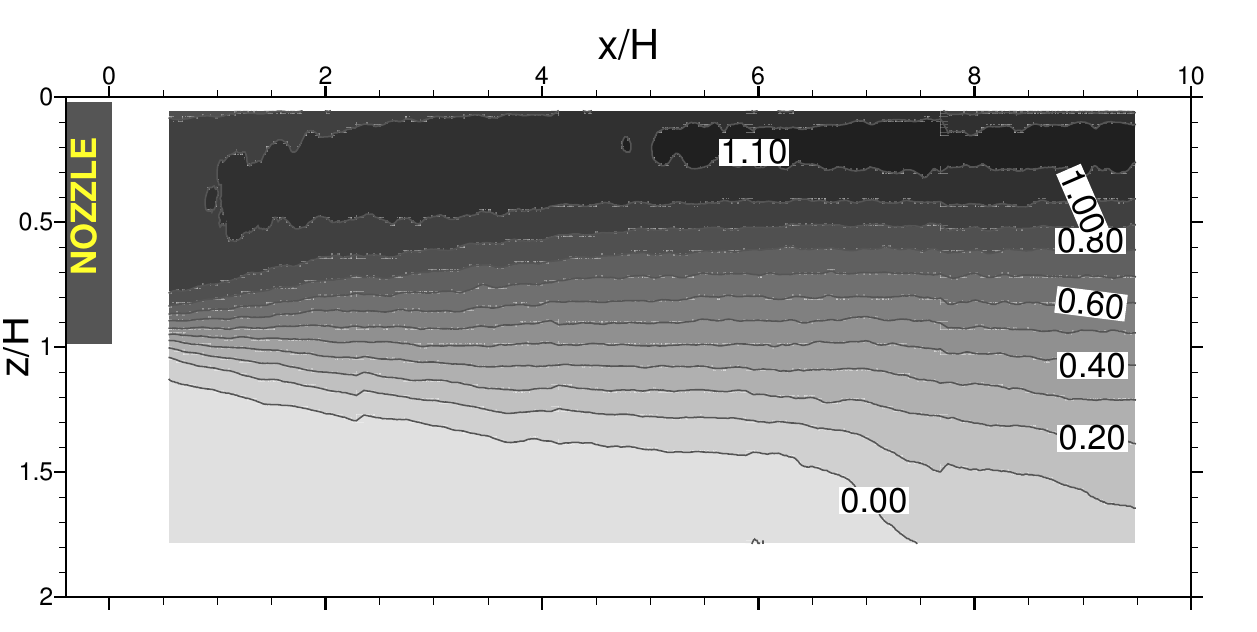}\includegraphics[height=.20\textheight]{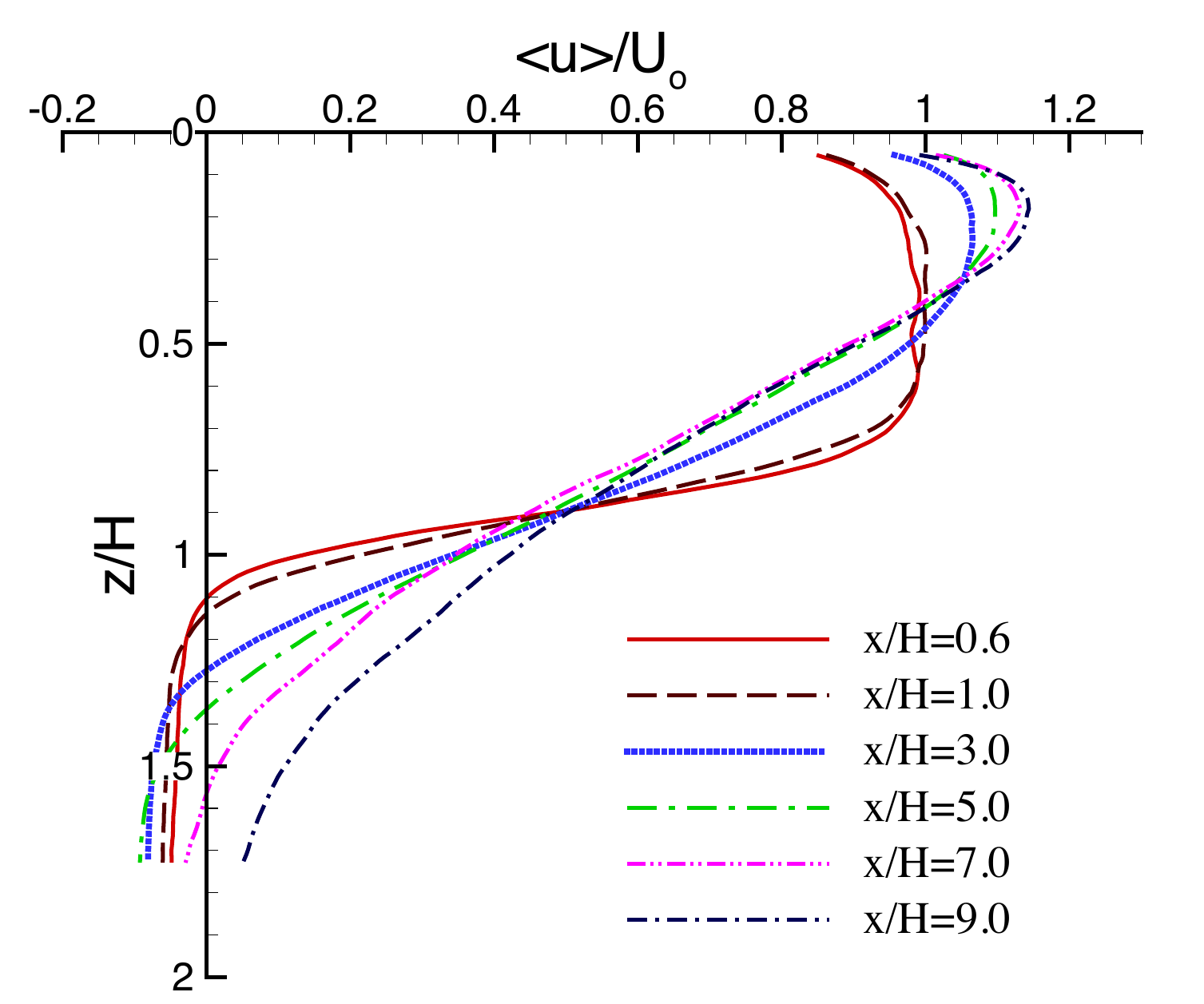}\\
\includegraphics[height=.20\textheight]{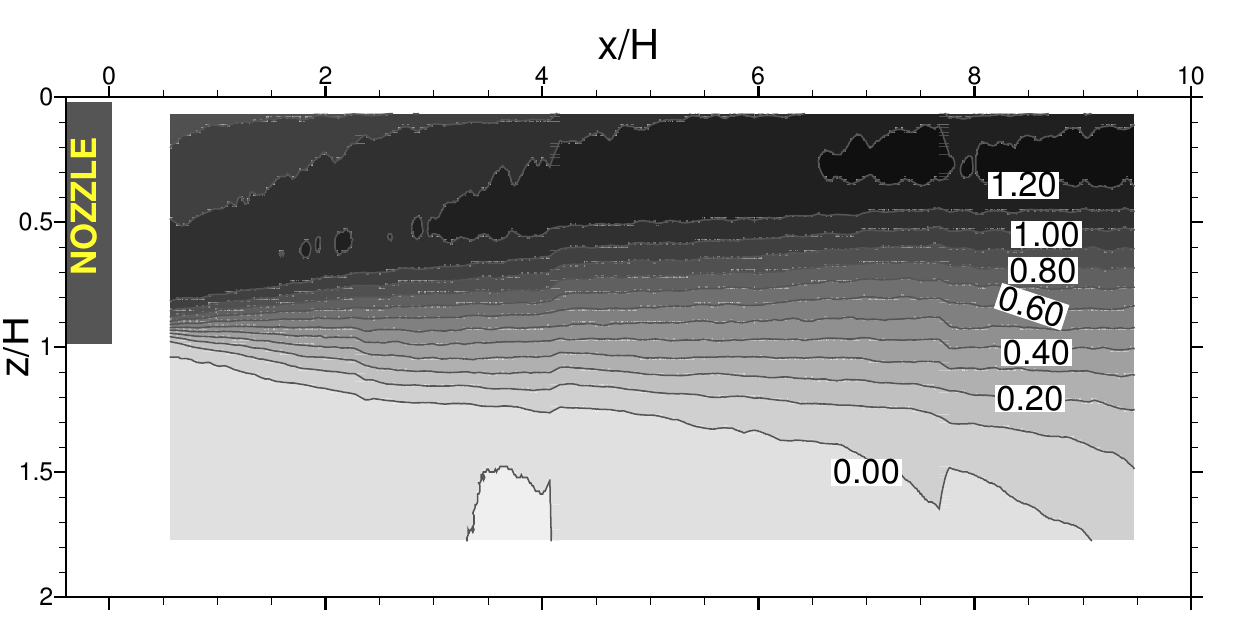}\includegraphics[height=.20\textheight]{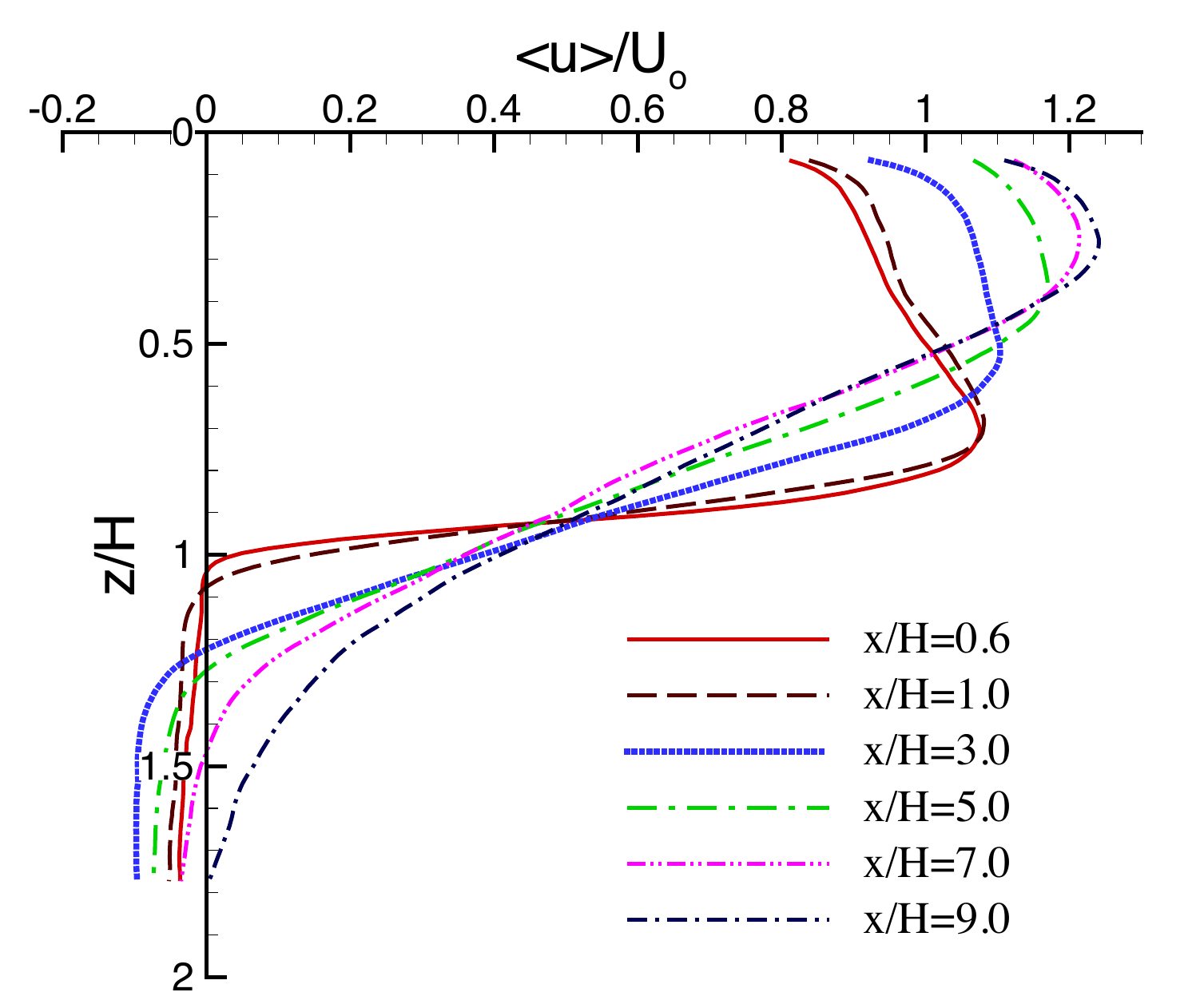}\\
\includegraphics[height=.20\textheight]{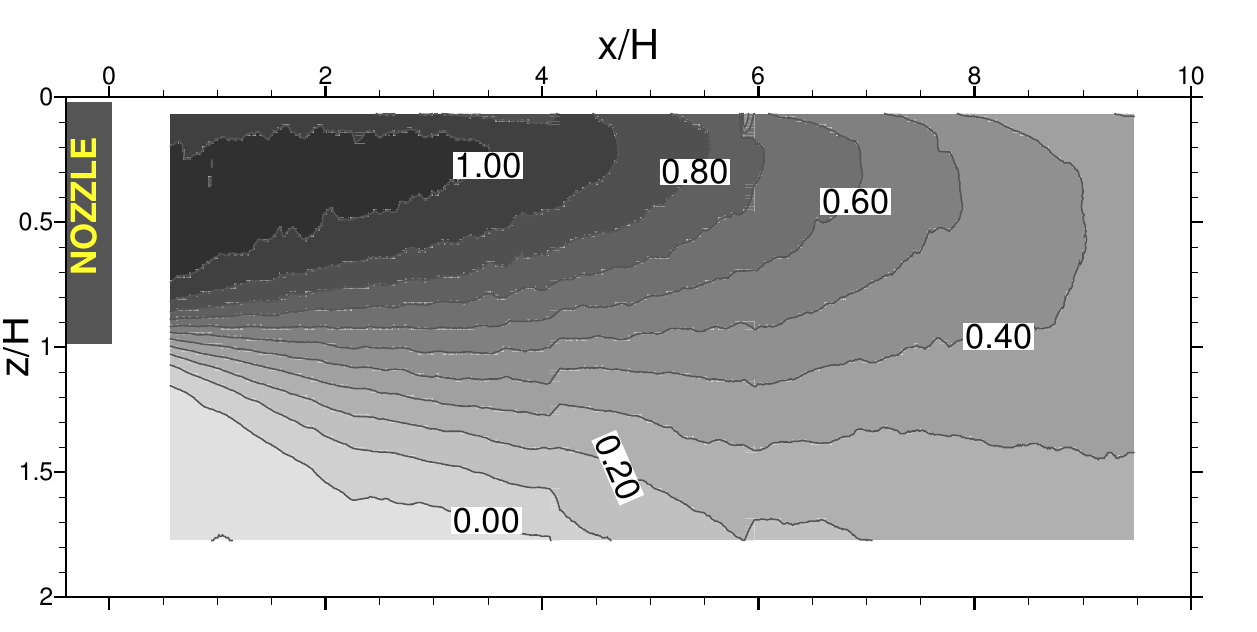}\includegraphics[height=.20\textheight]{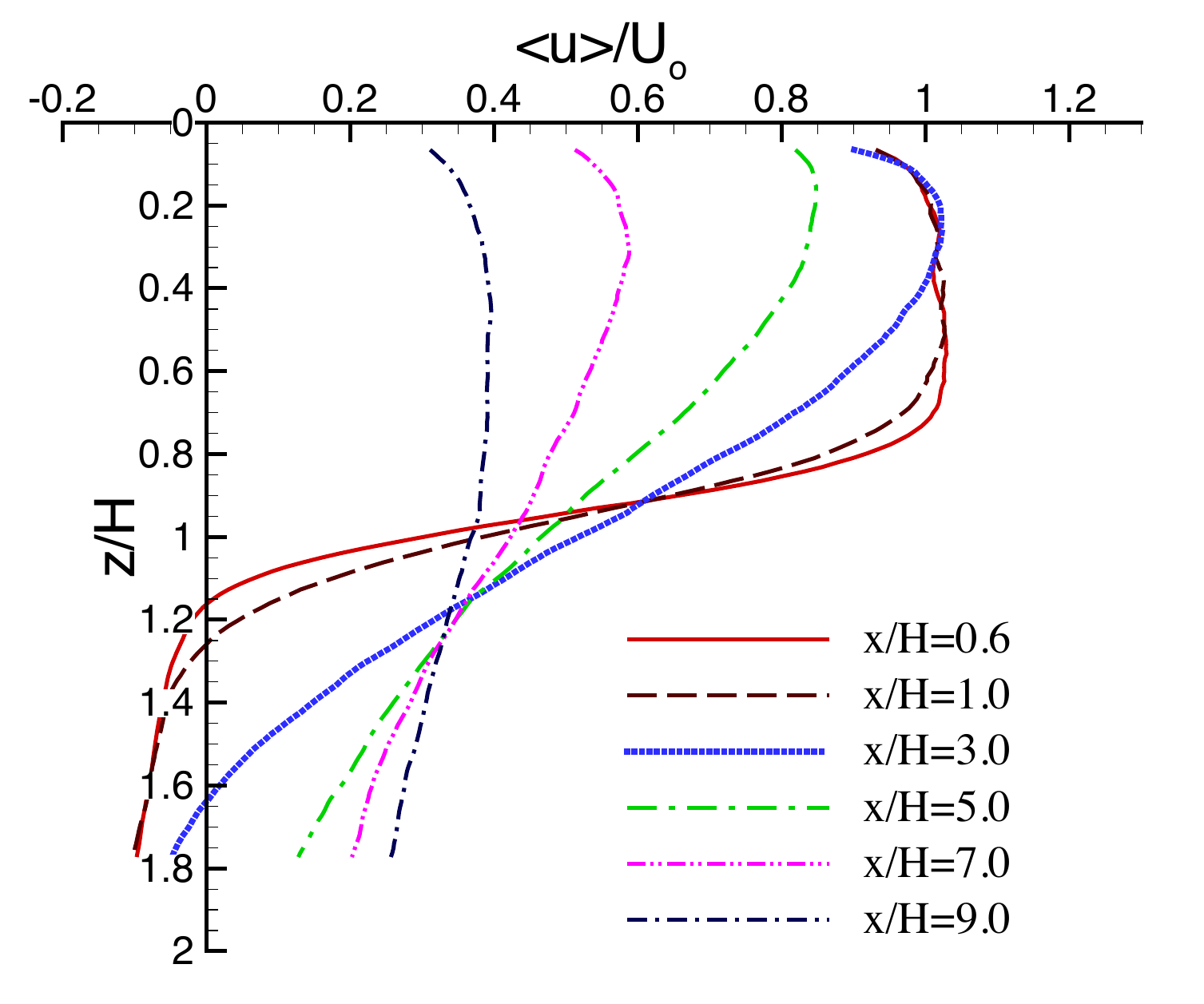}\\
\caption{\label{fig:meanU} Evolution of (left) mean velocity component $\langle u\rangle/U_0$ and (right) $z$-profiles of $\langle u\rangle/U_0$ at different downstream locations: (top) TS, (middle) LS (bottom) NS.}
\end{figure}

\begin{figure}
\centering
\includegraphics[height=.20\textheight]{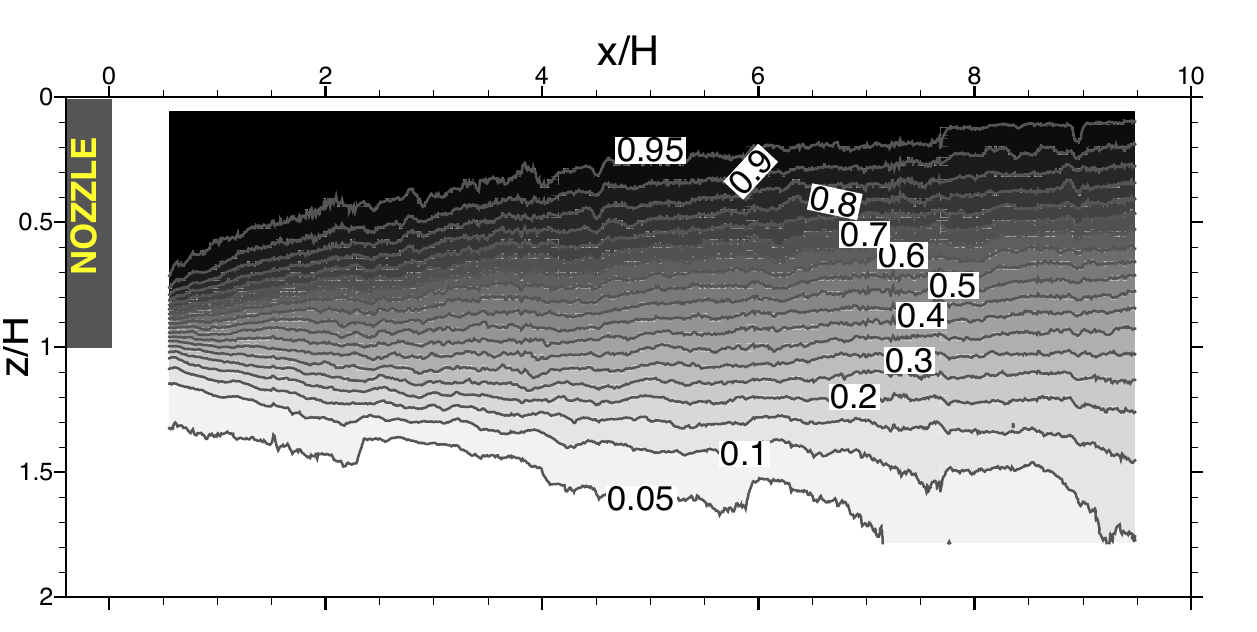}\includegraphics[height=.20\textheight]{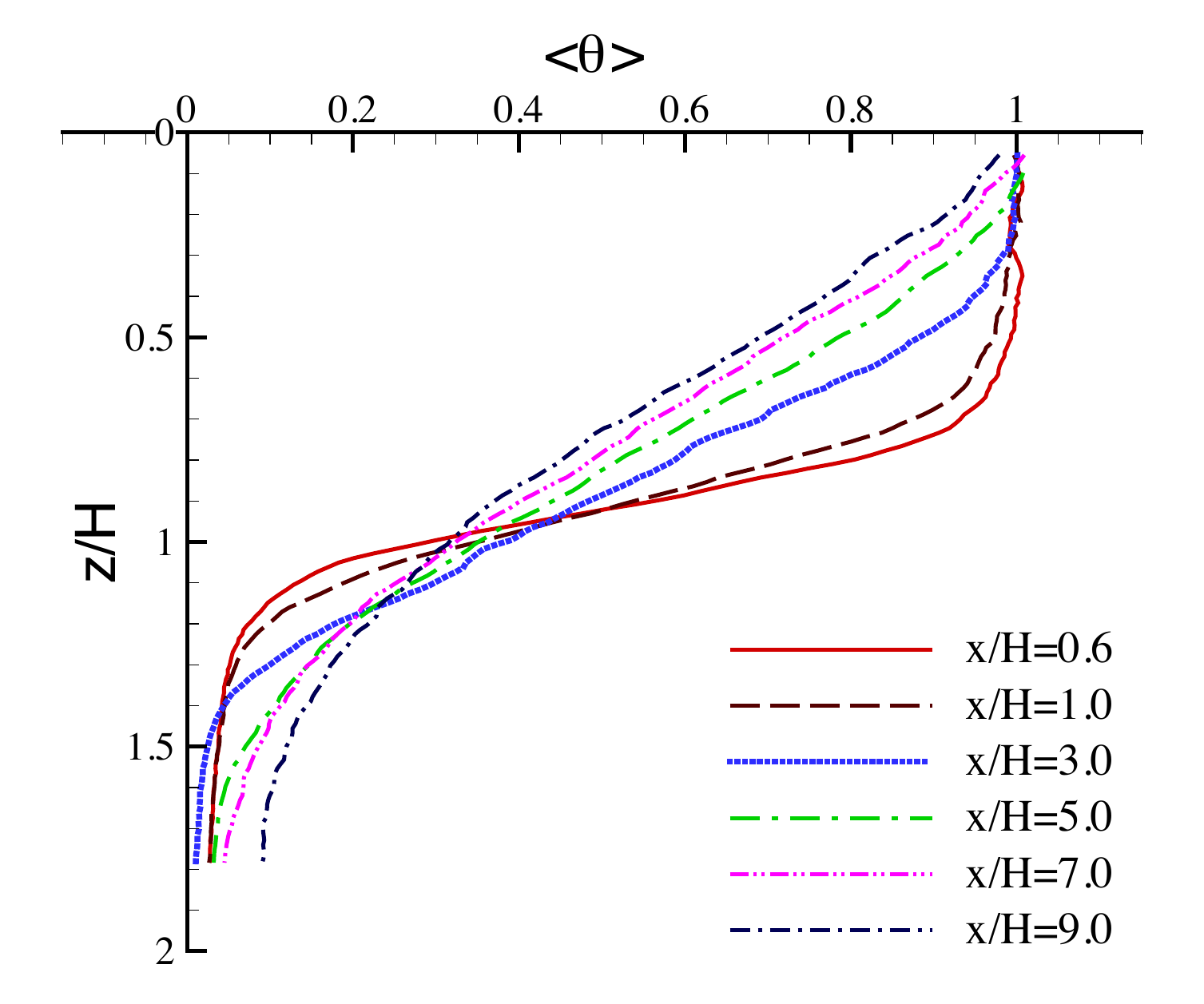}\\
\includegraphics[height=.20\textheight]{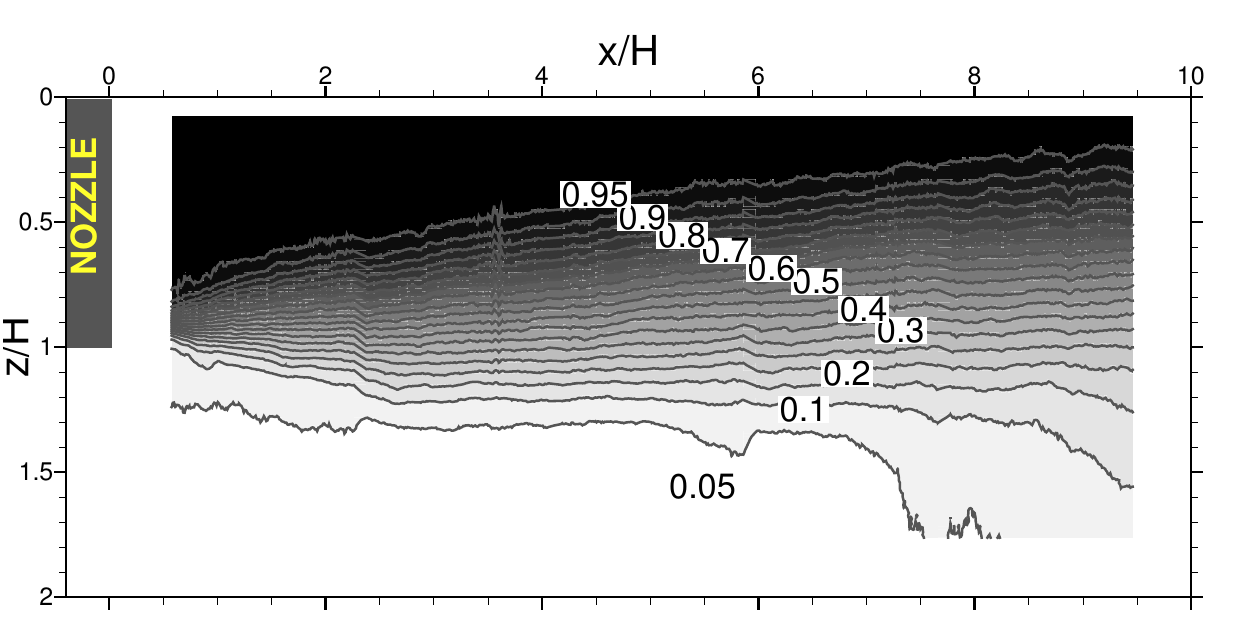}\includegraphics[height=.20\textheight]{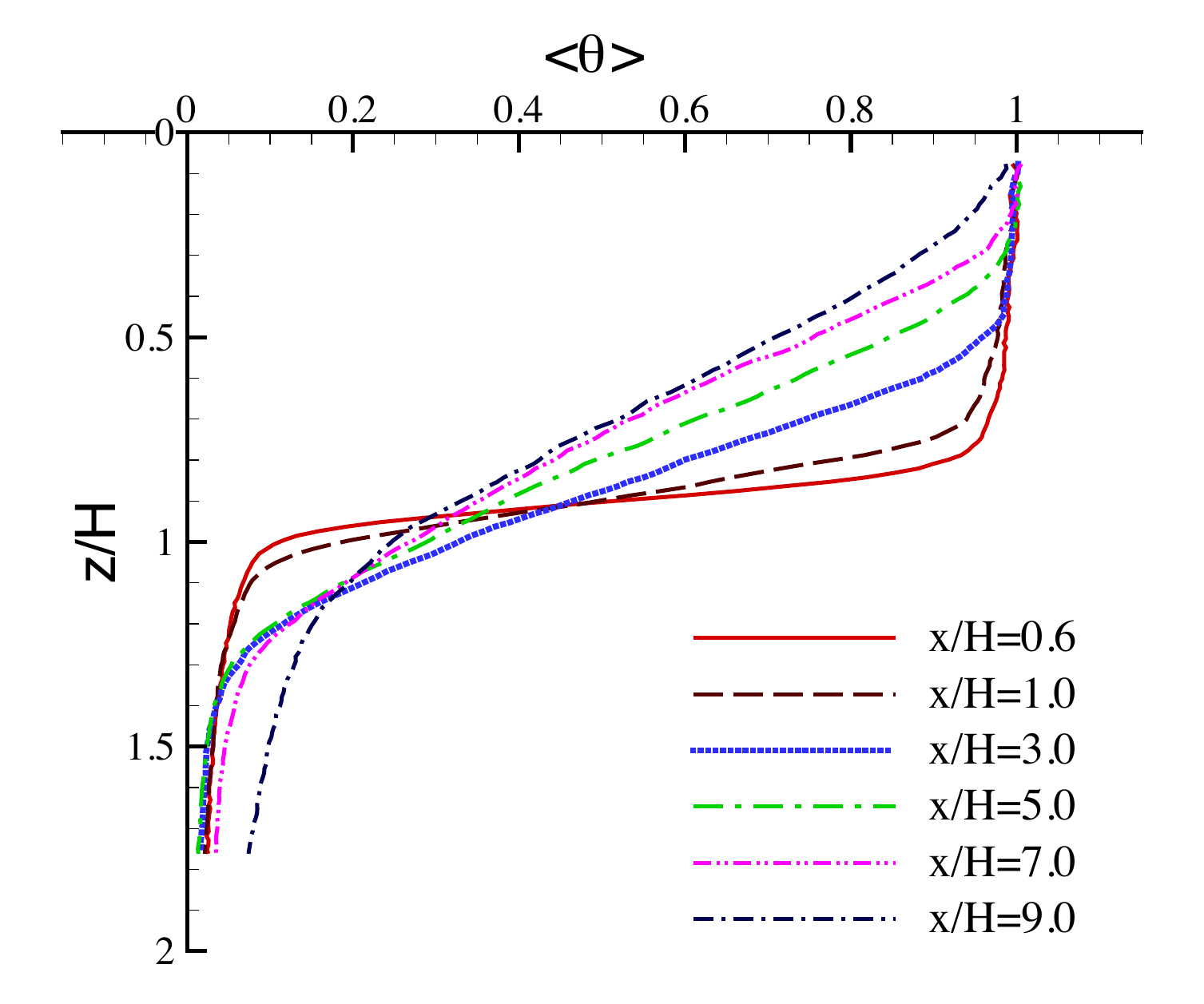}\\
\caption{\label{fig:meanRho} Evolution of (left) mean dynamic density $\langle \theta \rangle$ and (right) $z$-profiles of $\langle \theta \rangle$ at different downstream locations: (top) TS (bottom) LS.}
\end{figure}

The ensemble-averaged values of different parameters are shown in figures \ref{fig:meanU} and \ref{fig:meanRho}. Within the gravity current, the mean stream-wise velocity ($x$-component) is significantly larger than the $z$-component (not shown), i.e., $\langle u\rangle\gg\langle w\rangle$. From figure \ref{fig:meanU}, one sees the expansion of the gravity current owing to mixing and entrainment as it moves downstream, as reflected in the plug-like upper-stream profile being broadened as $x$ increases. The flow domain has four regions in order of their distance from the boundary plate: the plate boundary layer, the nearly undisturbed core region where $\langle u \rangle \approx$ constant, the mixing zone where $\partial \langle u\rangle/\partial z \approx$ constant, and the nearly undisturbed ambient region where $\langle u \rangle \approx 0$, . 
The boundary layer is not well resolved in frames $F_0$ to $F_4$ owing to interface laser reflections but its existence is seen in the rapid decrease of $\langle u\rangle$ as $z$ approaches 0. A self consistent estimate of the boundary layer structure yields a viscous boundary length of about 0.025 cm, a viscous sub-layer length of about 0.12 cm, and a log-layer over a narrow region at about 0.6 cm.

The nearly undisturbed core regime gradually shrinks with downstream distance owing to the development of the mixing zone. This regime disappears around $x/H \simeq 5$ in TS  and $x/H \simeq 7$ in LS. A noticeable increase (15 to 20~\%) of the peak value of $\langle u\rangle$ (denoted as $U_{max}$ thereafter) is observed as a function of downstream distance in both stratified cases, because the injection speed is slightly lower than the velocity imposed by buoyancy. Therefore, the current accelerates slightly and becomes thinner as its driving force changes from pump pressure gradient to buoyancy. 

For comparison, the development of $\langle u\rangle$ in NS is plotted in figure \ref{fig:meanU}(bottom), where the mean flow structure is significantly different. Because there is no gravitational forcing for the unstratified flow, $\langle u\rangle$ decreases and the mixing zone increases as functions of downstream distance over the entire measurement range. There is no increase of $U_{max}$, and the momentum decays quickly owing to mixing and dissipation. The peak value of $\langle u\rangle$ at $x/H=9.0$ is only $40\%$ of its value at $x/H=0.6$. The decay of the peak velocity is more rapid than for an initially laminar wall jet \citep{Wygnanski92}, presumably because of the initially turbulent conditions of the wall jet in our experiments. While for the stratified conditions, we catch the full field of interest over the whole range of downstream distances, in NS, the degradation of the sharp initial velocity gradient is only fully captured for upstream locations, $x/H<3.0$. Note that in the three cases, near the bottom part of the mixing zone,  $\langle u\rangle$ is noticeably negative.

The downstream expansion of the mixing zone, shown in figure \ref{fig:meanU}, can also be observed in figure \ref{fig:meanRho} which shows the density profiles. In the upper portion of the mixing zone, both $\langle u\rangle$ and $\langle\theta\rangle$ have nearly linear $z$-profiles in the mixing region. The majority of the expansion occurs upstream, $x/H \leq 5$  in TS , and is postponed to $x/H \simeq 7$ in LS. In these upstream regions, the magnitude of the velocity and density gradients decreases rapidly with $x$. Further downstream, they approach a non-zero asymptotic value. On the contrary, in NS, $\partial\langle u\rangle/\partial z$ continues to decrease and approaches zero. These observations indicate that without stratification the injected current fully mixes with ambient fluid whereas the nearly saturated structure in the stratified cases explains why this kind of gravity current can travel very long distances in nature.

\begin{figure}
\centering
\includegraphics[width=.32\textwidth]{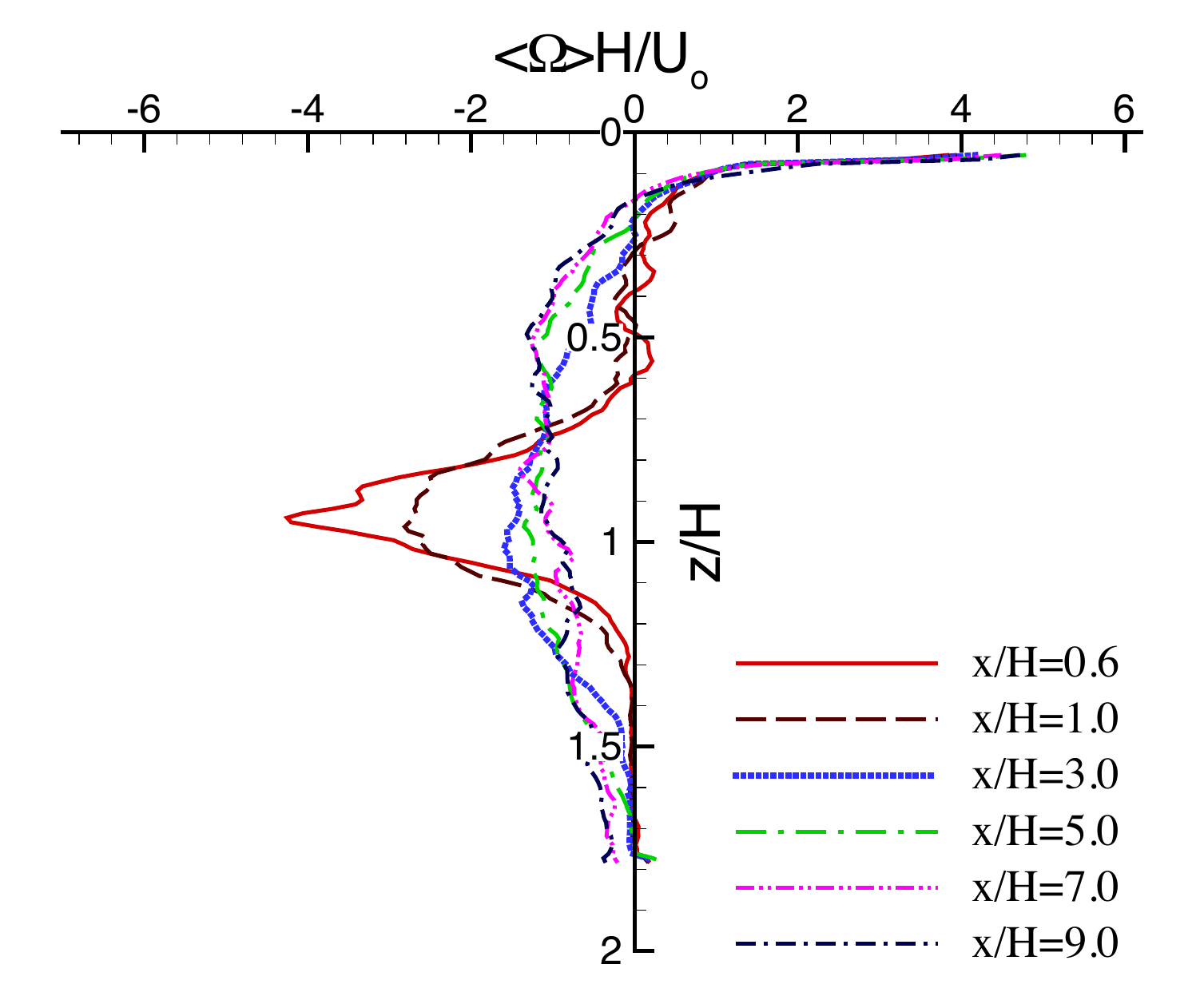}
\includegraphics[width=.32\textwidth]{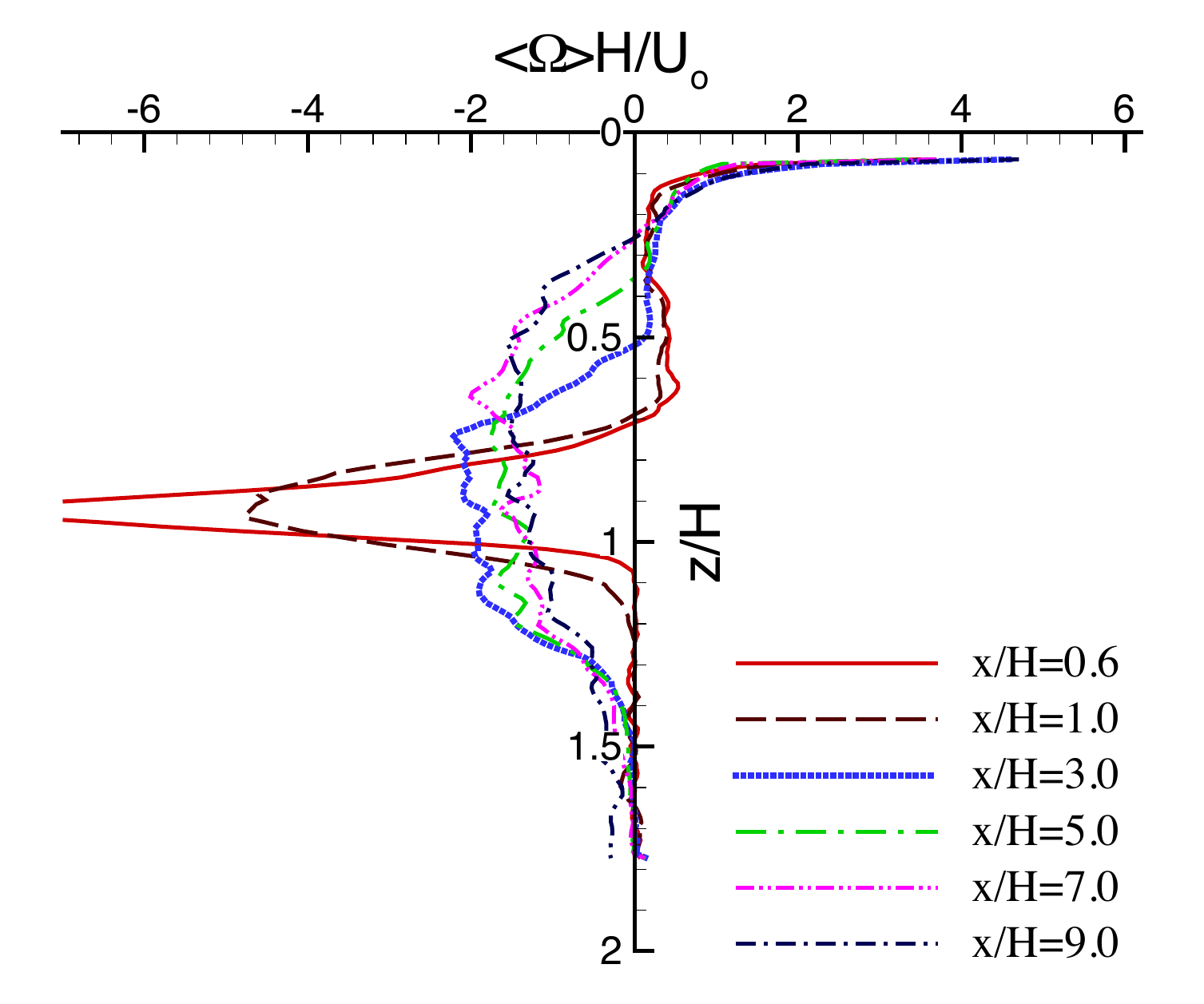}
\includegraphics[width=.32\textwidth]{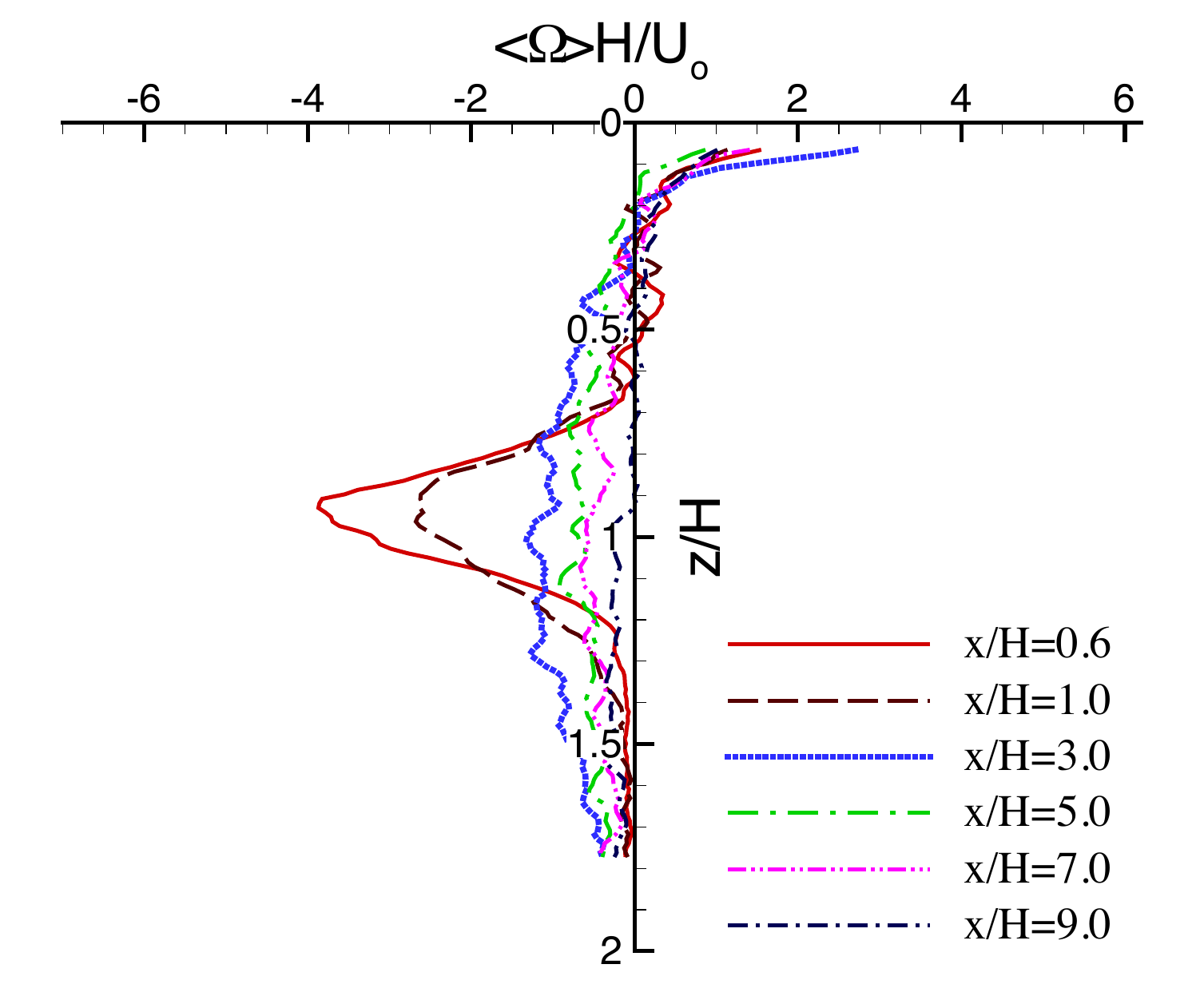}
\caption{\label{fig:meanVort} $z$-profiles of mean vorticity $y$-component $\langle\Omega_y\rangle H/U_0$ at different downstream locations: (left) TS, (middle) LS  (right) NS.}
\end{figure}

We next consider the development of the mixing zone arising from KH instability.  In particular, we compute the out-of-plane vorticity component ($y$-component), $\langle\Omega_y\rangle = {\partial \langle u\rangle}/{\partial z} -{\partial \langle w\rangle}/{\partial x}$ for the three cases considered here. The vertical shear $S=\partial \langle u\rangle/\partial z$ dominates the contribution to $\langle \Omega_y\rangle$, i.e., $S \gg \partial \langle w\rangle/\partial x$, so that $\Omega_y \approx  S$. The development of mean vorticity  $\langle \Omega_y \rangle$ is plotted in figure \ref{fig:meanVort}. In all three cases, $\langle \Omega_y \rangle$ is negative in the mixing zone region, $0.5 < z/H < 1.5$, corresponding to the strong mixing associated with KH instability, and is strongly positive for $z/H < 0.2$, consistent with wall boundary-layer shear.   The first upstream measured peak value of vorticity is highest for LS and is about the same for TS and NS. The downstream evolution of $\langle \Omega_y \rangle$ is very similar for the stratified cases, reaching values of about -1.2$U_0/H$ for  $x/H = 9.0$ whereas vorticity is rapidly dissipated for the unstratified flow, resulting in values of less than -0.2$U_0/H$ for $x/H = 9.0$.  Since the generation of vorticity is dominated by $S$, the similar upstream behavior of TS  and (turbulent) unstratified case is consistent with turbulent vertical momentum diffusion more rapidly dissipating the sharp velocity gradient near the exit nozzle as compared to the laminar flow.  Further downstream, the velocity gradient evolves to a similar shape for the stratified flows owing to the down-plane forcing of gravity whereas the lack of stabilization for the unstratified flow leads to a rapid and continual dissipation of momentum and associated vorticity.  Although KH instability plays an important role in degrading the shear gradient, its direct signature is difficult to see in the averaged vorticity $\langle \Omega_y \rangle$.

\begin{figure}
\centering
\includegraphics[width=0.38\textwidth]{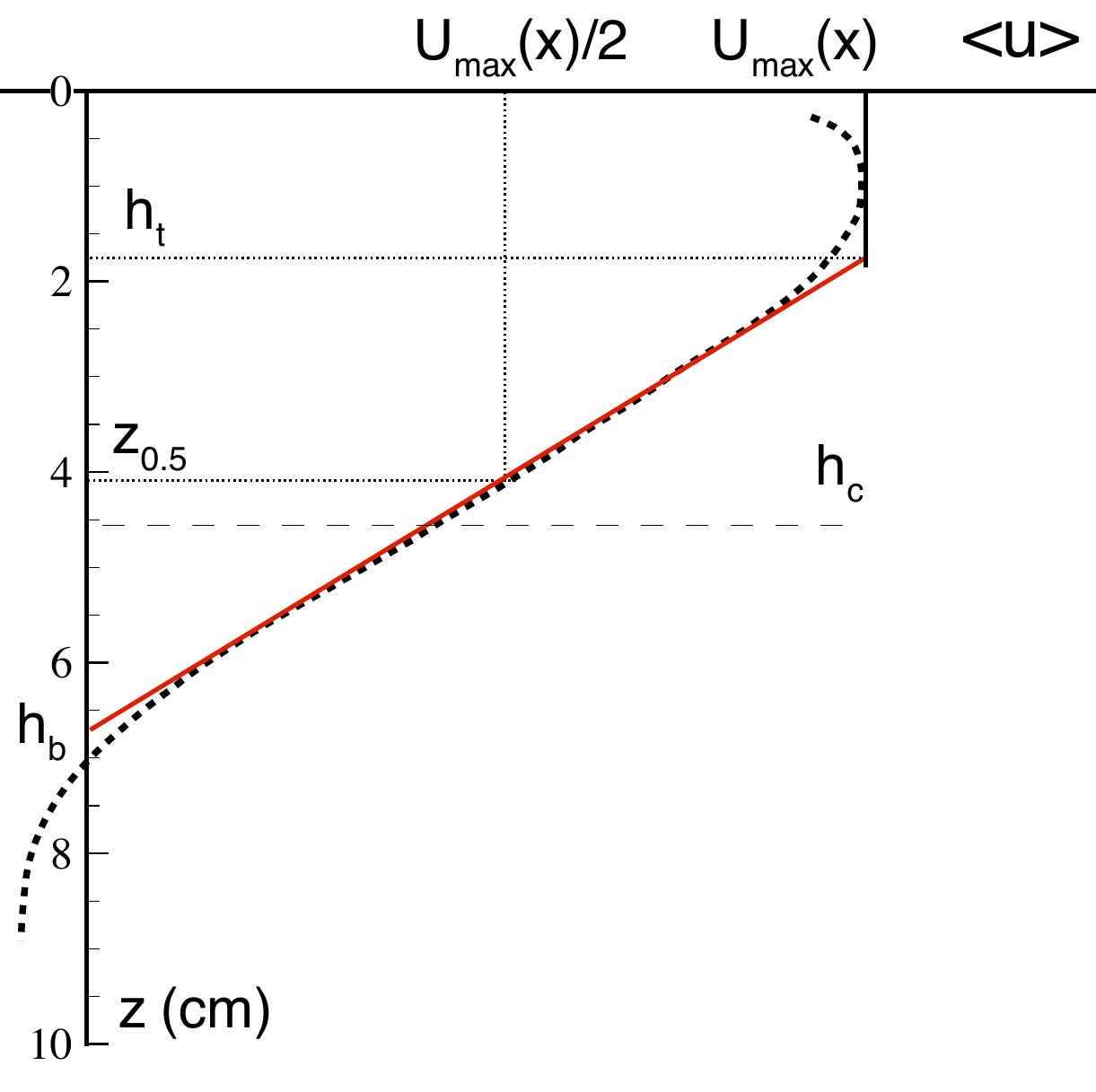}
\includegraphics[width=0.6\textwidth]{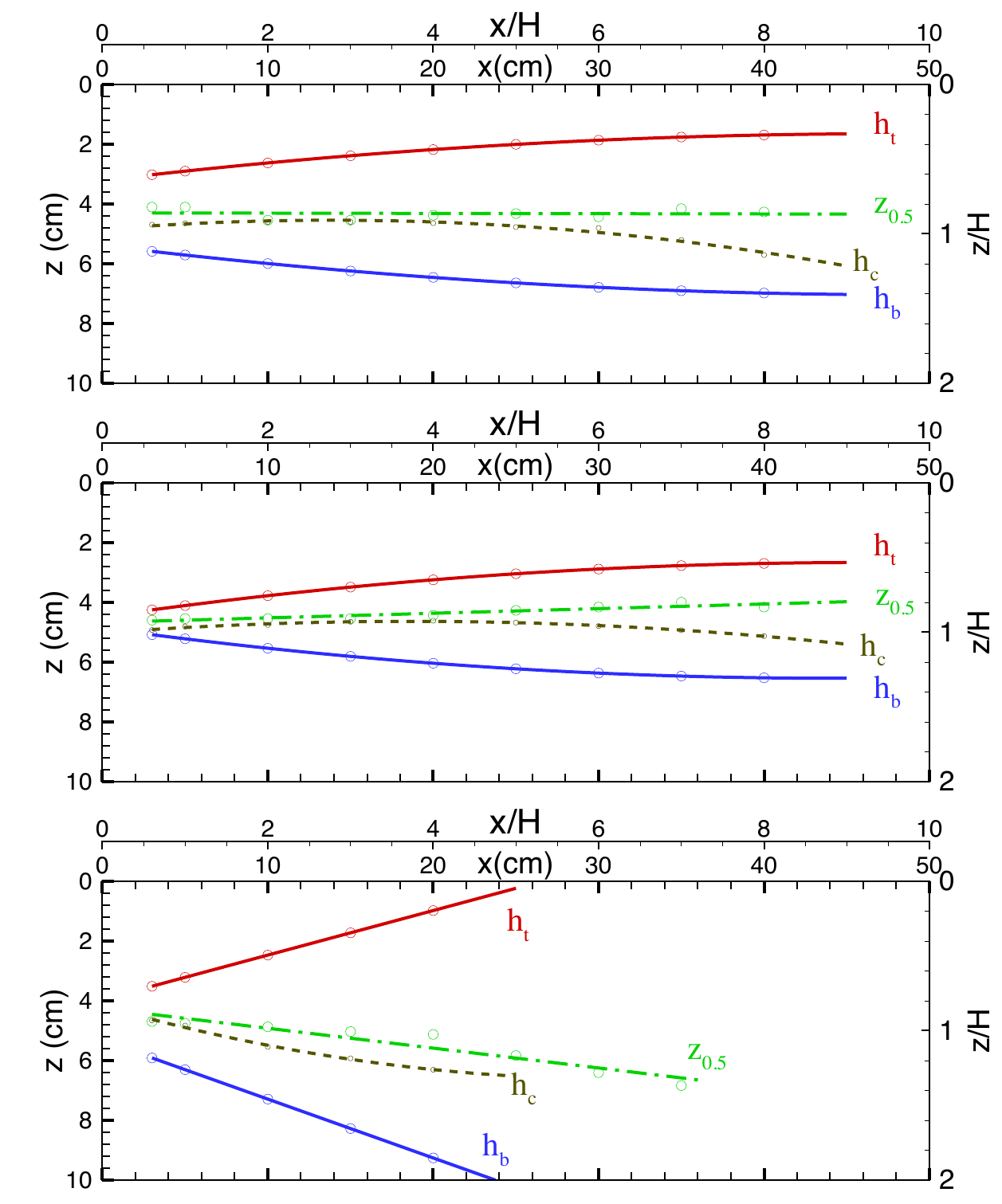}
\caption{\label{fig:mixingzone} {Development of $z_{0.5}$, $h_b$, $h_t$, and $h_c$ showing the spreading of the mixing zone and gravity current in (Top) TS, (Middle) LS (Bottom) NS. Top and left axis show dimensional quantities, whereas right axis and a second top axis show non-dimensional quantities. The sketch on the left illustrates the definitions of $z_{0.5}$, $h_b$, $h_t$, and $h_c$.}}
\end{figure}

To characterize the center line of the mixing zone, as explained in, for example, \cite{Pope00}, a cross-stream location $z_{0.5}(x)$ is defined such that $\langle u(x, z_{0.5})\rangle = 0.5 \cdot U_{\rm{max}}(x)$. Furthermore, to quantify the structural development of the mixing zone, we define envelopes of the mixing zone, $h_b$ and $h_t$, corresponding to the intercepts of the linear curve fitting of $\langle u\rangle$ at $z_{0.5}$ to vertical axis $\langle u\rangle =0$ and to the maximum mean velocity $\langle u\rangle = U_{\rm{max}}(x)$, respectively, as described in the left sketch of figure \ref{fig:mixingzone}. 

{\cite{Wells:JPO:10} used a different definition for the edge of the gravity current:}

\begin{equation}
h_c(x)=2\left(\int_0^\infty z\langle u(x, z)\rangle dz\right)/\left(\int_0^\infty \langle u(x, z)\rangle dz\right), 
\label{eq:hc}
\end{equation}

\noindent where integration is truncated at $z=9.0~$cm (bottom of field view).


The development of  $z_{0.5}$, $h_b$, $h_t$ and $h_c$ shows, figure \ref{fig:mixingzone}, the expansion of the mixing zone along the $x$-direction. The center of the mixing zone, well described by either $h_c$ or $z_{0.5}$, is nearly constant with $x$ for both stratified flows.  For NS, the flow spreads vertically at a rate $dz_{0.5}/dx = 0.07\pm0.01$, consistent with previous measurements of a wall jet with zero ambient velocity~\citep{Schwarz61,Launder83,Wygnanski92}.

\begin{figure}
\centering
\includegraphics[width=0.90\textwidth]{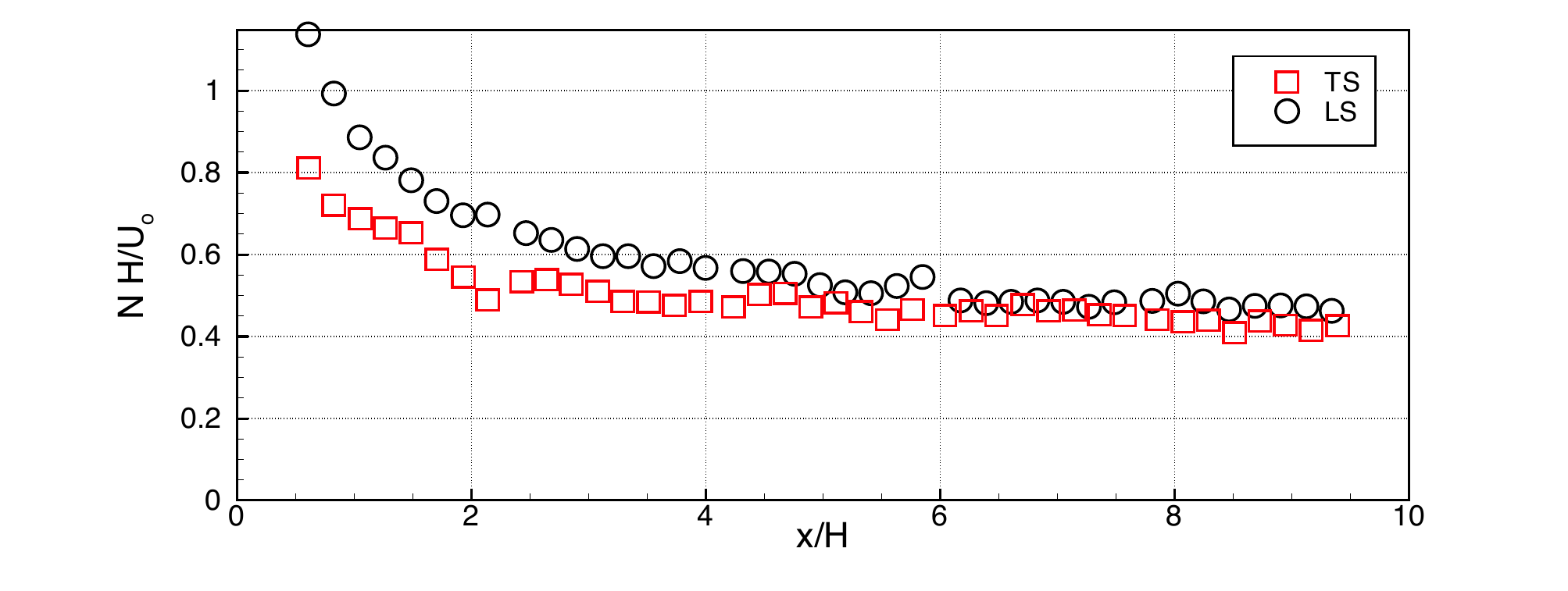}
\caption{\label{fig:BVFreq} Brunt-V\"ais\"al\"a frequency $N$ vs $x/H$ computed by averaging over the mixing zone.}
\end{figure}

Stratification competes with turbulence and shear to determine the shape and entrainment characteristics
of gravity currents. The degree of stratification is quantified by the Brunt-V\"ais\"al\"a frequency \cite[e.g.,][]{Knauss78} $N=\sqrt{g'\pt \theta/\pt z}$. $N$ decays gradually as $x$ increases, see figure \ref{fig:BVFreq}, and mixing reduces the density difference. In this figure, $N$ is computed using vertical averages of the density gradient within the mixing zone (between $z=h_b$ and $z=h_t$, see sketch in figure \ref{fig:mixingzone}). \footnote{Vertical averages are used throughout the paper. For quantities involving gradients of velocity and density, the averaging height should be limited to the mixing zone because the gradients vanish outside the mixing region. We define this zone using the heights $h_b$ and $h_t$. Vertical averages of the velocity and density, on the other hand, are defined by averaging over the whole height of the current.}  At short downstream distance, $N$ is larger for the LS case, owing to the sharper density gradient, which has not been eroded by the initial turbulence. The velocity gradient is, however, also initially steeper in the LS case for the same reason (see figure~\ref{fig:meanU}), inducing stronger KH instability, so that further downstream, $N$ becomes comparable between LS and TS cases. 

\section{Entrainment}\label{sec:entr}

In this section, we use mean quantities to compute the entrainment coefficient, commonly used in oceanographic applications to account for the ambient fluid entrained by a gravity current. We start by discussing in detail the definition of the main control parameter, the Richardson number.

\subsection{Richardson Number}\label{subsec:BVfreq}

The amount of ambient fluid entrained by the current depends on the Richardson number~\citep{Ellison59} but there are several ways to define this parameter, and we will explore these differences in detail. The bulk Richardson number $Ri_0$ ignores the evolution
of the flow with $x$. Inside the mixing zone, mixing is produced by an instability whose strength depends on the magnitude and structure of the shear in the $z$-direction of the streamwise velocity component \cite[see, e.g.,][]{Peltier03} as measured by the gradient Richardson number:

\begin{equation}
Ri_g=g'\frac{\partial\overline{\langle\theta\rangle}/\partial z}{\left(\partial\overline{\langle u\rangle}/\partial z\right)^2}=\frac{\overline N^2}{\overline S^2}.
\label{eq:GradRiNum}
\end{equation}

\noindent the $\overline{~\cdot~}$ notation indicates vertical averaging over the mixing zone, defined between the height $h_t(x)$ and $h_b(x)$.

Another way to measure the balance between stratification and shear effects is to define a local bulk Richardson number, using local quantities defined at the top of the mixing region:

\begin{equation}
Ri(x)=g'\frac{\langle\theta(x,z=h_t)\rangle \Delta h(x)}{\langle u(x, z=h_t )\rangle^2}
\label{eq:bulkRiNum}
\end{equation}

\begin{figure}
\centering
\includegraphics[width=0.90\textwidth]{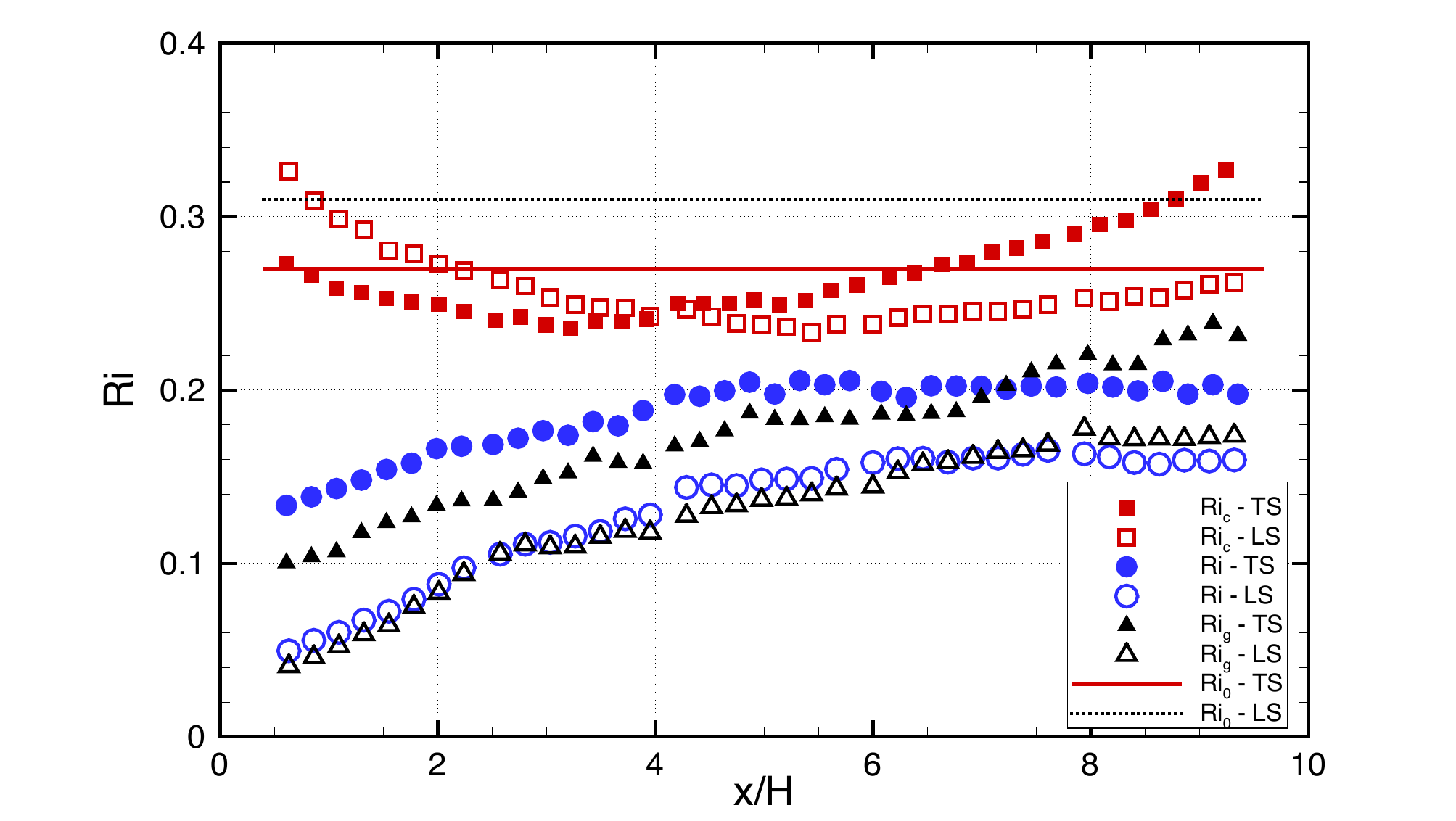}
\caption{\label{fig:RicNum} Definitions of Richardson Number for the gravity current vs $x/H$: $Ri_0$ (table \ref{tab:parameters}), $Ri_g$ (\ref{eq:GradRiNum}), $Ri$ (\ref{eq:bulkRiNum}) and $Ri_c$ (\ref{eq:Ric}). Same symbols indicate same definition, full symbols for TS case, open symbols for LS case}
\end{figure}

\noindent For idealized velocity and density profiles -- flat between the plate and $h_t$, linear between $h_t$ and $h_b$ and zero below $h_b$ -- the definitions~\ref{eq:GradRiNum} and~\ref{eq:bulkRiNum} are mathematically equivalent. The development of the Richardson numbers computed using both definitions is shown in figure \ref{fig:RicNum} for TS and LS. Both definitions give very similar values along the whole range of downstream distance. This observation would allow the use of definition~\ref{eq:bulkRiNum} in the computation of a local Richardson number in oceanic situations where access to accurate gradient quantities is sometimes difficult.

The Richardson number is always larger in TS compared with LS, implying a more stable current in TS with respect to KH instability. In figures~\ref{fig:meanU} and~\ref{fig:meanRho}, both the velocity and density gradients are lower in TS, owing to stronger erosion by turbulent fluctuations. Because the velocity gradient enters as the square in the definition of $Ri_g$, one expects a larger value for TS, as is observed.

$Ri$ and $Ri_g$ increase with $x$, for both TS and LS, attributable to the degradation of the initially strong mean velocity gradient by turbulent fluctuations dominating the weakening of the stabilizing density gradient. The current stabilizes downstream as indicated by nearly constant Richardson numbers for $x/H > 8$. Contrary to the convergence of $N$ with increasing $x$ between LS and TS, values of $Ri$ remain distinctly different at large $x$.

Because many existing experiments do not have detailed profile information, the Rich\-ardson number is often defined using accessible quantities averaged over the thickness of the gravity current, e.g.,

\begin{equation}
Ri_c=g'\frac{(1-\theta_c(x))\cdot h_c(x)}{U^2_c(x)}
\label{eq:Ric}
\end{equation}

\noindent where $U_c(x)$ and $\theta_c(x)$ are defined as: 

$$
U_c(x)  = \frac{1}{h_c(x)}\int_0^{h_c(x)}\langle u(x,z)\rangle dz\label{eq:meanU}~~~{\rm and}~~~
\theta_c(x)  = \frac{1}{h_c(x)}\int_0^{h_c(x)}\langle\theta(x,z)\rangle dz\label{eq:meanRhod},
$$

\noindent and averages in this case directly involve velocity and density, which do not vanish close to the inclined plate, so the averaging zone is defined differently from averages involving gradient quantities. The integrals are calculated from the inclined plate ($z=0$) to the center of the mixing region defined by $h_c$. The resulting $Ri_c$ is similar to $Ri_0$ with a small decrease at intermediate $x$, quite different from $Ri$ and $Ri_g$, see figure \ref{fig:RicNum}. This distinction is
important in determining the entrainment parameterization because different definitions produce different values of the control parameter, i.e., the Richardson number.

\subsection{Entrainment analysis}

The rate of fluid entrainment is an important parameter in many inhomogeneous turbulent flows including
gravity currents and oceanic overflows.  In large scale ocean simulations in which the mixing process is not fully resolved, entrainment is often parameterized using an entrainment assumption (i.e., $E=f(Ri)$) using some definition of $Ri$. There have been significant efforts to measure the $Ri$ dependence of $E$ using laboratory and field experiments \citep{Strang01, Strang04, Princevac05, Cenedese04}. \citet{Cenedese08} found a bulk Reynolds number dependence with increasing entrainment at higher Re. As discussed in section \ref{sec:introduction}, however, the bulk Reynolds number may not provide a good representation of the flow because of the relatively long distance it takes for the flow to reach steady state.  In our experiments, the active grids in the TS case accelerate the development of a fully turbulent current, compared to the LS case, providing a better understanding of the influence of turbulence in the gravity current on entrainment.

In~\cite{Odier:PhysD:12}, we studied entrainment-detrainment balance based on the correlation of local measurements of density flux and density fluctuations. Here, we compute a global measurement of the entrainment coefficient using two different approaches: 

$i)$ E is determined from the global balance of incoming and outgoing volume flux from a defined control volume assuming incompressibility, i.e., volume flux conservation. The entrainment is then defined using the volume flux crossing the bottom limit of the control volume (the complete derivation is in appendix~\ref{app:entr_vol}):

\be\label{equ:direct_W}
\vert W_v^d(x)\vert= \f{1}{x}\int_0^x \left(\la u\ra\sin\phi-\la w\ra \cos\phi\right)_{(s,h(s))} \d s \; ,
\ee

\noindent where an absolute value is used because the entrainment direction is opposed to the orientation of the $z$ axis. This notation $W$ is somewhat misleading - but widely used - since part of the incoming flux is related to the downstream component of velocity. 

A second, indirect estimate of entrainment velocity uses the difference between the input and output fluxes in the control volume:

\be\label{equ:indirect_W}
\vert W_v^i(x)\vert= \f{1}{x}\left(\int_0^{h(x)}\la u(x,z)\ra\d z-\int_0^H \la u(0,z)\ra \d z\right)~ .
\ee

$ii)$ The second approach uses the dilution of a scalar quantity entering and leaving the same control volume (scalar non-diffusive advection, i.e., dynamic density flux conservation). The direct estimate is (see appendix~\ref{app:entr_dil})

\be\label{equ:dilution_direct}
W_s^d=\f{1}{x\theta_c}\left\vert \theta_c\vert W_v\vert x-\int_0^x (-\la{\theta' u'}\ra\sin\phi+\la{\theta' w'}\ra\cos\phi)\vert_{(s,h(s))} \d s \right\vert\; ,
\ee

\noindent and the indirect estimate is

\be\label{equ:dilution_indirect}
W_s^i=\f{1}{x\theta_c}\left\vert \int_0^{h(x)}\la\theta\ra\la u\ra\vert_{(x,z)}\d z-\int_0^H \la\theta\ra\la u\ra\vert_{(0,z)}\d z\right\vert\; .
\ee

\noindent This effective entrainment velocity is not equal to the one defined from the volume conservation, since it is a balance between the volume entrainment defined earlier, and a turbulent transport contribution, with opposite effects as described in the appendix.\\

For $W_v$ or $W_s$, each direct and indirect estimate has a different source of error: in the direct case, the dominant contribution comes from the measured vertical velocity, which is the least accurately measured component of the velocity field. For the indirect case, the uncertainty arises from the small differences between two large quantities, i.e., input and output fluxes.

Since our measurements start at $x=3$ cm ($x/H=0.6$), the actual input side of the control volume has been defined at this value of $x$ and the actual origin of the $x$ coordinates has been redefined accordingly in the computation of equations~\ref{equ:direct_W} to \ref{equ:dilution_indirect}. For this reason all the curves that we present start at $x/H$=0.6.

For the volume flux conservation, we define the bottom contour using $u(x,z=h_u)=0.1\;U_{in}$, where $U_{in}$ is the vertically averaged velocity at $x=3$ cm (x/H=0.6):

$$U_{in} = \f{1}{H}\int_0^{H} \overline u(x/H=0.6,z)\;\d z~.$$

\begin{figure}
\centering
\includegraphics[width=.7\textwidth]{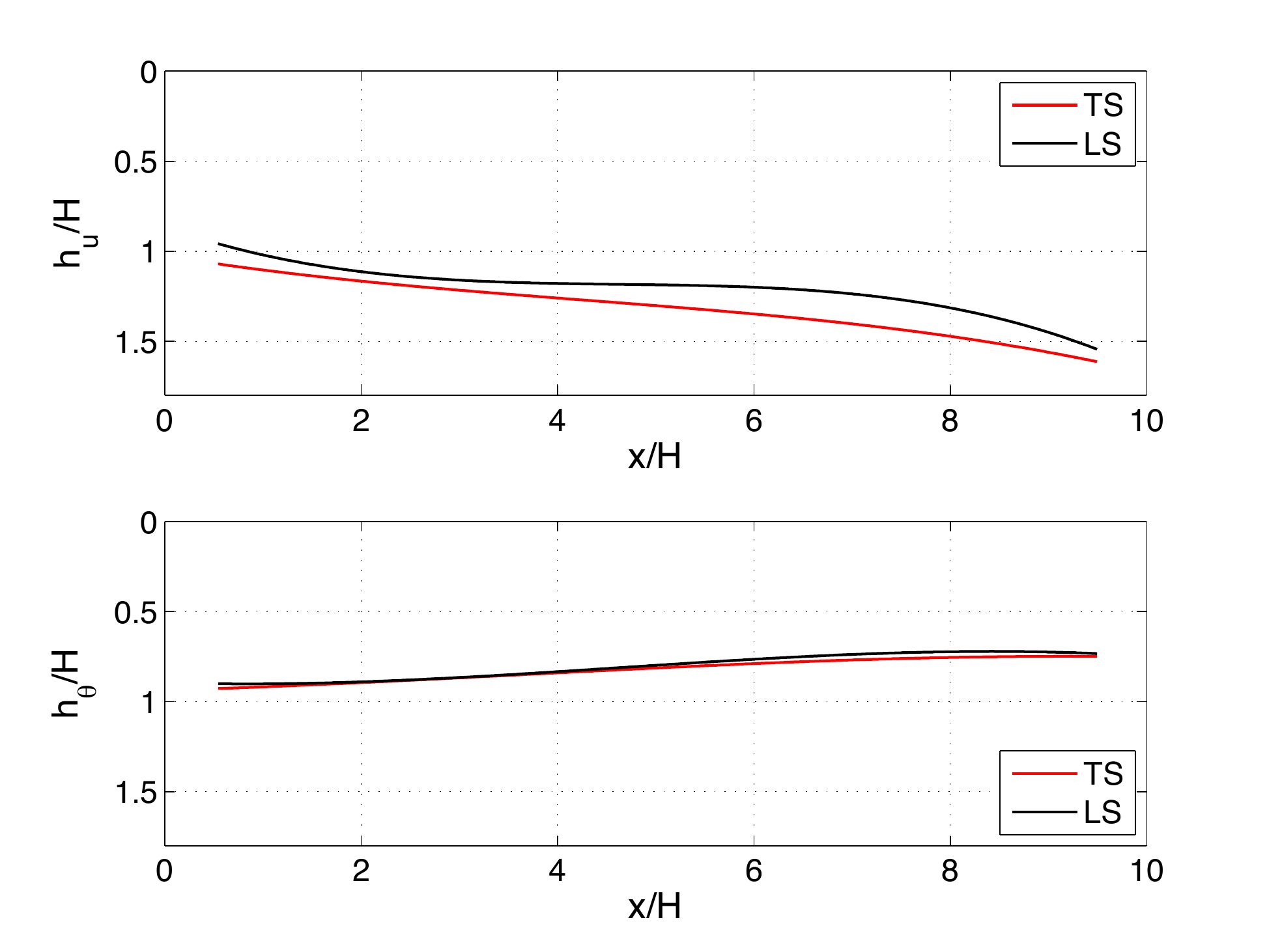}
\caption{\label{fig:box_bottom} (Top) bottom limit of the control volume used for the volume flux conservation study: $u(x,z=h_u)=0.1\;U_{in}$ (Bottom) bottom limit of the control volume used for the scalar dilution study: $\theta(x,z=h_\theta)=0.5$. Red: TS, black: LS.}
\end{figure}

We define the contour using a fraction of the averaged initial velocity and not an averaged local velocity because a local average cannot be defined until the contour is defined. The choice of a fraction equal to 10\% allows the control volume to encompass most of the current, with smaller measurement error. The resulting contour is shown in figure~\ref{fig:box_bottom} (top).

\begin{figure}
\centering
\includegraphics[width=.49\textwidth]{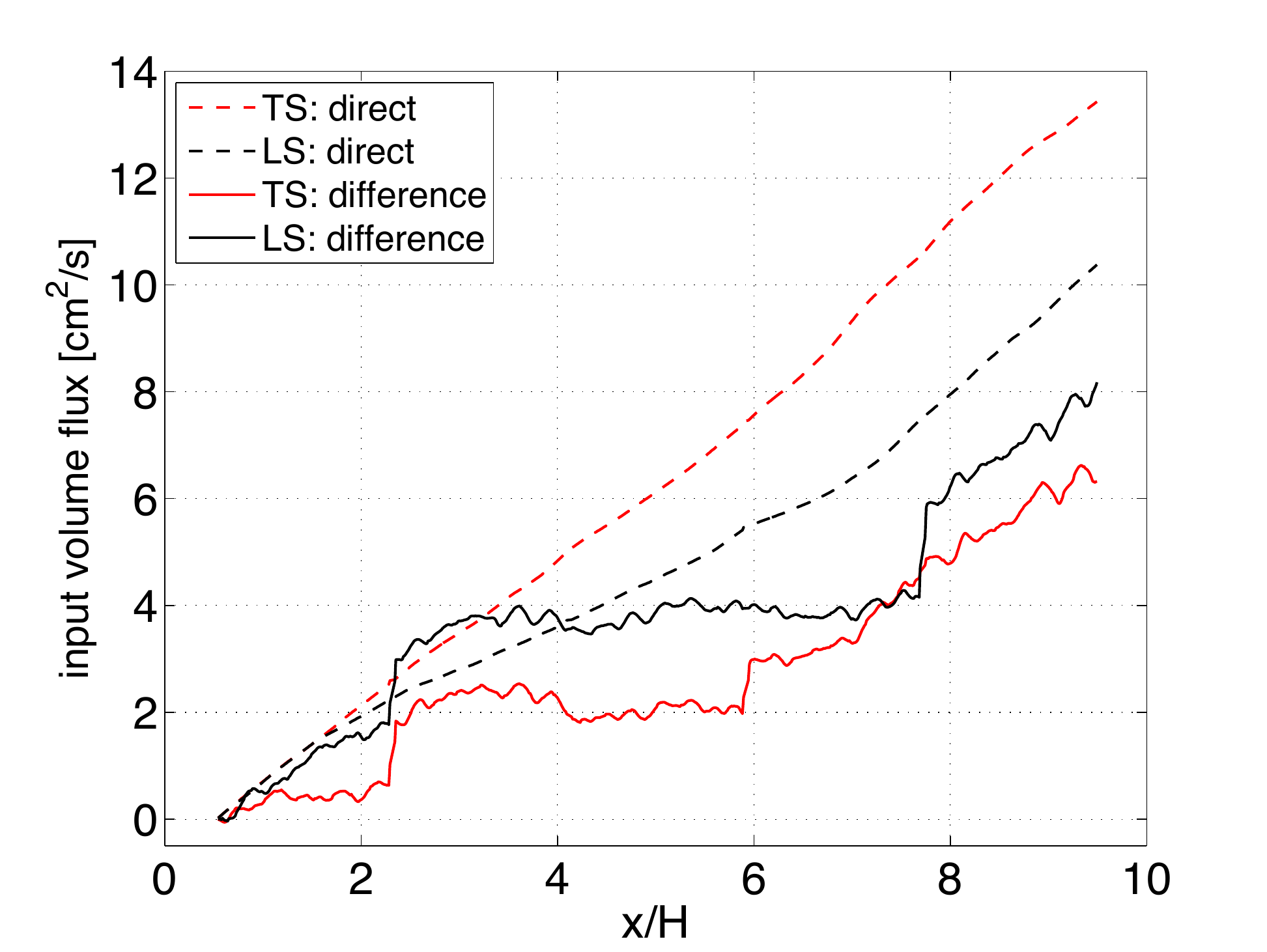}
\includegraphics[width=.49\textwidth]{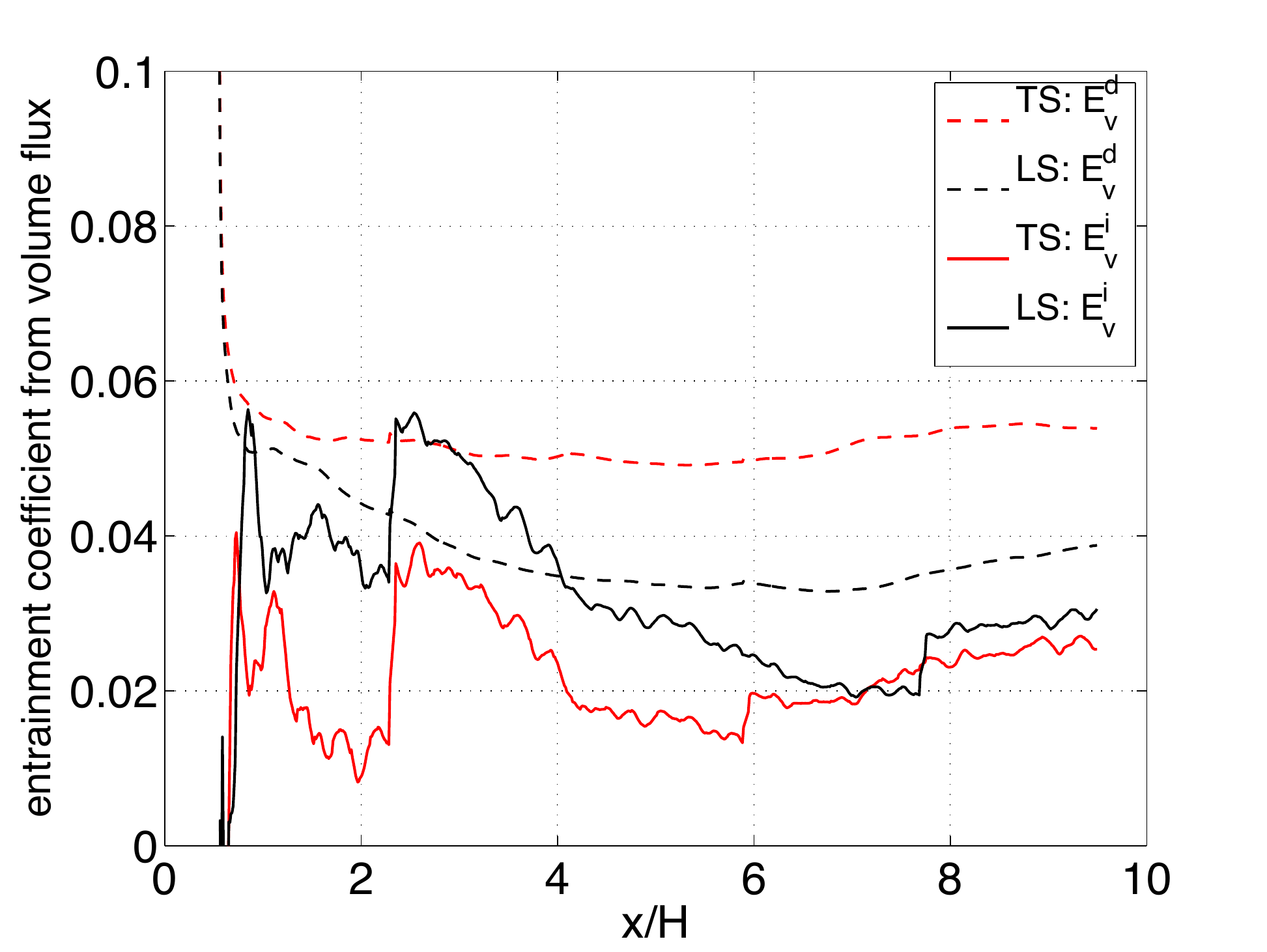}
\caption{\label{fig:volume_fluxes} (Left) Computation of the increase of volume flux in the control volume. (Right) corresponding entrainment coefficient as a function of downstream distance. Dashed lines: direct calculation, solid lines: calculation for the input/output difference. Red: TS, black: LS.}
\end{figure}

Figure~\ref{fig:volume_fluxes} (left) compares the two types of calculation of the increase of volume fluxes, i.e., the right- and left-hand sides of~\eqref{equ:volume_flux_int}, using this contour. There is a discrepancy between the estimates: the direct calculation gives larger values, with a more pronounced difference for TS. 

In addition to sources of error mentioned above, systematic errors may explain this difference: the velocity profiles are assumed to be uniform in the cross stream direction. This assumption is, however, only approximate because the initial lateral width ($y$ direction) of the flow exiting the nozzle is only 90\% of the full channel width and because there is a lateral boundary layer where the vertical velocity goes to zero. Therefore, the direct flow rate calculation yielding $W_v^d$, based on the vertical velocity measured in the center of the tank, overestimates the real flow rate from the bottom. On the other hand, $W_v^i$ is underestimated: the initial flow rate at the nozzle output is overestimated because of the narrower width at this position. Both effects tend to increase the difference between direct and indirect calculations.

The entrainment coefficient is then defined by dividing the entrainment velocities in equations~\ref{equ:direct_W} and~\ref{equ:indirect_W} by $U_{in}$ although another possible choice would be to divide by the local averaged velocity at $x$. Owing to the weak dependence of the $u$ profile with downstream distance, the difference between choices is smaller than our error bars. Figure~\ref{fig:volume_fluxes} (right) shows the entrainment computed from the direct and indirect estimates. For $x/H < 2$, strong fluctuations result from the division by $x$ in the expression for the entrainment velocity. We obtain an average entrainment coefficient by averaging over $x$ for $x/H> 2$. According to the discussion above about possible systematic errors, however, a fair approximation would be to take an average of both estimates (direct and indirect), with an error bar equal their difference divided by $2\sqrt{2}$. With this averaging we obtain $E=0.04\pm 0.01$ for TS and $E=0.032\pm 0.002$ for LS.  

\begin{figure}
\centering
\includegraphics[width=.49\textwidth]{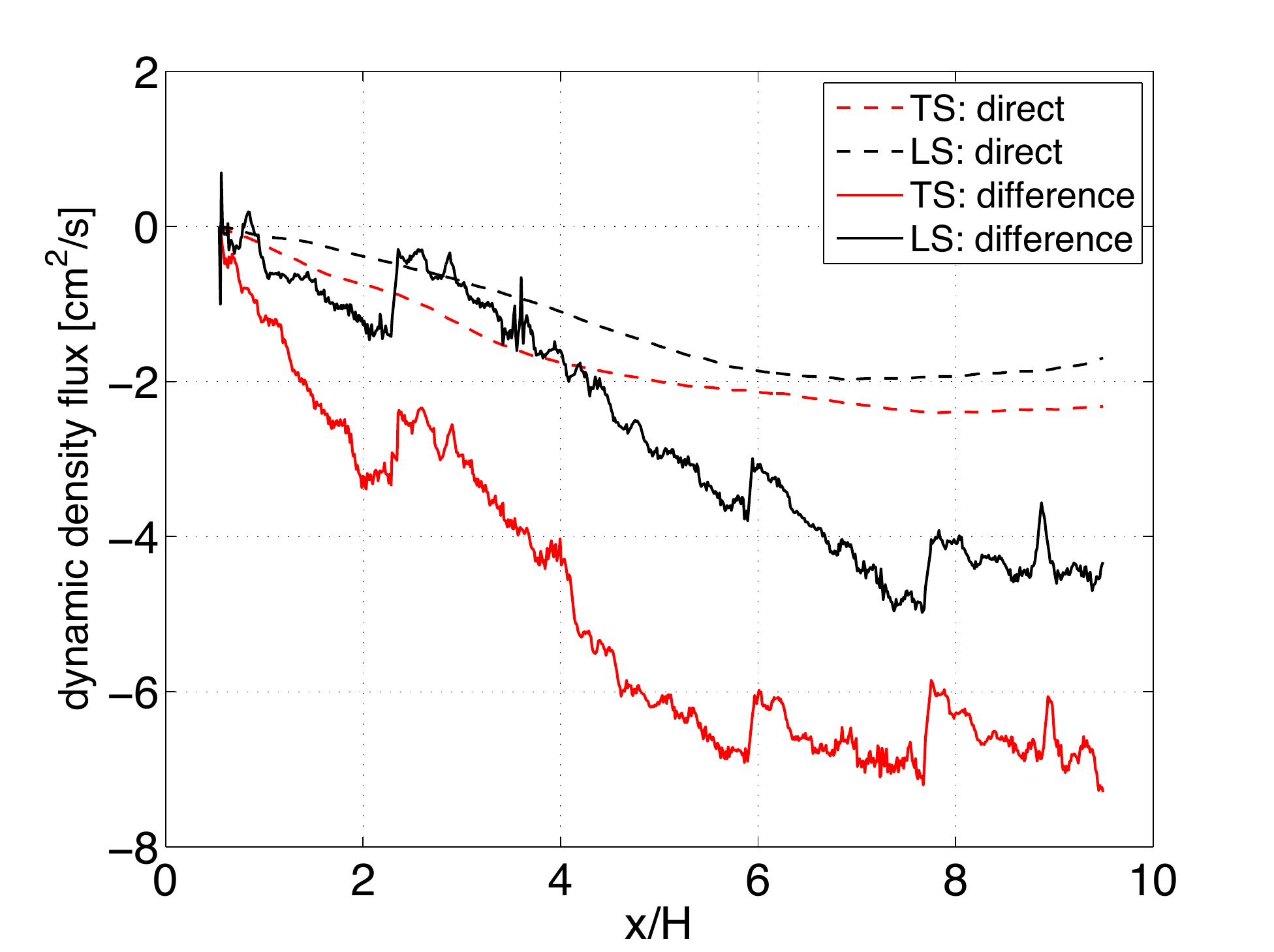}
\includegraphics[width=.49\textwidth]{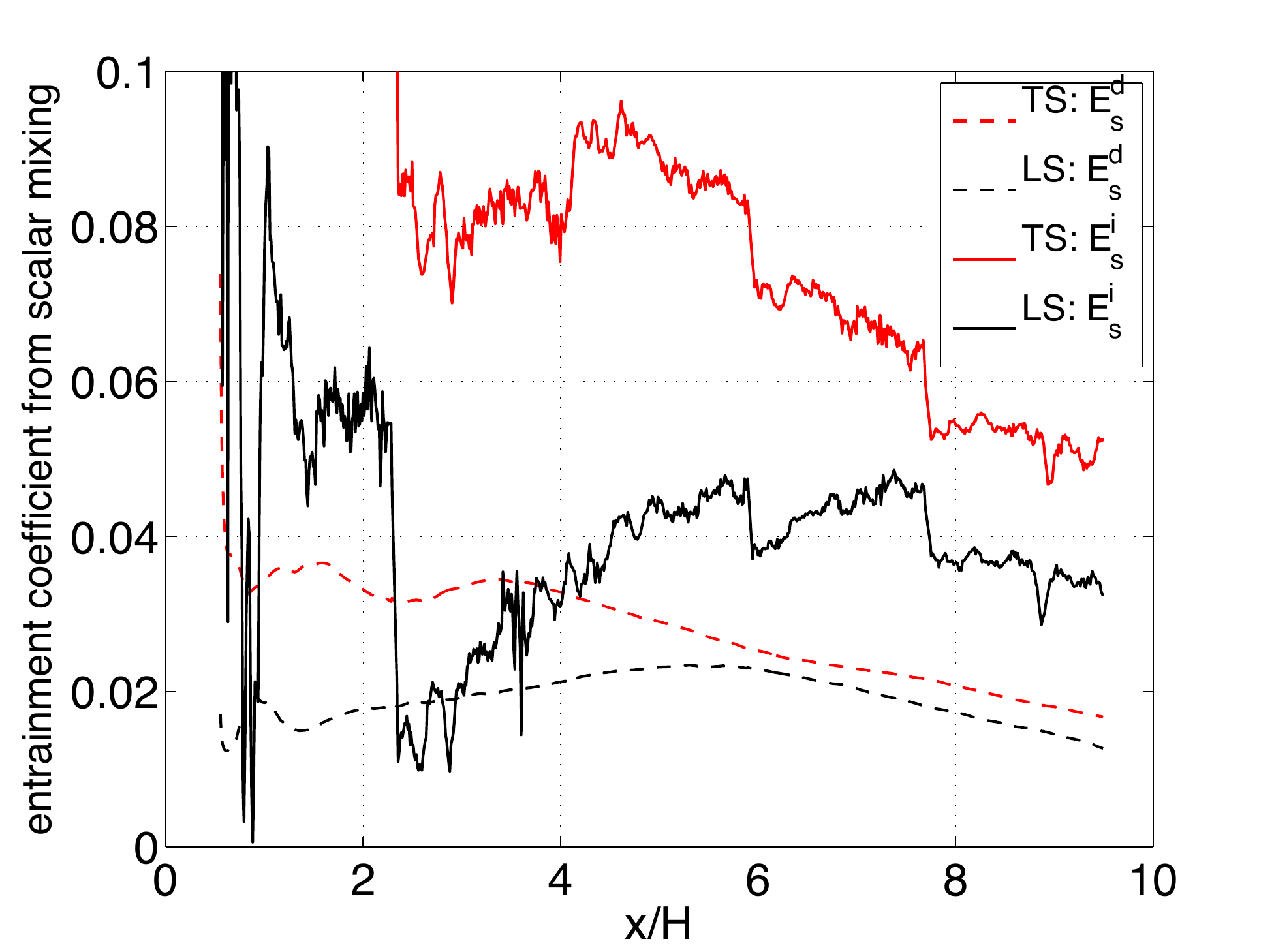}
\caption{\label{fig:dilution_fluxes} (Left) Two different computation of the decrease of dynamic density flux in the control volume. (Right) corresponding entrainment coefficient as a function of downstream distance. Dashed lines: direct calculation, solid lines: calculation for the input/output difference. Red: TS, black: LS.}
\end{figure}

Figure~\ref{fig:dilution_fluxes} (left) shows the variation of dynamic density flux computed directly and indirectly. This quantity decreases implying that the turbulent mixing contribution dominates over the dilution contribution (equation~\ref{dilution4}). There is also a difference between direct and indirect calculations, larger than for the volume fluxes, especially for TS. It is perhaps related to the correction from frame to frame made to the PLIF measurement of density (see section~\ref{subsec:PLIF2Den}). Although this correction seemed to give good results for local measurements in a given frame, it might prove more delicate to apply to the measurement of an integral quantity, over the whole downstream distance (the ``jumps'' observed in the figure are related to a remaining mismatch between the frames). In addition, the systematic errors of the volume flux measurements influence the dilution measurements as well. It is therefore more delicate to estimate the entrainment from these quantities. We show, however, in figure~\ref{fig:dilution_fluxes} (right) the entrainment coefficients computed from equations~\ref{equ:dilution_direct} and~\ref{equ:dilution_indirect}, again dividing by $U_{in}$. Values range between 2 and 8$\%$, with larger values for the indirect calculation.

\begin{figure}
  \centering
  \includegraphics[width=0.65\textwidth]{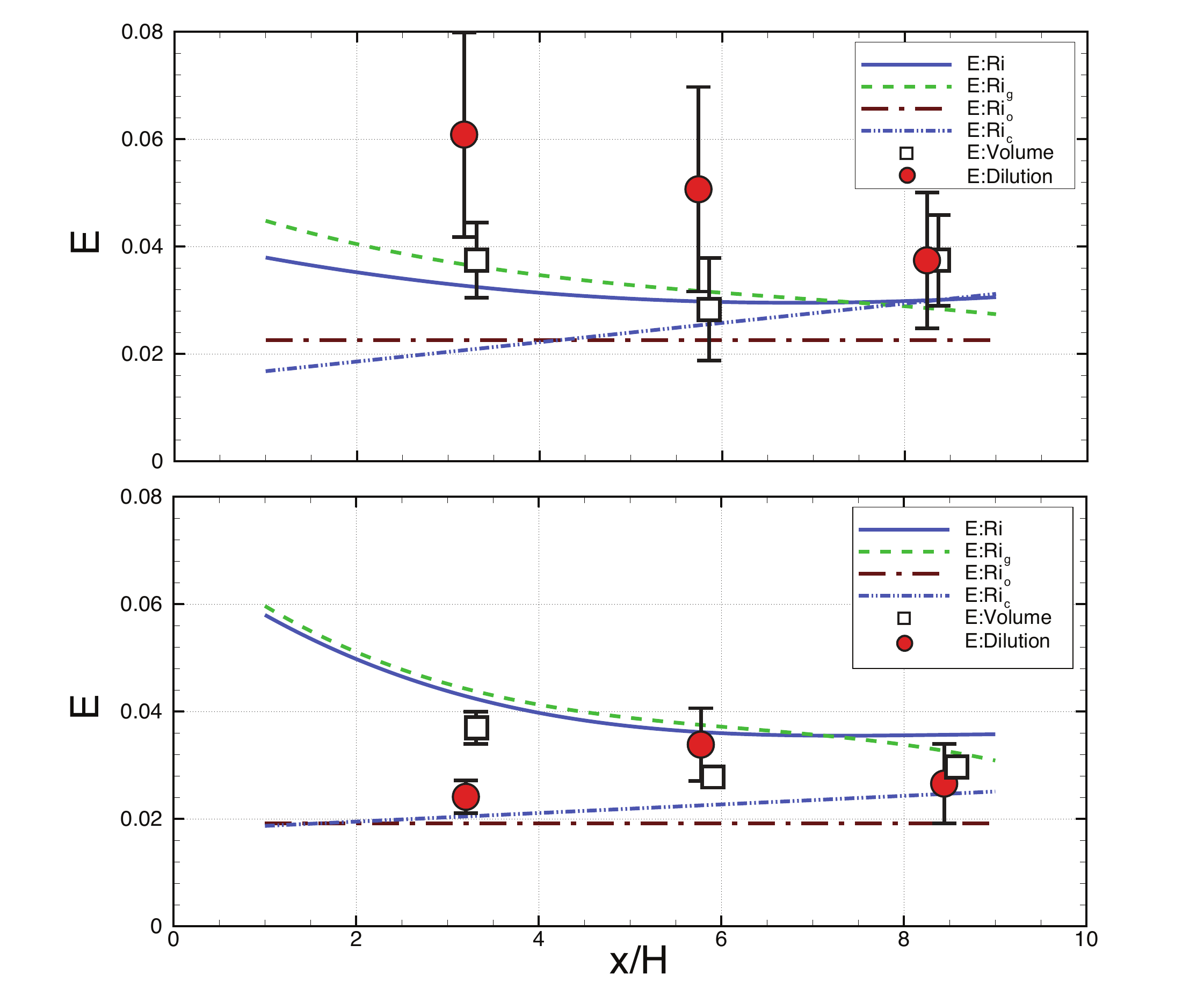}
 \caption{\label{fig:entrainConst} Entrainment constants at different downstream locations estimated from experimental results (average of direct and indirect measurements, from volume flux and scalar mixing). $E(Ri_g)$, $E(Ri_c)$, $E(Ri_0)$ and $E(Ri)$ using the Turner equation: (top) TS, (bottom) LS.}
\end{figure}

The entrainment constant can be calculated by plugging the measured $Ri_g$, $Ri$, $Ri_c$ and $Ri_0$ into the Ellison \& Turner equation, which was given in section~\ref{sec:introduction}. The results are shown in figure \ref{fig:entrainConst} together with our data, which have been averaged in 3 downstream sections of length $2.5H$. The Ellison \& Turner's entrainment constant determined using the gradient Richardson number $Ri_g$ (or the local Richardson number $R_i$) see figure \ref{fig:RicNum}, gives larger values than those obtained using $Ri_0$ or $Ri_c$. Our data, within our experimental error bars, fall generally in-between these two groups of values, except for the dilution estimate in TS, which is slightly larger. Our measurements are consistent with a recent study in a similar set-up~\citep{Krug:EF:13}.

\begin{figure}
  \centering
  \includegraphics[width=0.85\textwidth]{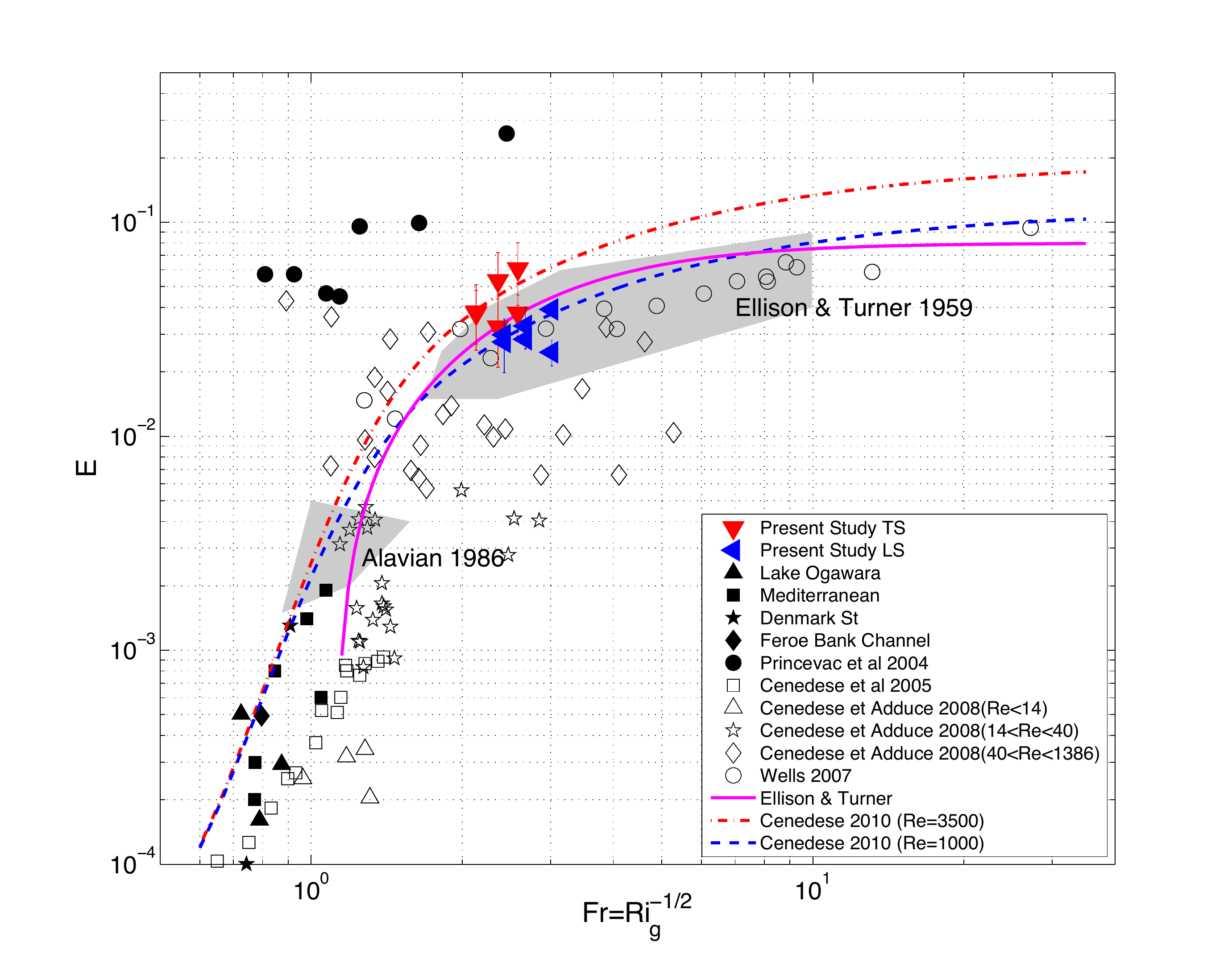}
 \caption{ \label{fig:plotCenedese} Relationship between $E$ and $Fr$ using the local value of $Ri$ to determine $Fr$.
Solid and open symbols are geophysical observational data~\citep{Princevac05,Wells:Euromech:07} and laboratory data~\citep{Cenedese04,Cenedese08}. Plot is generated using data from \cite{Cenedese08} with data points from the present study added. The solid curve is the Ellison and Turner classical parametrization and the other curves correspond to the~\cite{Cenedese:JPO:10} parametrization, with two different values of $Re$. The shaded areas represent the~\cite{Ellison59} and the \cite{Alavian:JHE:86} data.}
\end{figure}

\cite{Cenedese08} systematically described the Froude number ($Fr=1/\sqrt{Ri}$) dependence of $E$ using data from various sources including in-situ field experiments and laboratory experiments.  These data were compared with the predictions of entrainment parameterization schemes. Our measurements fall within the broad range of experimental values measured around the same Richardson number, see figure \ref{fig:plotCenedese},  but are in the upper range, probably owing to our relatively large Reynolds number. Within experimental error, our results for TS and LS indicate that $E$ depends on more than just $F_r$ and that a $Re$ dependence is needed to describe our data. \cite{Cenedese:JPO:10} propose such a parameterization which we apply to our data in  figure \ref{fig:plotCenedese} using our bulk value of Reynolds number ($Re=3500$) and our measured local Richardson number (red dashed-dotted curve). Our results for the TS data are compatible with this parametrization but the LS data lie below this curve. Adjusting the Reynolds number to $Re=1000$ in the \cite{Cenedese:JPO:10} parameterization produces a curve (blue dashed) compatible with our LS data. This lower Reynolds curve is also very close to the Ellison \& Turner parameterization which was obtained from data at lower Reynolds. We conclude that, as discussed in section~\ref{sec:introduction}, our active grid allows the turbulence to develop faster so that the flow coming out of the injection nozzle is representative of a developed channel flow at $Re=3500$. In the LS case, however, with the same bulk Reynolds number, the turbulence characteristics, at least in terms of entrainment, are closer to those of a channel flow at $Re=1000$.

\section{Characterizing stratified turbulence and associated mixing}\label{sec:mixing}

The entrainment of ambient fluid into the current is directly related to the local mixing induced by the turbulent boundary between the gravity current and its surrounding fluid. The characterization of the mixing properties of stratified turbulence is thus an important ingredient contributing to the understanding of geophysical flows. In this section, we present measurements of quantities related to the turbulent kinetic energy budget in our flow and then use these quantities to compute the mixing efficiency of the flow and to determine length scales associated with stratified turbulence. Finally, we compute the diapycnal turbulent diffusivity, a quantity widely used in oceanic or atmospheric numerical models, and compare it to experimental results, in-situ measurements, numerical simulations and associated parameterizations.

\subsection{Turbulent Transport and Energetics}\label{subsec:energetics}

The turbulent kinetic energy (TKE) is defined (as in table 1) using the measured velocity fluctuations:  $K=\frac{1}{2}\left(\langle u'^2\rangle+\langle v'^2\rangle+\langle w'^2\rangle\right)\simeq\frac{1}{2}\left(\langle u'^2\rangle+2\langle w'^2\rangle\right)$. With stratification, the governing equation for TKE in steady state~\cite[e.g.,]{Turner73} is

\begin{equation}
\langle{u_j}\rangle\frac{\partial K}{\partial x_j} +\frac{\partial T_j}{\partial x_j} = -\la u'_iu'_j\ra \frac{1}{2}\left(\frac{\pt\la u_i\ra}{\pt x_j}+\frac{\pt\la u_j\ra}{\pt x_i}\right) - g'\langle\theta'w'_g\rangle  -\epsilon \quad,
\label{eq:TurbKineEnergy}
\end{equation}

\noindent where $T_j = 1/2\langle{u_j'K}\rangle+1/\rho_0\langle{u_j'p'}\rangle -2\nu\partial\langle{Ku_j'}\rangle/\partial x_j$, and $w_g' = w' \cos \alpha + u' \sin \alpha$ represents the velocity fluctuations anti-parallel to the gravitational direction (along the $\hat{\mathbf{z}}_g$ direction in figure \ref{fig:facility}).

The first term on the right hand side of~\eqref{eq:TurbKineEnergy} is the turbulence production ${\cal P}$. When $\partial \langle u \rangle/\partial z$ is the dominant velocity derivative, ${\cal P}$  can be approximated using all resolved components, i.e.,
${\cal P}\simeq-\la u'^2\ra\pt\la u\ra/\pt x-\la u'w'\ra{\pt \la u\ra}/{\pt z}-\la u'w'\ra{\pt \la w\ra}/{\pt x}-\la w'^2\ra{\pt \la w\ra}/{\pt z}$, in which the first two terms contribute the most to ${\cal P}$,  and the last two terms are negligible.
The buoyancy flux ${\cal B}=g'\langle\theta'w'_g\rangle$, the second term in the right-hand side of~\eqref{eq:TurbKineEnergy}, only appears when density stratification is present. It quantifies the TKE production/destruction via mixing caused by degradation of local density gradients.  For stable (unstable) stratification, ${\cal B}$ has a positive (negative) value (sink (source) term in~\eqref{eq:TurbKineEnergy}) corresponding to the suppression (amplification) of turbulent fluctuations by stable (unstable) stratification and resulting in an upward (downward) mass flux. The last term on the right-hand side of~\eqref{eq:TurbKineEnergy} is the turbulent dissipation $\epsilon$. The vector $T_j$ sums the transport of TKE by turbulent velocity fluctuations $(1/2)\langle{u_j'u_i'u_i'}\rangle$, pressure gradient work $(1/\rho_0)\langle{u_j'p'}\rangle$, and viscous stresses $-2\nu\langle{u_i's_{ij}}\rangle$. The term ($-\partial T_j/\partial x_j$) represents the redistribution of energy from one point to another with no energy production or dissipation associated with it.

To examine the energetics and energy exchange that occur within the gravity current,~\eqref{eq:TurbKineEnergy} is vertically averaged over the mixing zone, between $h_t(x)$ and $h_b(x)$, as was done in section~\ref{sec:results} for $N$ and in section~\ref{subsec:BVfreq} for the computation of the gradient Richardson number. Denoting respectively ${\cal L}_c$ and ${\cal L}_T$ the vertical average of the first and second left-hand term of~\eqref{eq:TurbKineEnergy}, we obtain the averaged balance: ${\cal L}_c + {\cal L}_T = \overline{\cal P} - \overline{\epsilon} - \overline {\cal B}$.
%

\begin{figure}
\centering
\includegraphics[width=0.49\textwidth]{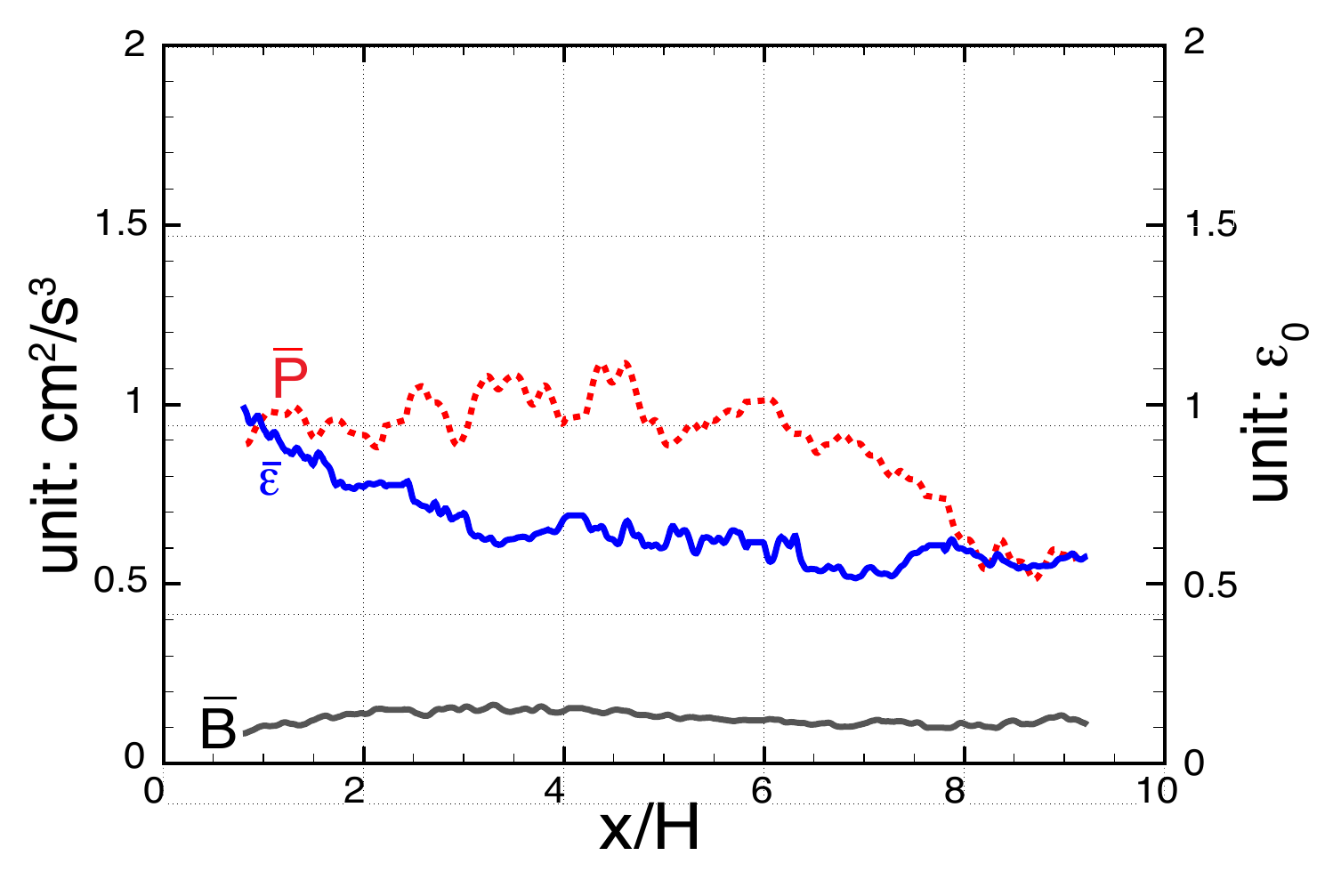}
\includegraphics[width=0.5\textwidth]{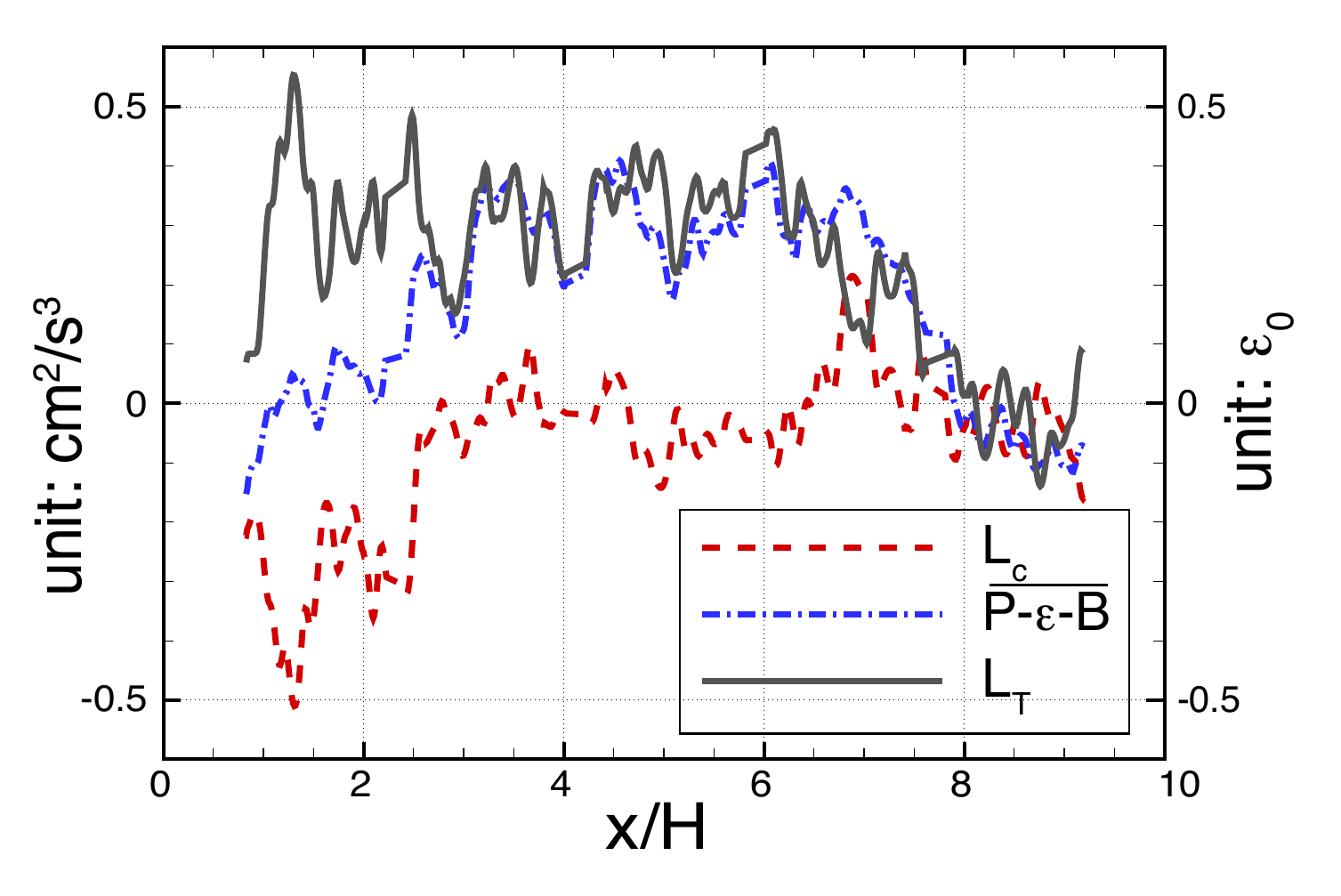}\\
\includegraphics[width=0.49\textwidth]{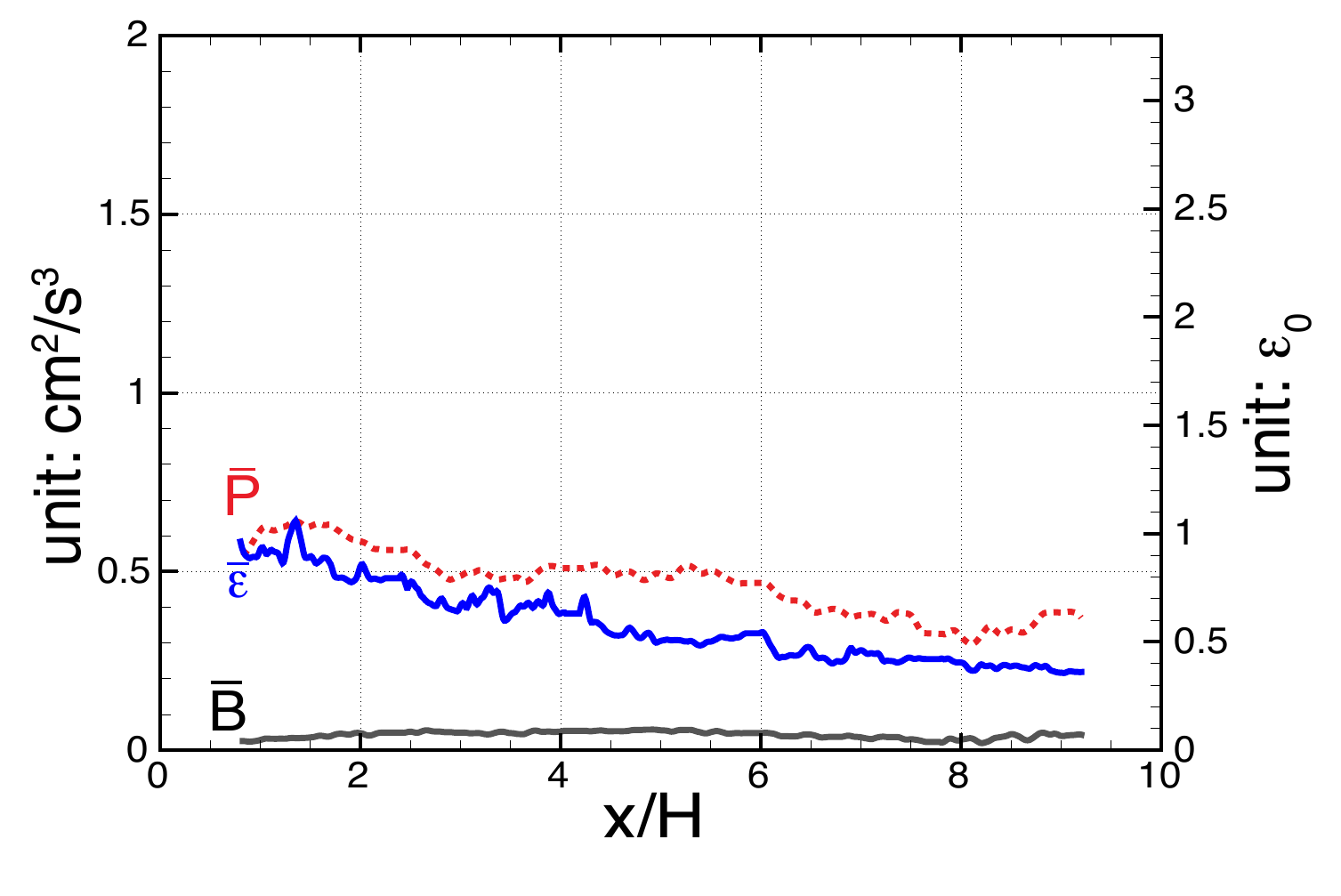}
\includegraphics[width=0.5\textwidth]{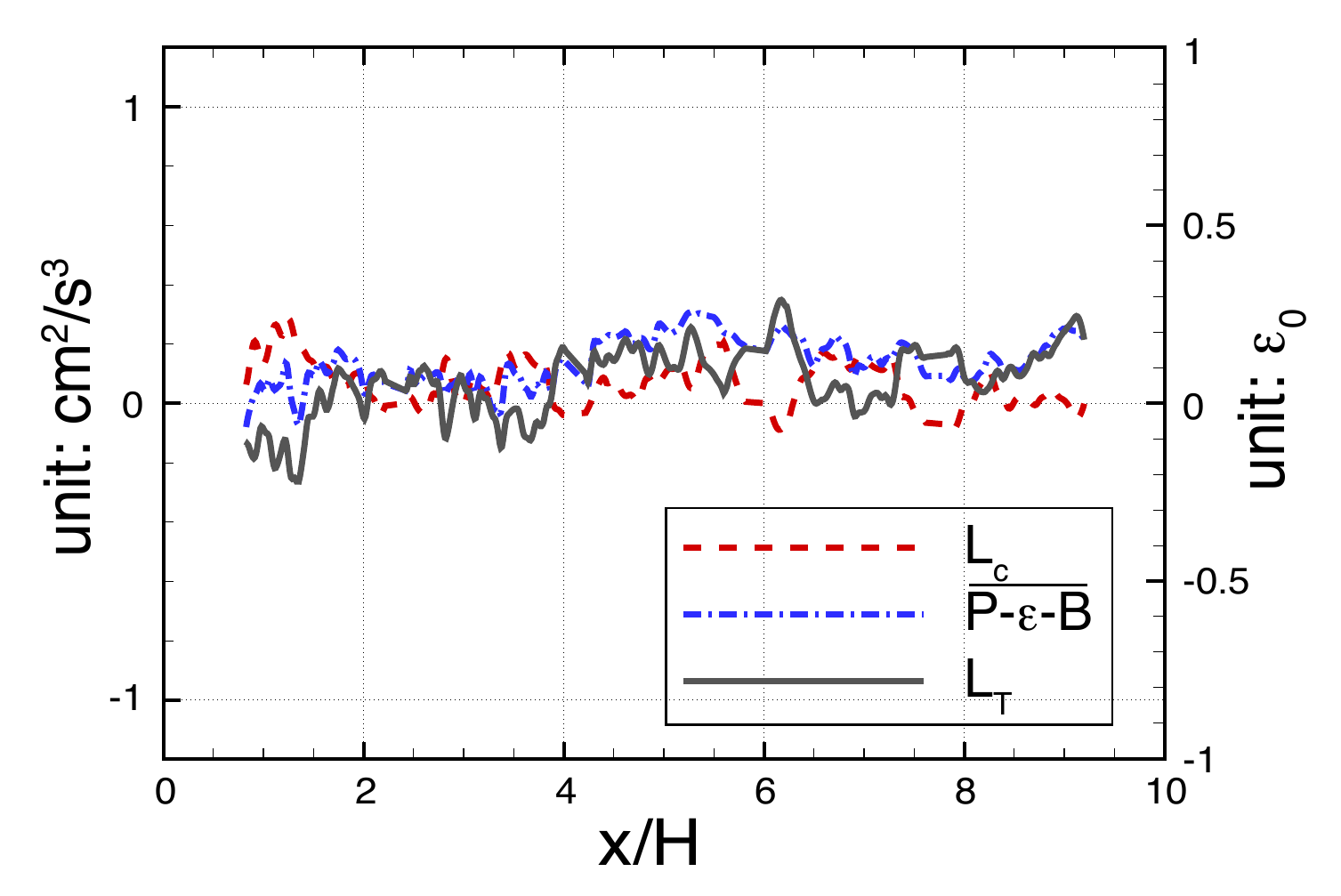}\\
\includegraphics[width=0.49\textwidth]{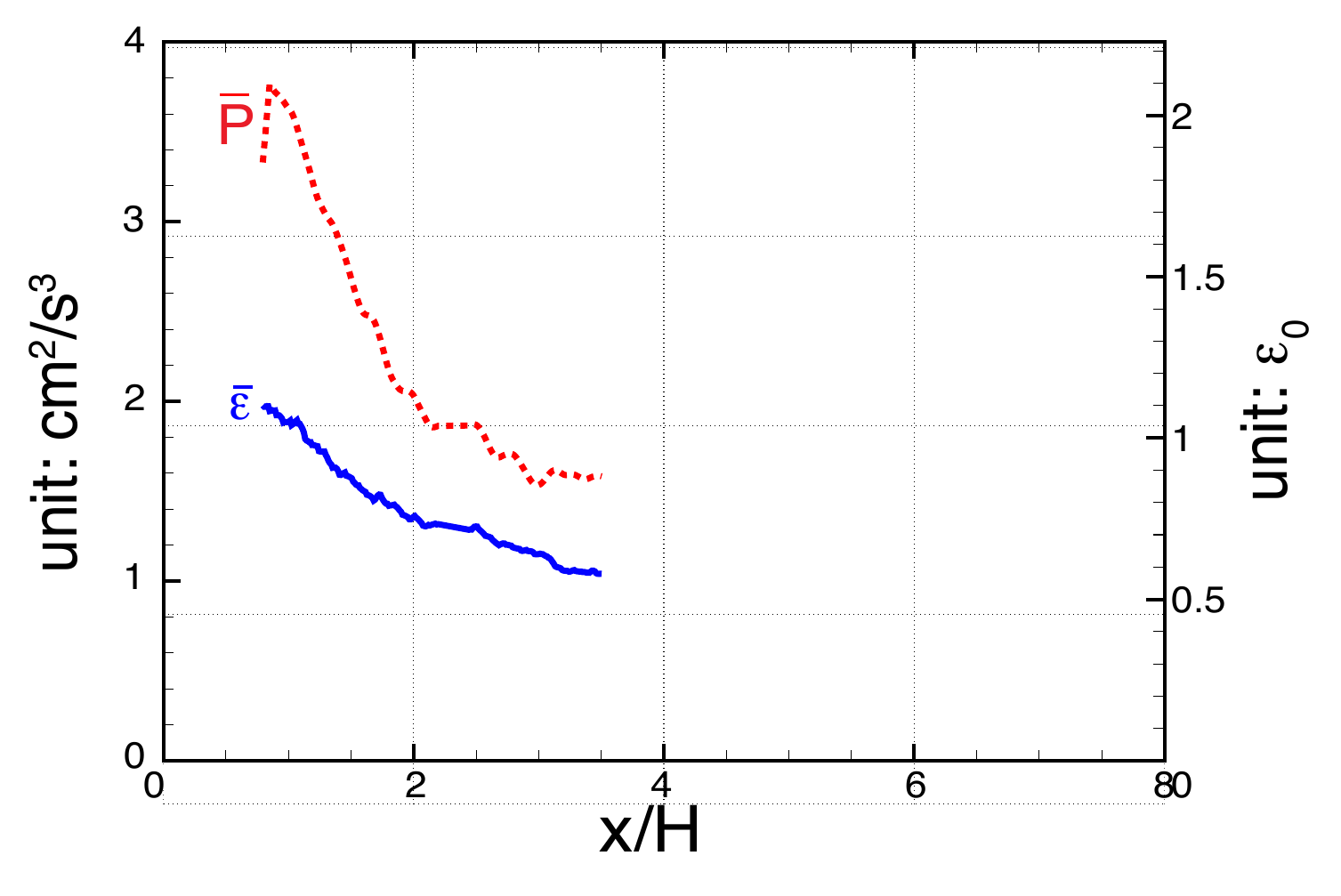}
\includegraphics[width=0.5\textwidth]{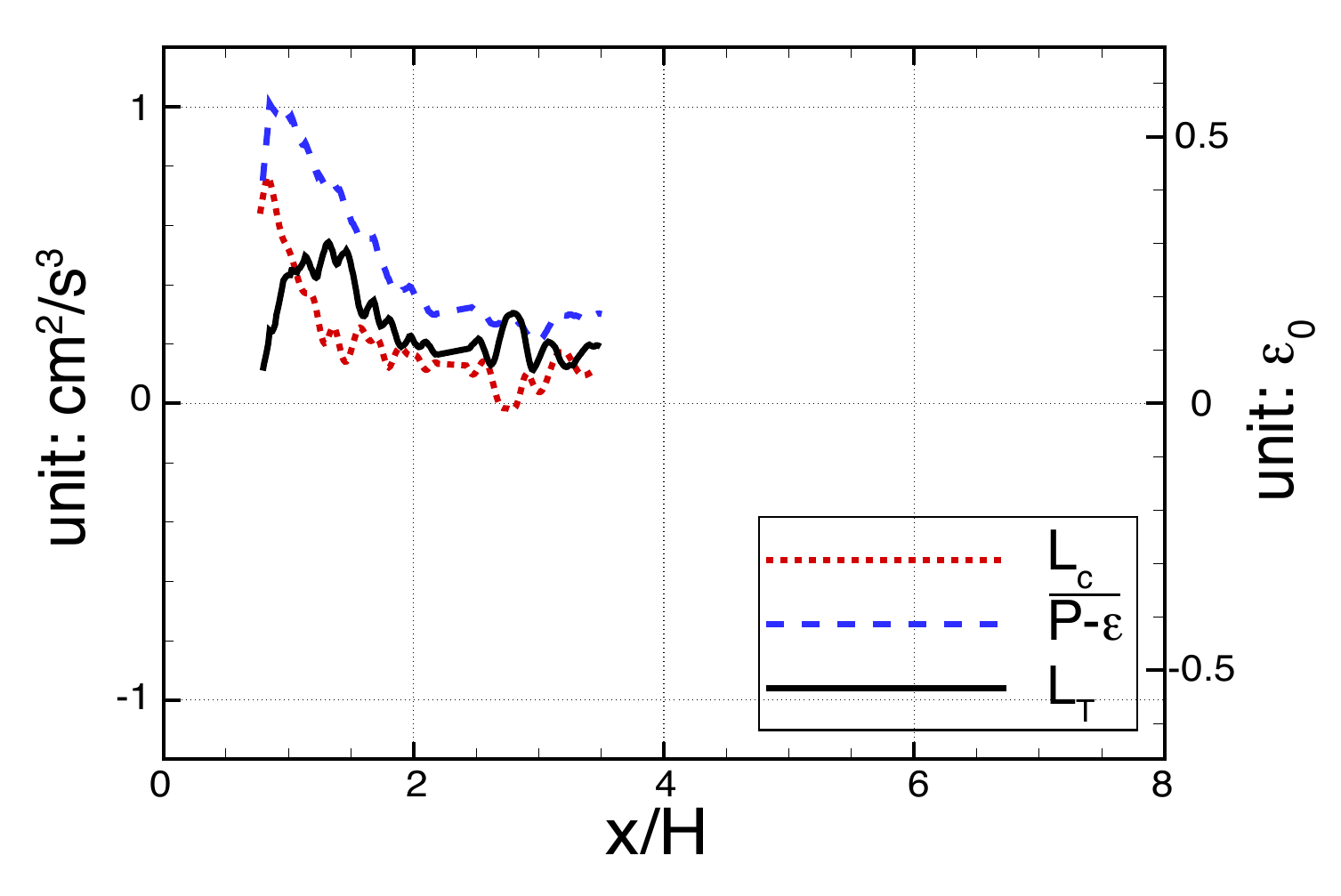}\\
\caption{\label{fig:energybudget} (Left) $\overline{\cal P}$, $\overline{\epsilon}$, and $\overline {\cal B}$ as functions of $x/H$. (Right) $\overline{\cal P}-\overline{\epsilon}-\overline {\cal B}$, ${\cal L}_c$, and their difference, equal to ${\cal L}_T$, vs $x/H$. The right axis indicates values normalized by $\epsilon_0$, with respectively $\epsilon_0$ = 1.1, 0.6, and 1.8 cm$^2$/s$^3$ from top to bottom, whereas the left axis shows unnormalized values. (top): TS; (middle): LS; (bottom): NS. (The vertical scale of the bottom plots (NS) is larger to accommodate the entire data range.)}
\end{figure}

The terms in this equation are plotted in figure \ref{fig:energybudget}. The buoyancy flux is small in both stratified cases compared to production and dissipation. Significantly more turbulence is produced for NS compared to TS in the upstream portion of the flow near the nozzle exit ($x/H < 0.6$) as indicated by the four-times larger $\overline{\cal P}$ despite the ostensibly equal initial conditions.
$\overline{\epsilon}$ is also significantly higher for NS than for TS. For NS,  $\overline{P}$ and $\overline{\epsilon}$ decrease rapidly as $x$ increases because the lack of stable stratification allows unimpeded dissipation by turbulent processes. On the other hand, in both stratified cases, $\overline{\cal P}$ and $\overline {\cal B}$ decrease more gradually.

Many turbulence closure models for geophysical problems are based on an assumption of balanced production and dissipation, i.e., $\overline{\cal P}\simeq\overline{\epsilon}+\overline{\cal B}$ \cite[see, e.g.,][]{Mellor82}. This balance is likely to occur in a closed system in steady-state conditions such as an ocean or lake thermocline, but is not expected in general for shear flows~\citep{Tennekes72}, and in particular for our experiment, since~\eqref{eq:TurbKineEnergy} was integrated over a cross-section through which TKE transport can take place, so the the transport terms ${\cal L}_c$ and ${\cal L}_T$ are not necessarily small. Indeed, the results in figure \ref{fig:energybudget} show that this balance is not satisfied for TS or NS, except at far downstream locations for TS. On the other hand, the LS case is closer to the balanced production and dissipation assumption.


Because the $T_j$ term in~\eqref{eq:TurbKineEnergy} involves pressure-fluctuation correlations and higher-order moments, it is not possible to obtain this term directly from experimental data. Nevertheless, we can indirectly evaluate its importance within the gravity current by calculating the difference between $\overline{\cal P} -\overline{\epsilon} -\overline {\cal B}$ and ${\cal L}_c$. The results are shown in figure \ref{fig:energybudget} (right). The ${\cal L}_c$ term, which represents the overall  advection of turbulent kinetic energy within the gravity current, is nearly zero for both stratified cases, except in the very upstream region for TS ($x/H<2$). For NS, this term is mostly smaller than ${\cal L}_T$ and $\overline{\cal P} -\overline\epsilon$ but is not negligible.  Thus, at least in the stratified cases, the unknown turbulent transport term ${\cal L}_T$ can be approximated, except just out of the nozzle, by the value of $\overline{\cal P} -\overline\epsilon -\overline{\cal B}$. In the stratified cases, it is non-zero upstream but tends toward smaller values with increasing $x/H$. This result is noteworthy since the $T_j$ term is generally neglected in most studies of the turbulent energy budget. For NS, ${\cal L}_T$ is smaller than in the stratified cases, except very close to the inlet. For NS, however, the accessible downstream range is limited to $x/H<3.5$ because the width of the mixing zone becomes larger than the field of view (see its evolution in figure~\ref{fig:mixingzone}).

%


\subsection{Flux Richardson number}\label{sec:Rif}

To quantify turbulent mixing in stratified flows, \cite{Linden:GAFD:79} introduced the flux Richardson number, defined in our case 
\begin{equation}
Ri_f=\frac{\overline{\cal B}}{\overline{\cal P}}
\label{eq:FluxRiNum}
\end{equation}
using the energetic quantities computed in the previous subsection. $R_f$ is the fraction of available turbulent kinetic energy which is converted into potential energy of stratification. Within the gravity current, its value is roughly constant for TS, about $0.11\pm0.01$. In the LS case, it is also almost constant ($0.1\pm0.01$) except in the very upstream region, where it decreases to 0.05.

For turbulent stratified geophysical flows, $\langle u'w'\rangle$ and ${\cal B}$, are often modeled using an eddy viscosity/diffusivity hypothesis. Thus, the flux Richardson number, which relates these two parameters, is of special significance. It is ideal to express $Ri_f$ in terms of $Ri_g$ which is related to the solvable unknowns ($\langle \theta\rangle$ and $\langle u_i\rangle$) in numerical simulations.
In a separate study using the same data \citep{Odier09}, we show that the turbulent momentum flux and the buoyancy flux are well described by a mixing length model with $\la u'w' \ra = L_m^2 \left(\la\pt u / \pt z\ra\right)^2$ and $\la \rho_d'w' \ra = L_{\rho}^2 \la\pt u / \pt z\ra~\la\pt \rho_d / \pt z\ra$, where $L_m$ and $L_{\rho}$ are characteristic lengths for the mixing processes. We also measured $L_m$ and $L_{\rho}$ in various configurations (changing injection velocity and/or initial density difference) and find that they are always close to one another~\citep{Odier:PhysD:12}. As a result, using the mixing length relations above in the expression for $Ri_f$ and the dominant term for $\cal P$, and assuming $\la \rho_d'w_g' \ra\simeq-\la \rho_d'w' \ra$, one can write :

\begin{equation}
Ri_f=\frac{\cal B}{\cal P}=\frac{g/\rho_0\;\la \rho'w' \ra}{\la u'w' \ra\la\pt u / \pt z\ra}
=\frac{g/\rho_0\;L_{\rho}^2 \la\pt u / \pt z\ra~\la\pt \rho / \pt z\ra}{L_m^2 \left(\la\pt u / \pt z\ra\right)^3}
=\frac{L_{\rho}^2}{L_m^2}Ri_g\simeq Ri_g
\end{equation}

\begin{figure}
\centering
\includegraphics[width=.73\textwidth]{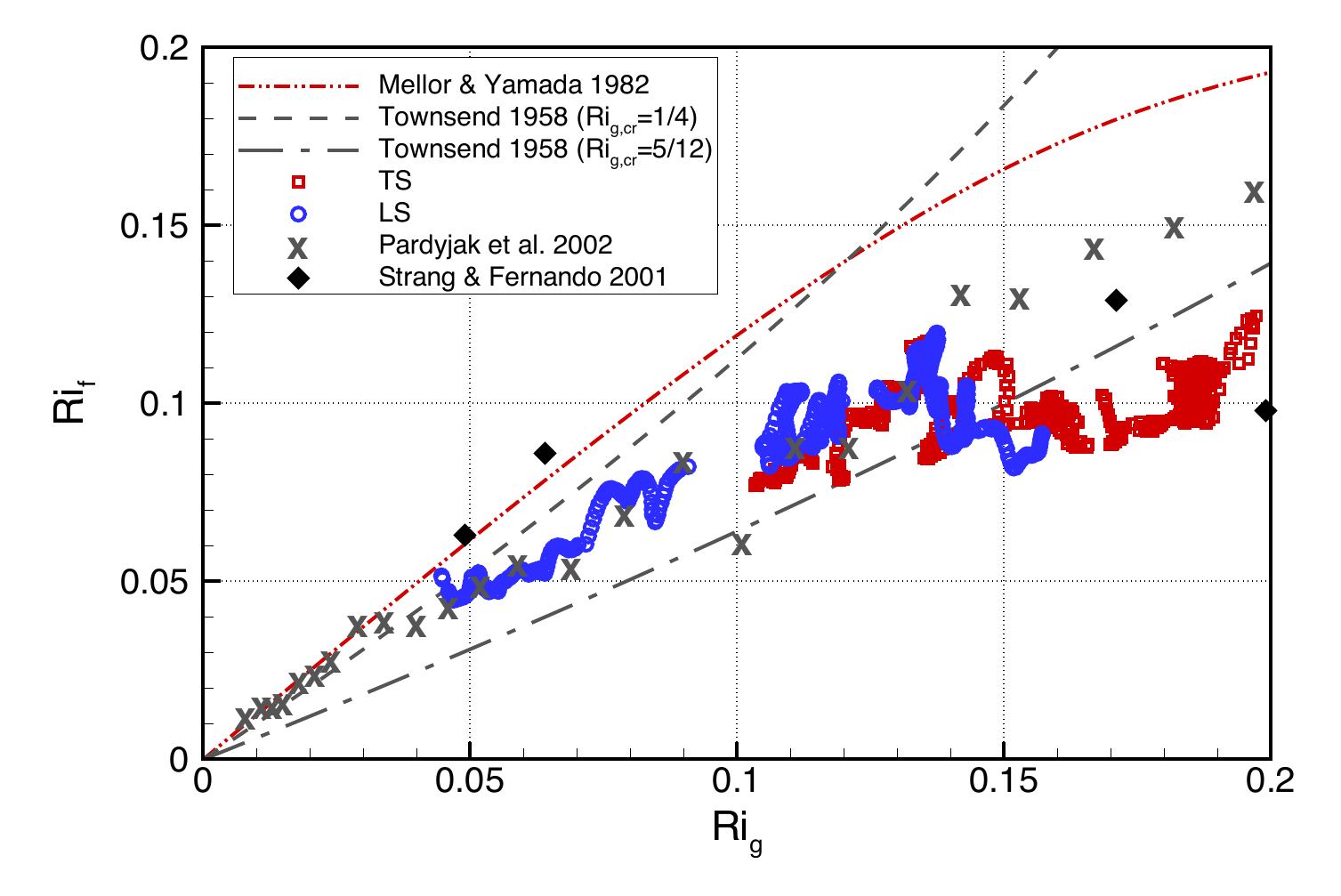}\\
\caption{\label{fig:RiGradNumRiFluxNum} $Ri_f$ versus  $Ri_g$, for our data, other experimental data, in situ measurements, and for various parametrizations.}
\end{figure}

\noindent This approximate equality between $Ri_f$ and $Ri_g$ is observed for low values of $Ri_g$ ($Ri_g<0.1$) in figure~\ref{fig:RiGradNumRiFluxNum}, showing our measurements of $Ri_f$ as a function of $Ri_g$, for TS and LS.\\

In turbulence closure schemes, the relationship between $Ri_f$ and $Ri_g$ has been considered in more detail by several groups. \cite{Mellor74,Mellor82} propose 

\be\label{eq:Mellor79}
Ri_f=0.659\left(Ri_g+0.178-\sqrt{Ri_g^2-0.322 Ri_g+0.0316}\right),
\ee

\noindent whereas \cite{Townsend58} derives the relation:

\begin{equation}
Ri_f=\frac{1}{2}\left(1-\sqrt{1-{Ri_g}/{Ri_{g,cr}}}\right),
\label{eq:Townsend58}
\end{equation}

\noindent where $Ri_{g,cr}=1/12\eta$ is a critical gradient Richardson number, with $\eta$ a coefficient related to the ratio of $\theta'w'$ to $u'w'$. Townsend assumes $\eta\simeq 1$, yielding a critical $Ri$ of 1/12. This prediction is, however, larger than our experimental data by about a factor 4. Another possibility is choosing a critical Richardson number $Ri_{g,cr}=1/4$, which is the standard value for the onset of KH instability. Finally, a direct calculation of the factor $\eta$ using our data gives $\eta\simeq0.2$, yielding $Ri_{g,cr}=1/(12\cdot0.2)=5/12$.

Figure~\ref{fig:RiGradNumRiFluxNum} shows our experimental data and the predictions from equations~\ref{eq:Mellor79} and~\ref{eq:Townsend58} (the latter using critical values $Ri_{g,cr}=1/4$ and $Ri_{g,cr}=5/12$). The Mellor and Yamada prediction, as well as the Townsend prediction using $Ri_{g,cr}=1/4$, are larger than our results by roughly 50 $\%$. The Townsend prediction using $Ri_{g,cr}=5/12$, falls close to our measurements, although the curvature in the evolution of $Ri_f$ with $Ri_g$ is different.

For comparison, we also plot in figure~\ref{fig:RiGradNumRiFluxNum} two other data sets in our $Ri_g$ range: a laboratory experiment (\cite{Strang:JPO:01}) and an atmospheric gravity current experiment (\cite{Pardyjak92}).  Our data are consistent with the other experimental measurements, except at large $Ri_g$ where they differ slightly from the Pardyjak data. The consistency at small $Ri_g$ with the data from \cite{Pardyjak92} is noteworthy, since these data comes from measurements performed in a macroscopic large scale atmospheric flow (Reynolds number of about $10^7$) whereas ours is a laboratory experiment.  This observation indicates that the Reynolds number dependence of the flux Richardson number is weak, at least in a range of Reynolds number starting around our values which are amongst the largest obtained for a laboratory gravity current.


\subsection{Turbulent length scales}

One way to describe stratified turbulence is to compute length scales which characterize the typical sizes of the turbulent cascade in the stratified flow. \cite{Istweire:JPO:93} defined 3 length scales which are used in several recent studies based on laboratory and in-situ measurements~\citep{Hult:JGR:11b,Bouffard:DAO:13,Bluteau:JGR:13}:

\begin{equation}
L_O=\left(\frac{\epsilon}{N^3}\right)^{1/2} \quad ; \quad L_E=\frac{\rho_{rms}}{\partial\rho/\partial z}  \quad ; \quad \eta_K=\left(\frac{\nu^3}{\epsilon}\right)^{1/4}
\end{equation}\label{equ:lengths}

The first, the Ozmidov scale $L_O$, is the scale at which the buoyancy forces become of the same order of magnitude as inertial forces, i.e., the largest overturning scales in the flow. The second, $L_E$, introduced by~\cite{Ellison:JFM:57}, represents the typical vertical distance travelled by a fluid particle before either returning towards its equilibrium level or mixing. The last is the Kolmogorov scale $\eta_K$ which gives the smallest overturning scale not strongly dissipated by viscosity.

\begin{figure}
\centering
\includegraphics[width=.5\textwidth]{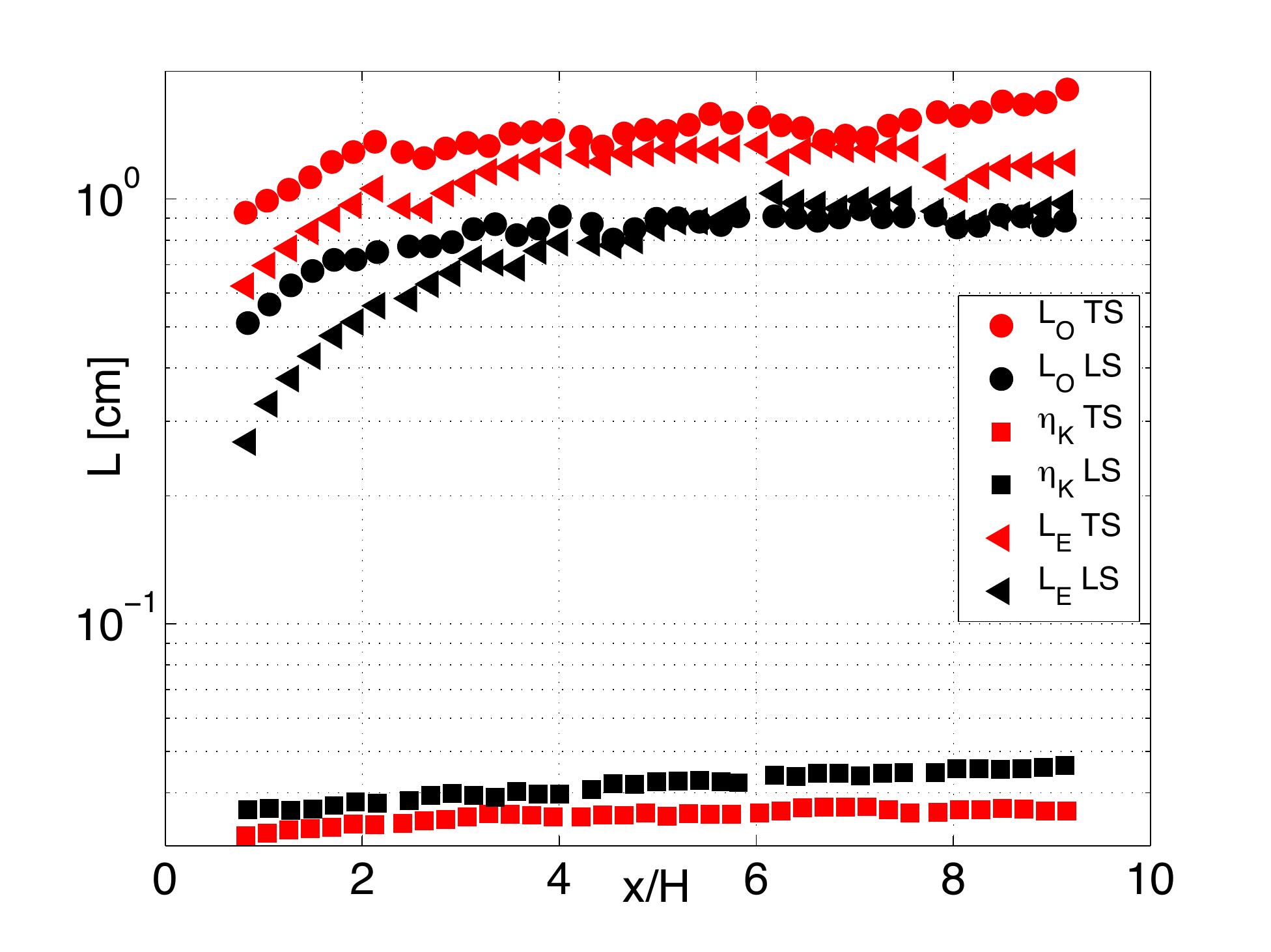}
\includegraphics[width=.48\textwidth]{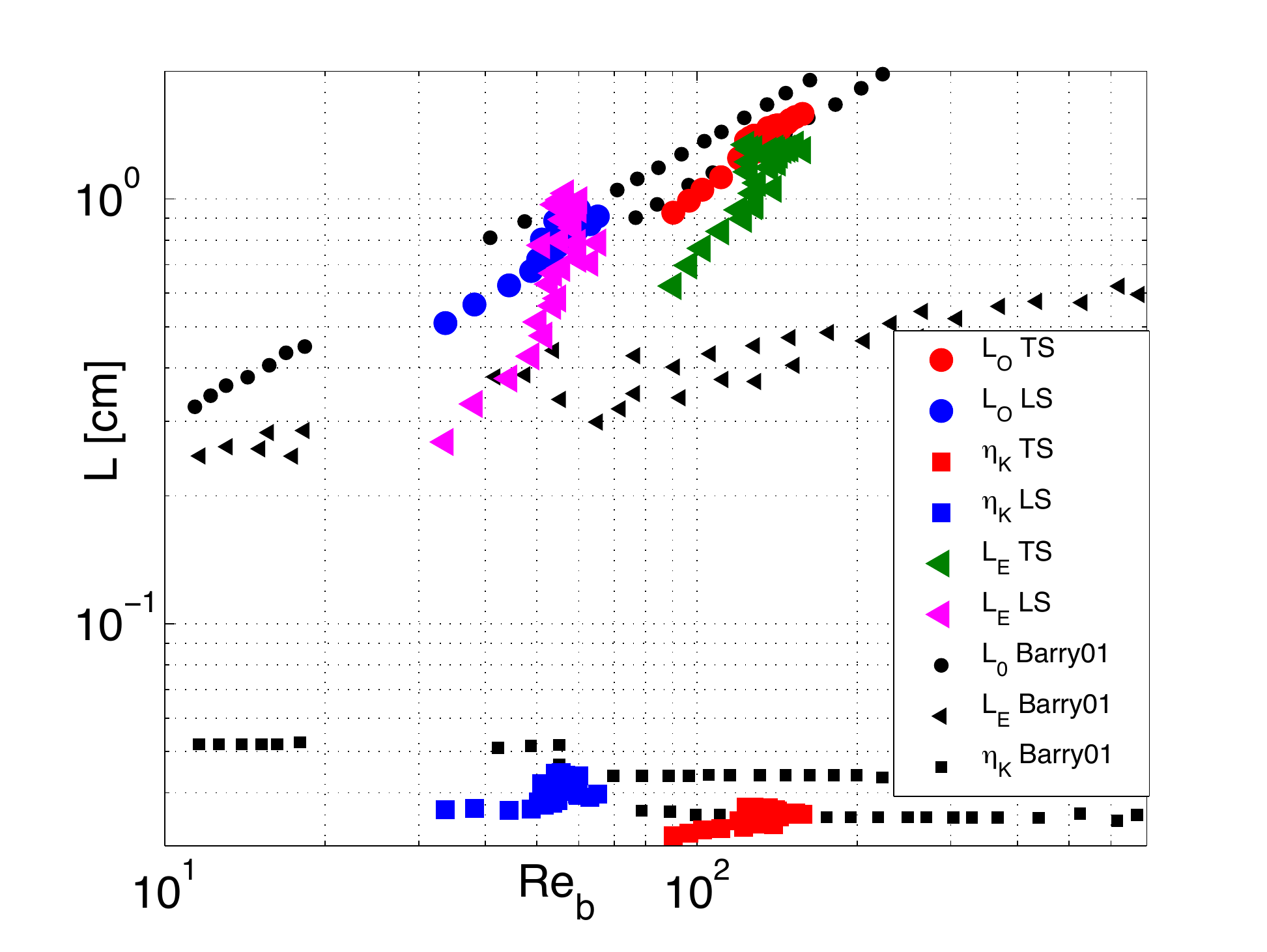}\\
\caption{\label{fig:lengths} (left) Length scales $L_O$, $L_E$ and $\eta_K$ vs $x/H$. (right) Length scales as a function of $Re_b$, compared with results from an oscillating grid experiment in a stratified fluid~\citep{Barry:JFM:01}.}
\end{figure}

Figure~\ref{fig:lengths} (left) shows these three length scales, measured for TS and LS, plotted as functions of $x/H$. As expected, $L_E$ falls between $L_O$ and $\eta_K$, since buoyancy does not permit larger scales than $L_O$ whereas viscosity prevents scales smaller than $\eta_K$. $L_O$ and $L_E$ increase with $x/H$, especially close to injection, because the density gradient decreases as one goes downstream (the reduction in $\epsilon$ with $x/H$ has a smaller effect because it appears only to the 1/3 power in the Ozmidov scale).

The intensity of turbulence can be characterized by comparing the largest to smallest overturns, i.e., $L_O$ to $\eta_K$, typically in terms of a buoyancy Reynolds number:

\begin{equation}
Re_b=\left(\frac{L_O}{\eta_K}\right)^{4/3}=\frac{\epsilon}{\nu N^2}.
\end{equation}

\noindent which can also be interpreted as the ratio between the typical timescale associated with buoyancy, $1/N$, and the timescale for a turbulent event to fully develop, $\sqrt{\epsilon/\nu}$. Figure~\ref{fig:lengths} (right) shows the evolution of the three lengths scales as functions of $Re_b$. $Re_b$ is concentrated around a narrow range between 55 and 60 for LS and between 125 and 160 for TS. Only the region of the flow at small $x$ (lower values of $L_O$) correspond to lower $Re_b$ (down to 30 in LS and 90 in TS). {These smaller values may arise from the non steady-state nature of the flow for small $x$.} Because of the small variation of $\eta_K$ in our flow, the dependence of $L_O$ with $Re_b$ is naturally linear on a log scale, with an expected 4/3 slope. $L_E$ has a similar dependence on $Re_b$ with a larger slope, especially for the LS data. For comparison, data from a grid turbulence experiment in a stratified fluid~\citep{Barry:JFM:01} is shown in figure~\ref{fig:lengths} (right). For $L_O$ and $\eta_K$, the values and behaviors obtained in the grid experiment are very comparable to ours. For $L_E$, their values are slightly smaller. A turbulent Froude number can also be defined using the ratio of the Ozmidov to the Ellison lengths: $Fr_T=\left({L_O}/{L_E}\right)^{2/3}$. Using our data, we find values in the range 1-1.5 for LS and 1-1.3 for TS, concentrating around a value of 1 for most of the data.

\subsection{Diapycnal turbulent diffusivity}

At small scale, entrainment is directly related to the mixing of the fluids of different density. Another common way to parameterize this mixing in ocean simulations is to use a linear relation between buoyancy flux and density gradients, in the same way an eddy viscosity is used in turbulence closure schemes to parameterize Reynolds stress. This relation defines a turbulent eddy diffusivity $K_{\rho}$, also known as diapycnal diffusivity:

\begin{equation}
\langle\rho'w'\rangle=-K_{\rho}\left\langle\frac{\partial \rho}{\partial z}\right\rangle
\end{equation}

\begin{figure}
\centering
\includegraphics[width=.85\textwidth]{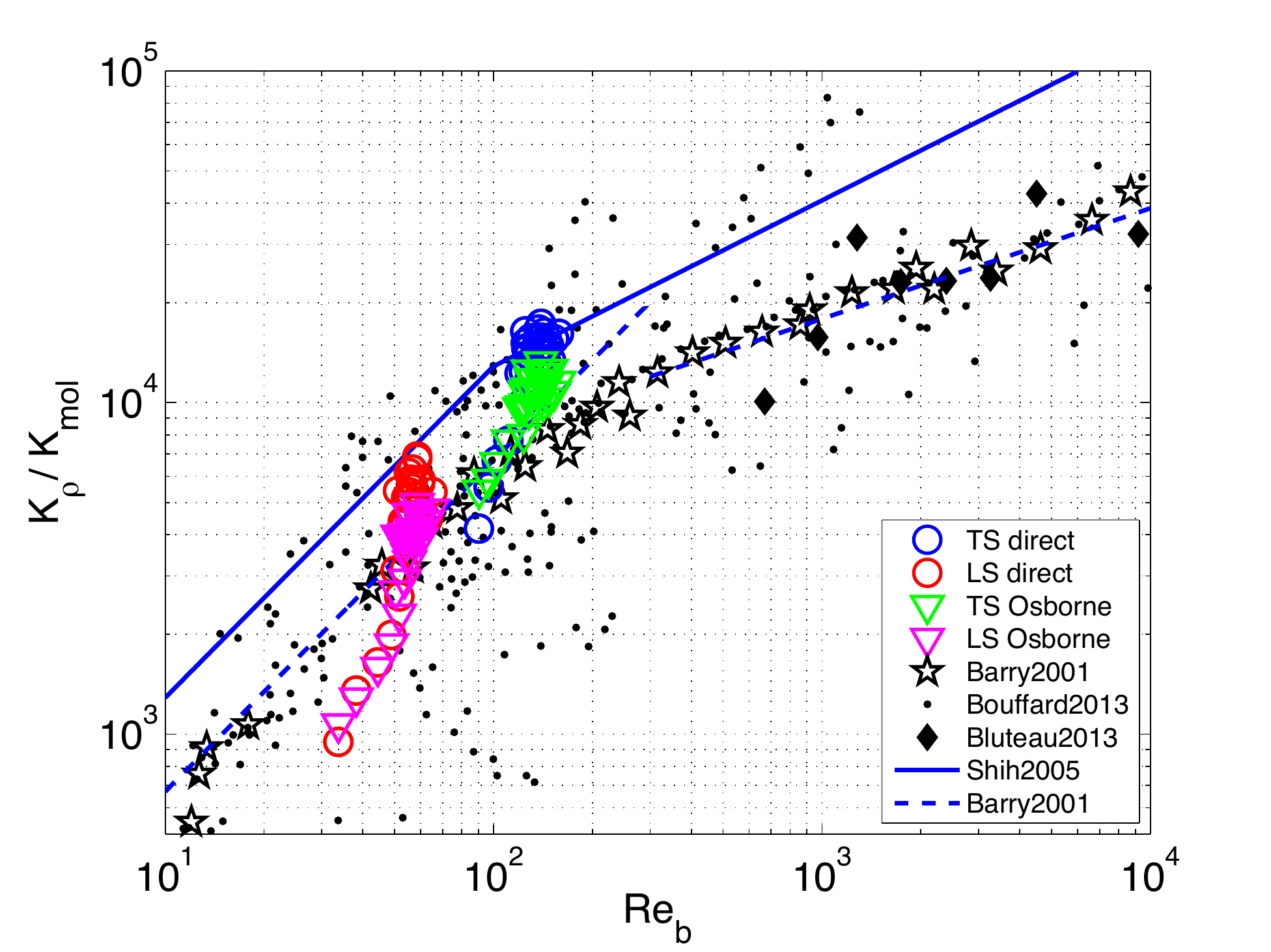}\\
\caption{\label{fig:Krho} Turbulent diffusivity, normalized by the molecular diffusivity of salt in water, vs~$Re_b=\epsilon/\nu N^2$.}
\end{figure}

Using our data, the diapycnal diffusivity is computed directly: $K_{\rho}=\rho_0\overline{\cal B}/g\overline{\langle{\partial\rho/\partial z}\rangle}$. Figure~\ref{fig:Krho} shows our direct measurements versus $Re_b$, together with experimental data from other sources. For TS, $K_\rho$ is generally above the oscillating grid experiment~\citep{Barry:JFM:01} by about a factor of 2; on the contrary, our low $Re_b$ events, corresponding to measurements close to injection, are very close to the Barry measurements. In the LS case, our data in most of the current is closer to the Barry data, except close to the inlet, which it is lower. Our measurements are compatible with ocean data~\citep{Bouffard:DAO:13}, which has a rather large scatter. Data by~\cite{Bluteau:JGR:13} are slightly higher than ours, at larger values of $Re_b$.

It is often difficult to perform a direct measurement of $K_{\rho}$ in the ocean so several parameterizations have been developed. Assuming that the flow is in steady state, the balance between sources and sink terms of the TKE equation, ${\cal P}={\cal B}+\epsilon$, implies writing $K_{\rho}$ in terms of the flux Richardson number:

\be
K_{\rho}=\frac{Ri_f}{1-Ri_f}\frac{\epsilon}{N^2} \label{equ:Osborn}
\ee

Ocean models often assume a constant value of 0.17 for $Ri_f$, yielding a constant value of 0.2 for the mixing efficiency defined by the ratio $\Gamma=Ri_f/(1-Ri_f)$~\citep{Osborn:JPO:80}. Our flux Richardson numbers yield a lower value of $\Gamma$, equal to $0.12\pm0.015$ for TS and $0.11\pm0.01$ for LS except close to the injection nozzle where it goes down (in the LS case) to 0.05. In figure~\ref{fig:Krho}, we show $K_{\rho}$ computed from~\eqref{equ:Osborn} using our measured values of $Ri_f$. It is slightly lower than our direct estimate. The difference is attributable to the lack of  TKE balance between production and sink terms in our flow, particularly at small $x/h$. The difference ${\cal P}-{\cal B}-\epsilon$ being positive (see figure~\ref{fig:energybudget}(right)) explains the lower estimate of $K_\rho$ from~\eqref{equ:Osborn}. 

Several studies, experimental~\citep{Barry:JFM:01} and numerical~\citep{Shih:JFM:05}, have proposed parameterizations of the relation between diapycnal diffusivity and buoyancy Reynolds number:

\begin{equation*}
{\rm Barry~et~al,~2001} \left \lbrace
\begin{array}{lccc}
Re_b<300 : & K_{\rho} & = & 0.9\nu^{2/3}\kappa^{1/3} Re_b \\
300<Re_b : & K_{\rho} & = & 2\nu^{2/3}\kappa^{1/3} Re_b^{1/3} \\
\end{array}
\right.
\end{equation*}

\begin{equation*}
{\rm Shih~et~al,~2005} \left \lbrace
\begin{array}{lccc}
7<Re_b<100 : & K_{\rho} & = & 0.2\nu Re_b \\
100<Re_b : & K_{\rho} & = & 2\nu Re_b^{1/2} \\
\end{array}
\right.
\end{equation*}

The Barry parameterization, similar to the Barry data on which it is based, falls slightly below our data. On the other hand, the Shih parameterization {agrees very well with our TS data and relatively well with our LS data for $x/H > 4$}. The low branch of the Shih parametrization corresponds to the Osborn model with $\Gamma=0.2$. The change of behavior predicted by this parametrization occurs at $Re_b=100$, which is between the ranges of our TS and LS data. An extension of any of the branches in the other region would lead to an overestimate of the diffusivity, confirming the need for a two-regime parameterization.


\section{Conclusions}\label{sec:conclu}

This experimental study explores the structural development and dynamics of a wall-bounded gravity current characterized by wall boundary layer turbulent processes, mixing zone mechanisms, and stratification effects.  The indices of refraction of light fluid and dense fluid are matched while maintaining a density difference of 0.26\%. A combined PIV-PLIF system is used to simultaneously measure the planar velocity and density fields along the central plane of the test section, covering a total area of $9.0\times45.0$~cm$^2$ in five adjacent downstream locations.  The initial turbulence level of the gravity current is controlled by four active grids with little modification of the mean flow.  Three experimental conditions were examined for understanding the interaction of turbulence and stratification:  turbulent-stratified (TS) case ($Ri_0=0.27$, $R_{\lambda 0}=100$),  laminar-stratified (LS) case ($Ri_0=0.31$, $R_{\lambda 0}=42$), and unstratified (NS) case  ($Ri_0=0.0$, $R_{\lambda 0}=120$).

The evolution of the gravity current for our three cases has many features consistent with past measurements.  In particular, in NS the spreading of the mixing zone and the deceleration of the current after the velocity profile has eroded from its initial plug shape reflects the vertical turbulent momentum transport of a wall jet \citep{Wygnanski92} with approximately the same spreading rate but with faster decay of peak velocity, the latter apparently the result of an initially turbulent wall jet.  The flow with stable stratification reflects the negative buoyancy of the current with respect to the quiescent fluid in that it accelerates slightly after exiting the nozzle and the broadening
of the mixing zone slows appreciably for $x/H >  5$. $d\langle u\rangle/dz$ approaches a non-zero asymptotic value at far downstream locations in both stratified cases whereas it rapidly decreases for NS. 

We carefully consider different definitions of the Richardson number and its downstream evolution.  A Richardson number defined in terms of the initial parameters $Ri_0$ overestimates $Ri_g$ by from 1.5 to 3 with the ratio decreasing as the current broadens. To better reflect $Ri_g$, we define a local bulk Richardson number $Ri$ in terms of the width of the mixing zone instead of the whole current height.  The behaviors of $Ri$ and $Ri_g$ with respect to downstream distance are similar, and this new definition should be easier to incorporate into simulations than $Ri_g$. 

Analysis of the entrainment coefficient $E$ using both volume conservation and scalar mass anomaly conservation yields values that are compatible within our error bars, although in the TS case, the scalar conservation is systematically higher. 
Compared to the Ellison \& Turner parameterization, the LS data give estimates comparable with these predictions, computed using our different definitions of the Richardson number, whereas the TS estimate of entrainment is slightly higher. Our results for TS are consistent with the~\cite{Cenedese:JPO:10} parameterization using our bulk Reynolds number, but for LS one needs a Reynolds number 3 times smaller.

The measured velocity-density data is used to analyze the budget of turbulent kinetic energy. In both stratified cases one has $\overline{{\cal B}}/\overline{{\cal P}}\simeq10\%$, and the classic assumption of balanced production and dissipation ($\overline{{\cal P}}\simeq\overline{\epsilon}+{\cal B}$) in turbulence closure models does not hold, particularly in upstream locations. A number of mixing and entrainment properties, however, do not seem to be sensitive to this imbalance except perhaps at upstream regions with $x/H < 2$.

We investigate the flux Richardson number, various turbulent lengths scales of the flow and the diapycnal turbulent diffusivity, several quantities that are widely used for modeling stratified turbulence in the ocean. The mixing efficiency in the gravity current is explored through the dependence of $Ri_f$ on $Ri_g$. Our data show that the classic model proposed by \cite{Townsend58}, as well as the \cite{Mellor74,Mellor82} model, over predicts the value of $Ri_f$ in both stratified cases. Our results are quantitatively consistent with measurements of an atmospheric gravity current \citep{Pardyjak92} and another laboratory stratified shear flow \citep{Strang01}. 

We then compute various length scales associated with stratified turbulence and find them in good agreement with other experimental data in a different set-up (oscillating grid turbulence), except for the Ellison scale, which is slightly higher in our case. Finally we study the dependence of the diapycnal turbulent diffusivity with the ratio of the Ozmidov to the Kolmogorov scale, known as the buoyancy Reynolds number. Our results show reasonable agreement with both in situ oceanic measurements and other experimental data.

Overall, the behavior of the stratified and unstratified flows are well explained.  In addition, the role of turbulence in the gravity current has only partially been clarified.  In ocean flows, the gravity current Re is much higher so the gravity current is much more turbulent relative to turbulence generated via Kelvin Helmholtz instability, quite different from what happens here where the KH part is large (or of equal weight) compared to the initial current turbulence conditions. Further analysis of gravity currents using detailed experimental probes would be useful in continuing the characterization of these complex flows.

\begin{acknowledgments}

This work was carried out under the auspices of the National Nuclear Security Administration of the U.S. Department of Energy at Los Alamos National Laboratory under Contract No. DE-AC52-06NA25396. The authors thank Michael Rivera for helpful discussions and experimental assistance.  We also thank one of the anonymous referees for numerous very useful suggestions.
\end{acknowledgments}

\appendix
\appendixpage

\section{Consistency checks}\label{app:consist}

We often estimate quantities using formulas based on the homogeneous, isotropic turbulence assumption.  On the other hand, we demonstrated that the isotropic assumption for turbulent energy dissipation overestimates a calculation that includes all the measured spatial gradients in the flow, i.e.,~\eqref{eq:TurbDissipationApx3}.  Here we examine the internal consistency of our parameterization.  Our main independent measurements are the velocity fluctuations, $u'$ and $w'$, the dissipation rate $\epsilon$, and the Taylor micro-scale $\lambda$ obtained from the velocity autocorrelation function.  
The combination of $u'$ and $\epsilon$ yields the integral length $\ell = u'^3/\epsilon$ and the large scale Reynolds number Re$_\ell = u' \ell/\nu$ (ignoring small factors such as computing the mean fluctuations using both $u'$ and $w'$). The Taylor Reynolds number is determined from the combination of $u'$ and $\lambda$, $R_\lambda = u' \lambda/\nu$.
The isotropic homogeneous assumption leads to several relationships among these quantities, for example $R_\lambda = \sqrt{15 Re_\ell}$ and $\ell/\eta = Re_\ell^{3/4}$ \citep[e.g.,][]{Tennekes72, Pope00}. A consistency check can thus be made by computing the ratios $Re_\lambda/\sqrt{15 Re_\ell}$ and $(\ell/\eta)Re_\ell^{-3/4}$, which should yield order 1 factors near the nozzle outlet where the turbulence is nearly isotropic and homogenous. For the former ratio, we find 1.1 (TS), 1.2 (LS), and 1.2 (NS). The latter ratio is equal to 1.2~$\pm$~0.05 for all three flows. We thus observe that using quantities derived from independent measurements yields a self consistent parameterization (within an order one coefficient of about 1.2) despite the apparently anisotropic nature of $\epsilon$.


\section{Derivation of the entrainment expressions}

\begin{figure}
\centering
\includegraphics[width=.6\textwidth]{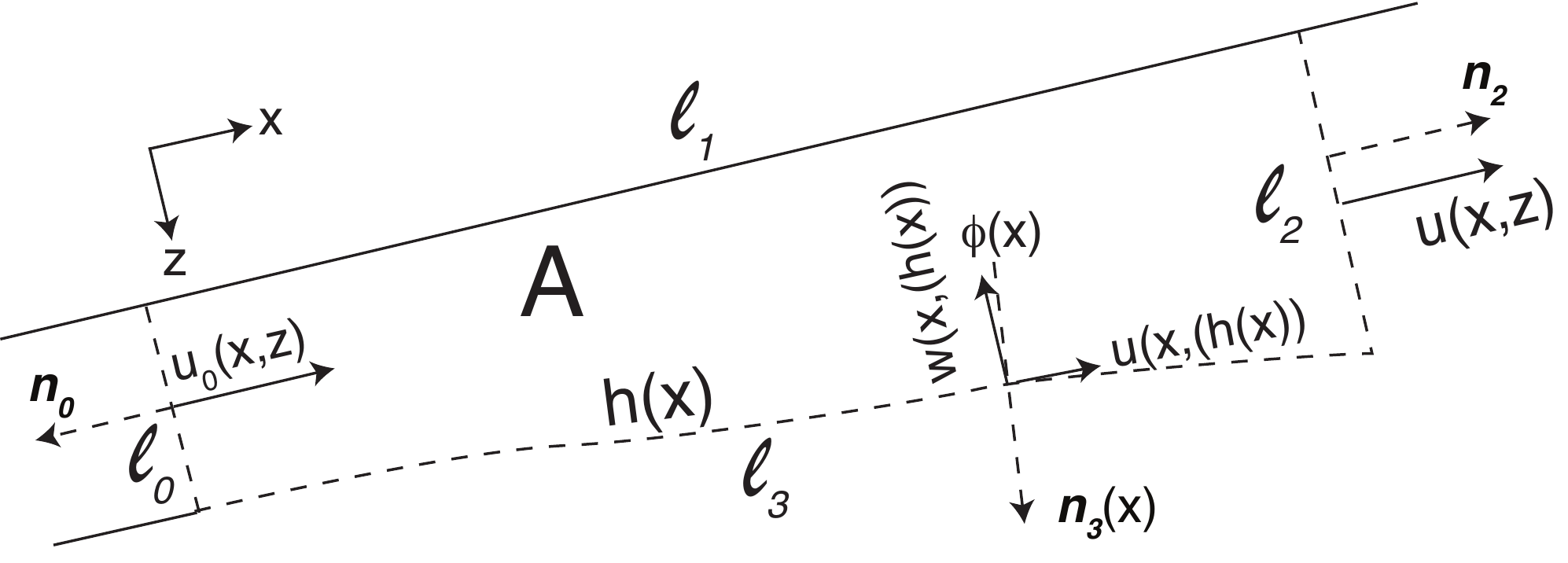}
\caption{\label{fig:control_volume} Schematics of the control volume (area) defined by the plate (${\cal \ell}_1$), two cross-sections perpendicular to the plate, ${\cal \ell}_0$ at $x=0$, ${\cal \ell}_2$ at $x$, and the bottom contour of the current, ${\cal \ell}_3$ at $z=h(x)$.}
\end{figure}

We base our analysis on the assumption that there is no cross stream transport and perform a 2-D analysis so that computed quantities are considered per unit length along the cross-stream direction. We use a control volume (actually a control area) defined in figure~\ref{fig:control_volume}.  The bottom limit is defined as a line of constant velocity $\la u\ra$ for the volume conservation and constant $\la\theta\ra$ for the dilution study, defining in each case a height $z=h(x)$, as shown in~figure~\ref{fig:control_volume}. The exact choice of the bottom contour will be discussed later.

\subsection{Volume flux conservation}\label{app:entr_vol}

Using the notation of figure~\ref{fig:control_volume}, the integration over the control volume of the incompressibility condition, $\bm{\nabla \cdot u}=0$,  gives :

\be\label{cons_vol}
\int_{{\cal \ell}_0}\bm{u\cdot n_0}\;\d {\bm{\ell}}_0+\int_{{\cal \ell}_2}\bm{u\cdot n_2}\;\d {\bm{\ell}}_2+\int_{{\cal \ell}_3}\bm{u\cdot n_3}\;\d {\bm{\ell}}_3=0
\ee

The normal vectors are defined in the following way: $\bm{n_0}=-\bm{x}$ and $\bm{n_2}=\bm{x}$. In addition, if $\phi$ is the angle between $\bm{n_3}$ and the downward pointing vector $\bm{z}$ perpendicular to the inclined plate, we have $\bm{n_3} = -\sin\phi\;\bm{x}+\cos\phi\;\bm{z}$ and $\bm{u} = u\;\bm{x}+w\;\bm{z}$, so that: $\bm{u\cdot n_3}=-u\sin\phi+w\cos\phi$. Time averaging~(\ref{cons_vol}), we get :

\be\label{equ:volume_flux_int}
-\int_0^H \la u(0,z)\ra \d z+\int_0^{h(x)}\la u(x,z)\ra\d z=\int_0^x \left(\la u\ra\sin\phi-\la w\ra \cos\phi\right)_{(s,h(s))} \d s,
\ee

\noindent where $s$ is a variable along the current bottom line defined as $z=h_c(s)$. Denoting the right hand side of~\eqref{equ:volume_flux_int} as $\left | W^d_v(x)\right| x$ and the left hand side as $\left | W^i_v(x)\right| x$, we obtain two estimates for $E=W/U$ based on volume conservation (\eqref{equ:direct_W} and \eqref{equ:indirect_W}). The estimation of entrainment velocity depends on the choice of the bottom contour of the control volume: if one chooses a contour such that the angle $\phi$ (angle between the normal to the contour and the vertical direction) is defined as $\tan \phi=w/u$, $\left | W^d_v(x)\right|$ vanishes by definition. In other words, such a choice would result in a contour everywhere tangent to the velocity, so that no fluid crosses this contour (on average). By continuity, a less steep contour yields a positive entrainment velocity, whereas a steeper contour yields a negative velocity.


\subsection{Scalar mixing}\label{app:entr_dil}

We now calculate entrainment based on scalar dilution and mixing. The advection-diffusion equation for $\theta$ is (\ref{equ:scal}), which, in steady state, neglecting molecular diffusion, and using $\bm{\nabla\cdot u}=0$, can be written as $\bm{\nabla\cdot}(\theta {\bm u})=0$. Again this equation can be integrated over the control volume, yielding:

\be\label{equ:dilution}
\int_{\ell_0}\theta\bm{u\cdot n_0}\;\d {\bm \ell_0}+\int_{\ell_2}\theta\bm{u\cdot n_2}\;\d \bm{\ell_2}+\int_{\ell_3}\theta\bm{u\cdot n_3}\;\d {\bm \ell_3}=0
\ee

As before, we assume no cross stream dependence and transform~\eqref{equ:dilution} into a single integral equation. Since time averaging is used, the Reynolds decomposition gives $\la{\theta u}\ra=\la\theta\ra\la u\ra+\la{\theta' u'}\ra$ and similarly for $\la{\theta w}\ra$ so we get:

$$
-\int_0^H \la\theta\ra\la u\ra\vert_{(0,z)}\d z+\int_0^{h(x)}\la\theta\ra\la u\ra\vert_{(x,z)}\d z+\int_0^x (-\la\theta\ra\la u\ra\sin\phi+\la\theta\ra\la w\ra\cos\phi)\vert_{(s,h(s))} \d s
$$
\be\label{dilution2}
+\int_0^x (-\la{\theta' u'}\ra\sin\phi+\la{\theta' w'}\ra\cos\phi)\vert_{(s,h(s))} \d s=0
\ee

\noindent The bottom limit of the control volume is defined as a line of constant $\la\theta\ra=\theta_c$ so that the factor $\la\theta\ra$ can come out of the integral of the third term of~\eqref{dilution2}. We then obtain:

\be\label{dilution4}
\int_0^{h(x)}\la\theta\ra\la u\ra\vert_{(x,z)}\d z-\int_0^H \la\theta\ra\la u\ra\vert_{(0,z)}\d z=\theta_c\vert W_v\vert x-\int_0^x (-\la{\theta' u'}\ra\sin\phi+\la{\theta' w'}\ra\cos\phi)\vert_{(s,h(s))} \d s
\ee

\noindent The left-hand terms of~\eqref{dilution4} represent the variation of the flux of dynamic density, between the injection upstream and any position downstream. In other words, it represents the mixing, which can have two different sources: the first right-hand term of~\eqref{dilution4} represents the contribution to mixing from the volume of fluid entrained in the current, causing a dilution of the initial scalar concentration. The second right-hand term contains the contribution to the mixing owing to turbulent fluctuations. Even in the absence of entrainment as defined in the previous subsection (no net variation of the volume flux), mixing takes place because of turbulent fluctuations.
This turbulent quantity is not present in the volume flux conservation equation (\ref{equ:volume_flux_int}) because there is no correlation term that remains after averaging. But, of course, the influence of the turbulence is implicitly present in~\eqref{equ:volume_flux_int}, since the velocity field is a result of the turbulent mixing of momentum, as represented by the Reynolds stress.

In~\eqref{dilution4}, the dominant contribution in the turbulent mixing term comes from the product $\la \theta'w'\ra$, which is related to ${\cal B}$ defined in section~\ref{subsec:energetics}. It is positive, corresponding to a sink term in the energy equation. Therefore, the turbulent mixing term in~\eqref{dilution4}, taking into account the minus sign in front of the integral, is negative. It thus contributes to decrease the flux of dynamic density since it provides a decrease of $\theta$ by mixing with no added volume flux. Its effect is then opposed to the increase of flux of dynamic density by the contribution of the term $\theta_c \vert W_v\vert x$. For any non zero value of $\theta_c$, one can then define an effective entrainment velocity based on scalar dilution, $W_s$, using either side of~\eqref{dilution4}. This yields definitions for $W^d_s$ and $W^i_s$ in~\eqref{equ:dilution_direct} and \eqref{equ:dilution_indirect}, respectively.
\clearpage
\bibliographystyle{jfm}
\bibliography{JFM09}

\end{document}